\documentclass[traditabstract]{aa}

\usepackage{graphicx}

\newcommand{\kms}{\ifmmode {{\rm km\,s^{-1}}}                 
                  \else {\hbox{{\rm km$\,$s$^{\rm -1}$}}}\fi}

\begin{document}

\title{Deep optical imaging of AGB circumstellar envelopes
  \thanks{Based on observations made at the European Southern
   Observatory, Chile (programs 078.D-0102, 082.D-0338 and
   0.84.D-0302) and on de-archived data obtained with the ESO Very
   Large Telescope  and with the NASA/ESA {\it Hubble Space Telescope}} }

\author{N. Mauron\inst{1}, P.J. Huggins \inst{2}, C.-L. Cheung\inst{2}}

\offprints{N. Mauron}

\institute{Laboratoire Univers et Particules de Montpellier, UMR 5299
 CNRS \& Universit\'e de Montpellier II, 
 Case CC72, Place Bataillon, F-34095 Montpellier Cedex 5, France
 \email{nicolas.mauron@univ-montp2.fr} 
\and
Physics Department, New York University, 4 Washington Place, New York
 NY 10003, USA} 

\date{Received xxx/  Accepted xxx}

\abstract {
We report results of a program to image the extended circumstellar
envelopes of asymptotic giant branch (AGB) stars in dust-scattered
Galactic light. The goal is to characterize the shapes of the
envelopes to probe the mass-loss geometry and the presence of hidden
binary companions. The observations consist of deep optical imaging of
22~AGB stars with high mass loss rates: 16 with the ESO 3.6~m NTT
telescope, and the remainder with other telescopes. The circumstellar
envelopes are detected in 15 objects, with mass loss rates $\ga
2\times 10^{-6}$~$M_{\odot}$\,yr$^{-1}$. The surface brightness of the
envelopes shows a strong decrease with Galactic radius, which
indicates a steep radial gradient in the interstellar radiation
field. The envelopes range from circular to elliptical in shape, and
we characterize them by the ellipticity ($E$ = major/minor axis) of
iso-intensity contours. We find that $\sim$50\% of the envelopes are
close to circular with $E \la 1.1$, and others are more elliptical
with $\sim$20\% with $E \ga 1.2$. We interpret the shapes in terms of
populations of single stars and binaries whose envelopes are flattened
by a companion. The distribution of $E$ is qualitatively consistent
with expectations based on population synthesis models of binary AGB
stars. We also find that $\sim$50\% of the sample exhibit small-scale,
elongated features in the central regions. We interpret these as the
escape of light from the central star through polar holes, which are
also likely produced by companions.  Our observations of envelope
flattening and polar holes point to a hidden population of binary
companions within the circumstellar envelopes of AGB stars. These
companions are expected to play an important role in the transition to
post-AGB stars and the formation of planetary nebulae. }


\titlerunning{Imaging AGB circumstellar envelopes}
\authorrunning{N. Mauron et al.}
\maketitle
\section{Introduction}


\begin{table*}[!t]
\caption[]{Properties of the AGB stars}

	\begin{center}
        \begin{tabular}{llrrlrllrrl}
        \noalign{\smallskip}
        \hline
	\hline
        \noalign{\smallskip}

 IRAS & other name &   $l$ & $b$  & Chem. & $d$~~  & $V_{\rm exp}$ &
      ~~~$\dot{M}$ & $f_{12}$~ & $f_{12}/f_{25}$ & Refs. \\
      &           &        &      &     & (pc) & (\kms) &(10$^{-5}$
      M$_{\odot}$\,yr$^{-1}$) &  (Jy)  &                 &\\  
\noalign{\smallskip}
\hline
\noalign{\smallskip}
\noalign{\smallskip}

 00193$-$4033 &\object{AFGL 5017} & 326 & $-$75    & O &  360 & 17.3 &
 0.10    & 312  & 2.07 &  1 \\

 01037$+$1219 & \object{IRC+10011} &  128 & $-$50    & O &  740 & 21.6
 & 2.2   & 1155 & 1.19 &  2 \\ 

 04575$+$1251 &  \object{AFGL 5134} & 188 & $-$18 & O & 1200 & 12.9 &
 1.0  & 133 & 1.09 &  3, 4, 9\\

 05418$-$3224 &  & 237 & $-$27 & C & 4300 & 10.0 & 1.1  & 63 & 2.47 &
 5 \\

06012$+$0726 &\object{AFGL 865}  & 201 & $-$07 &  C & 1470 & 16.6 &
1.8 & 320 & 1.42 &  5 \\

06255$-$4928 &  & 258 & $-24$ & O & 1100  & 15.0 &  0.31  & 66 & 1.53
&  4, 9, 10 \\

 07454$-$7112 & \object{AFGL 4078}  & 283 & $-$21   & C &  940   &
 13.5     & 1.6    & 613  & 1.99 &  5 \\ 

 09116$-$2439 & \object{AFGL 5254} & 253 & $+$16   & C & 940  & 12.8 &
 1.3  & 737  & 1.85 &   5 \\

 09452$+$1330 &\object{IRC+10216}  &  221 & $+$45  & C  & 120  & 14.7
 & 1.75  & 47500 & 2.06 &  6 \\ 

 09521$-$7508 &\object{AFGL 4098}  & 292 & $-$16 & C & 1300 & 12.8 &
 1.1  & 345 & 1.86 &  5 \\ 

 10323$-$4611 &  & 280 & $+$10  & O  & 580 &  20.4 &  1.0   & 537  &
 1.46 & 4, 9\\ 

 12384$-$4536  &  & 301 & $+$19   & O & 1100 & 8.1  &  0.20  & 159  &
 2.25 & 1 \\ 

14591$-$4438   &  & 326 & $+$12   & O  &  600 & 15.2 & 0.25  & 304 &
1.47    & 1 \\

 16029$-$3041 & \object{AFGL 1822} & 345 & $+$16     &  O & 1900 &
 16.8 & 1.10  & 141 & 0.53 & 1 \\ 

 17319$-$6234 & \object{OH 329.8$-$15.8} &  330 & $-$16 &  O & 1200 & 17.5 &
 0.84  & 252 & 0.85 &  1 \\ 

 18240$+$2326 & \object{AFGL 2155} & 52 & $+$16   & C & 920 & 15.1 &
 0.46  &  731  & 1.63 &  5 \\ 

 18467$-$4802 & \object{OH\,348.2$-$19.7} & 348 & $-$20& O & 1200 & 12.3 & 0.78
 &  285  &0.83&  1 \\ 

 19178$-$2620 & \object{AFGL 2370} & 12 & $-$18   & O & 2200 & 18.0 &
 1.3 &   78  & 0.81 &  1 \\ 

 20042$-$4241 & \object{AFGL 5578} &  358 & $-$31 & O & 460 & 14.4 &
 0.15  & 221 & 1.41 &  7, 8 \\

 20077$-$0625 & \object{AFGL 2514} &   36 & $-$20   & O &  760 & 16.0
 & 1.7  & 1255  & 1.18 &  2 \\ 

 23257$+$1038 & \object{AFGL 3099}  &  92 & $-$47   & C & 1610 & 10.5
 & 0.85  &  190  & 1.34 &  5 \\ 

 23320$+$4316  & \object{AFGL 3116} &  108 & $-$17   & C &  780 & 14.7
 & 0.70   &  959  & 2.04 &  5 \\ 

\noalign{\smallskip}
\noalign{\smallskip}
\hline
\end{tabular}
\end{center}
{\small Refs:\\
(1) $d$ , $V$, and $\dot{M}$  from Loup et al.\ (\cite{loup93})  \\ 
(2) $d$, $V$, and $\dot{M}$ from Olivier et al.\ (\cite{olivier01}) \\
(3) $V$ from OH observations, David et al.\ (\cite{david93}) \\
(4) $\dot{M}$ estimated through a correlation of $\dot{M}$ with the $K
- $[12] color, where $K$ is from the 2MASS catalog, and [12] = $-$2.5
log($f_{12}/28.3$), with $f_{12}$ from the IRAS point-source catalog \\
(5) $d$ , $V$, and $\dot{M}$ from Groenewegen et al.\
(\cite{groenewegen02}) \\
(6) $d$ from Ramstedt et al.\ (\cite{ramstedt08}), $\dot{M}$ from
Sch\"{o}ier \& 
Olofsson (\cite{schoier01}), $V_{\rm exp}$ from Loup et al.\ (\cite{loup93}) \\
(7) $V$ from Loup et al.\ (\cite{loup93}) \\
(8) $d$ and $\dot{M}$ from Whitelock et al.\ (\cite{whitelock94})\\
(9) $d$ estimated assuming $L=3500$~$L_\odot$ (e.g., Jackson et al.\ \cite{jackson02})\\
(10) No $V_{\rm exp}$ available, 15~\kms\ adopted value
}
\end{table*}


It is well established that stars lose a significant fraction of their
mass during evolution on the asymptotic giant branch (AGB).  The mass
loss forms expanding envelopes of molecular gas and dust around the
stars, and eventually recycles the stellar material back into the
interstellar medium. The circumstellar envelopes are important for the
study of many astrophysical processes, including circumstellar
chemistry and dust formation. They are also important for stellar and
galactic evolution, especially in connection with the formation of
planetary nebulae.  For reviews on AGB envelopes and their properties
see, e.g., Habing \& Olofsson (\cite{habing04}) and Kerschbaum et al.\ (\cite{kershbaum11}).

Despite extensive study, there are many aspects of AGB envelopes that
are not well understood. Here we focus on one of the most basic
properties, the geometry of the envelopes. On large size scales the
morphology is known to be affected by interaction with the interstellar medium 
(e.g. Cox et al. 2012), but the shaping mechanisms close to the star are not 
well understood. It is not known if the mass
loss of single stars is closely spherically symmetric, or if it is
influenced by factors such as the stellar magnetic field. Binary
companions are expected to influence the geometry, but information on
the presence of companions is problematic. Apart from symbiotic
systems, only a few dozen AGB binaries are known (Jorissen \cite{jorissen08}), even
though they are expected to be common. The reason so few are detected
is because of the luminosity and variability of the AGB stars, and the
opacity of their circumstellar envelopes. There must be a large
population of hidden companions that can interact with the envelopes.

The effects that stellar companions can have on AGB envelopes is
suggested by hydrodynamical simulations (Theuns \& Jorissen \cite{theuns93};
Mastrodemos \& Morris \cite{mastrodemos99}; Gawryszczak \cite{gaw02}). There are two main
interactions.  First, the reflex motion of the AGB star forms a spiral
pattern in the envelope, and this has been detected in the case of
AFGL 3068 (Mauron \& Huggins \cite{mh06}, hereafter MH06; Morris et al.\
\cite{morris06}). Second, the large scale geometry of the envelope is expected to
be flattened by a companion, and this has been discussed by Huggins et
al.\ (\cite{huggins09}) as a possible probe of the presence of binary
companions. Indirect shaping effects on the envelopes by jets launched
from accretions disks around a companion are also possible, as seen in
the case of Mira (Josselin et al.\ \cite{josselin00}).

There is a lack of detailed observational information on the structure
of circumstellar envelopes because the relatively cool material in the
envelopes is difficult to image at high resolution.  A review of
techniques is given by Marengo (\cite{marengo09}).  Recent advances include
millimeter interferometry of the molecular gas (e.g., Castro-Carrizo
et al. \cite{castro10}) and imaging of the thermal dust emission, in the
mid-infrared with ground-based observations (e.g., Lagadec et al.\
\cite{lagadec11}) 
and in the far infrared with Herschel (Decin et al.\ \cite{decin11}). In this
paper, we report progress using a different technique: observing the
envelopes in dust-scattered Galactic light. This technique makes use
of deep optical imaging, so that high resolution is more readily
attained than at longer wavelengths.

Our initial observations using deep optical imaging produced the first
wide field image of the circumstellar envelope of the AGB archetype
IRC+10216 at arcsecond resolution (Mauron \& Huggins \cite{mh99}, \cite{mh00}).
The core of the envelope is illuminated by light from the central
star, but the extended envelope is illuminated by the ambient Galactic
radiation field, which reveals the detailed structure. We subsequently
carried out pilot observations of other, more distant AGB stars, and
found that the technique can provide useful information on the shapes
of the circumstellar envelopes (MH06). In this paper, we report more
extensive observations using the New Technology Telescope (NTT) at the
European Southern Observatory at La Silla (Chile), with arcsecond
resolution and exposures up to several hours.  The goal is to extend
the sample of AGB stars with resolved envelopes to determine their
geometry.

Section 2 describes the observations; Sect.~3 presents the images
and information on the shapes, together with those of our earlier
observations in the same format; and Sect.~4  discusses the
results and their implications.


\begin{table}[!th]
\caption[]{Details of NTT observations}

	\begin{center}
        \begin{tabular}{lllllll}
        \noalign{\smallskip}
        \hline
	\hline
        \noalign{\smallskip}

 IRAS & Instr.$^a$ & Band \& Exposure &  $\Delta \theta_\mathrm{im}^b$ \\
      &            &    (min)       &     (\arcsec)   \\

\noalign{\smallskip}
\hline
\noalign{\smallskip}
\noalign{\smallskip}

00193$-$4033  & II & $B$ (210), $V$ (150), $I$ (5)  & 1.9 ($V$) \\

04575$+$1251 & II & $B$ (150), $V$ (90), $I$ (1)   & 1.2 ($V$) \\

05418$-$3224 & II & $B$ (60), $I$ (20s)          & 2.2 ($B$) \\

06012$+$0726 & II & $V$ (42), $I$ (10)           & 1.6 ($V$) \\

06255$-$4928  & II & $B$ (70), $I$ (20s)          & 2.2 ($B$) \\

07454$-$7112  & II & $B$ (210), $V$ (30)          & 1.3 ($B$) \\

09116$-$2439  & I  & $B$ (30), $I$ (6)$^c$        & 1.0 ($B$) \\

09521$-$7508  & I  & $B$ (30)                   & 1.1 ($B$) \\

10323$-$4611  & II & $B$ (120), $V$ (30)          & 1.0 ($B$) \\

12384$-$4536  & II & $B$ (270), $V$ (30)          & 1.1 ($B$) \\

14591$-$4438  & II & $B$ (90)                   & 1.1 ($B$) \\

16029$-$3041  & II & $B$ (30), $V$ (180)           & 1.2 ($V$) \\

17319$-$6234  & II & $B$ (60), $V$ (30)            & 1.0 ($V$) \\

19178$-$2620  & II & $B$ (240), $V$ (90), $I$ (10s) & 1.7 ($B$) \\

20042$-$4241  & I  & $B$ (60)                   & 1.4 ($B$) \\

23257$+$1038  & II & $V$ (510), $I$ (6)           & 1.6 ($V$) \\

\noalign{\smallskip}
\noalign{\smallskip}
\noalign{\smallskip}
\hline
\end{tabular}
\end{center}
{\small $^a$ I = EMMI, II = EFOSC2 \\
$^b$ FWHM of field stars in the images  \\
$^c$ $I$-band image obtained with EFOSC2  \\
}

\end{table}

\section{Observations}

\subsection{The sample}
The deep imaging technique is well suited for investigating the shapes
of the circumstellar envelopes of AGB stars with high mass loss
rates. The thickest envelopes produce the maximum surface brightness
and the largest envelope sizes for observation in dust scattered
Galactic radiation.  The high opacity of these envelopes also reduces
or eliminates the stellar light contribution, which may be affected by
structure close to the star and can mask the faint, external illumination.

The AGB targets for observations were selected from the catalog of
Loup et al.\ (\cite{loup93}), on the basis of strong IRAS fluxes and large
infrared excesses (small $f_{12}/f_{25}$ ratios), as well as
declinations accessible from La Silla. Preference was given to stars
at relatively high galactic latitude, in order to minimize the effects
of interstellar extinction and crowded star fields.

Details of all the AGB stars discussed in the paper are given in
Table~1. The list includes new targets observed with the NTT (Section
2.2) and earlier observations which we include for completeness
(Section 2.3). Table~1 gives the name, the Galactic co-ordinates, the
chemistry (oxygen-rich or carbon-rich), distance $d$, expansion
velocity $V_{\rm exp}$, mass-loss rate $\dot{M}$, and relevant IRAS
fluxes from the point source catalog. It can be seen that they
represent AGB stars in the solar neighborhood (within a few kpc)
with high mass loss rates, comparable in many cases with IRC+10216.

\subsection{NTT observations}

The observations with the NTT were carried out in four observing
runs. Two were made with the EMMI instrument in imaging mode on 12--14
October 2006 and 16--18 March 2007, and two were made with the EFOSC2
instrument on 20--23 February 2009 and 15--19 October 2009.  The
angular field is $5\farcm7 \times 5\farcm7$ with a pixel size of
$0\farcs3608\pm 0\farcs0003$ for EMMI, and $4\farcm1 \times 4\farcm1$
with a pixel size of $0\farcs2406\pm0\farcs0002$ for EFOSC2. The first
two observing runs were affected by poor weather, so most of the data
reported are with EFOSC2, which also has greater sensitivity and
better sampling of the PSF.


\begin{table}[!t]
\caption[]{Observations from earlier work}

	\begin{center}
        \begin{tabular}{lllllll}
        \noalign{\smallskip}
        \hline
	\hline
        \noalign{\smallskip}

IRAS  & Tel. & Band \& Exposure &  $\Delta \theta_\mathrm{im}^a$ \\
      &            &    (min)       &   (\arcsec)     \\

\noalign{\smallskip}
\hline
\noalign{\smallskip}
\noalign{\smallskip}

01037$+$1219 & VLT  & $U$ (4), $B$ (2), $V$ (2) & 1.1 ($V$)  \\ 
             & HST  & $F814W$ (11.3)        & 0.12 ($F606W$)  \\
09452$+$1330 & VLT  & $V$ (120)             & 0.7 ($V$)  \\
18240$+$2326 & OHP  & $V$ (300)            & 2.5 ($V$)  \\
18467$-$4802 & VLT  & $U$ (8), $B$ (2), $V$ (2) & 1.1 ($V$)  \\
             & HST  & $F606W$ (11.6)        & 0.12 ($F606W$) \\
20077$-$0625 & DT   & $V$ (120)           & 2.3 ($V$)  \\
             & HST  & $F606W$ (11.6)        & 0.12 ($F606W$)\\
23320$+$4316 & OHP  & $B$ (150), $V$ (60)     & 2.1 ($B$)  \\

\noalign{\smallskip}
\noalign{\smallskip}
\noalign{\smallskip}
\hline
\end{tabular}
\end{center}
{\small $^a$ FWHM of field stars in the images
}

\end{table}

Details of the observations including the instrument, the filters, and
the exposure times are listed in Table~2.  In many cases, we first
secured an $I$-band exposure to ascertain the precise position of the
target. The deep $B$ or $V$ images to detect the faint envelopes were made
with a series of individual exposures, typically of 30~min each, with
a small shift in $\alpha$ or $\delta$ between them to improve the
quality of the final image. The data were reduced using standard
procedures with the MIDAS software: bias subtraction, flat-fielding
with flats made on the dome or the sky, and the removal of cosmic-rays
hits.  The individual frames were then shifted and stacked. In some
cases, we corrected residual gradients in the background level using
division by a normalized polynomial of order 1 or 2.

The photometric calibration of the EFOSC2 observations was made using
photometric sequences close to the star T~Phe and in the field SA
98-642 (Snodgrass \cite{snodgrass09}). For these observations, the relation between
magnitudes and ADU counts is found to be:
\[B=25.80 -0.20 X -2.5 \log (F/t) \]
for the $B$ band, and:
\[V=25.86 -0.11 X -2.5 \log (F/t) \]
for the $V$-band, where $X$ is the air-mass, and $t$ is the exposure
in seconds. For each relation, $B$ or $V$ is the stellar magnitude if $F$
is the spatially integrated ADU count for the star, or the
surface brightness if $F$ is the ADU count per arcsec$^2$ measured on
an extended object.  The coefficients of $X$ are the usual extinction
coefficients for photometric nights at La Silla: the quoted values are
from the documentation of the 1.54 Danish telescope. Overall the
calibrations are expected to have an uncertainty of 0.1 mag, and the
results are found to be consistent within 0.05 mag when comparing the
two EFOSC2 runs.

The photometric calibration for the EMMI observations was made by
considering un-saturated stars and their $B$-band magnitudes taken from
the USNO-B1.0 catalog (Monet et al.\ \cite{monet03}); in this case the
photometric uncertainty is $\sim$ 0.4 mag (1-$\sigma$). Comparisons of
the calibration with the USNO catalog and the EFOSC2 photometric
calibration are consistent with this.

The envelopes illuminated by the ambient Galactic radiation field are
faint, typically a few percent of the sky background. Hence the
limiting factor for envelope detection is the sky noise. The exposures
reported here are relatively long, and all the runs were made in dark
time. The results are therefore representative of the quality of the
data that can be obtained with this technique, using an intermediate
sized, ground-based telescope.


\begin{table*}[!th]
\caption[]{Observations of the circumstellar envelopes}

	\begin{center}
        \begin{tabular}{llllccccccl}
        \noalign{\smallskip}
        \hline
	\hline
        \noalign{\smallskip}

 IRAS & {Tel.$^a$} &
      \multicolumn{2}{l}{Detection$^b$}  &
      \multicolumn{2}{c}{Envelope S.B.$^c$} &
       Ellipticity  &   P.A. &
      $a$ & $ t_\mathrm{exp}$ & Core$^d$\\
      &  &  &  &  ($B/V$) & ($B_0$) &
      ($a/b$) & ($\degr$)& (\arcsec) & (yr) \\

\noalign{\smallskip}
\hline
\noalign{\smallskip}
\noalign{\smallskip}

00193$-$4033  & NTT & $\ldots$ & $\ldots$ & $<$27.7
 & $<$27.6 & $\ldots$  & $\ldots$ & $\ldots$ & $\ldots$ & $\ldots$ \\

01037$+$1219  & VLT & st &  env   & 24.3$^e$ & 25.2 &
$1.07\pm0.03$  & 89 & 5.2  & 840 & pc \\

04575$+$1251  & NTT & $\ldots$ & $\ldots$ & $<$27.2
 & $<$25.5 & $\ldots$ & $\ldots$ &   $\ldots$ & $\ldots$ & $\ldots$ \\

05418$-$3224  & NTT &  st & $\ldots$ & $<$26.9$^f$ & $<$26.8
& $\ldots$ & $\ldots$ &  $\ldots$ & $\ldots$ & $\ldots$ \\

06012$+$0726  & NTT & $\ldots$     & $\ldots$ &
$<$25.6$^e$ & $<$24.7 &  $\ldots$ &  $\ldots$ &  $\ldots$ & $\ldots$ & $\ldots$ \\

06255$-$4928  & NTT & st    &$\ldots$  & $<$26.8$^f$ & $<$26.6
&  $\ldots$ &  $\ldots$ &  $\ldots$ & $\ldots$ & $\ldots$ \\

07454$-$7112  & NTT & st  & env & 25.8 & 25.1 & $1.12\pm
0.03$ & 5 & 3.9 & 1290 & pc \\

09116$-$2439  & NTT & $\ldots$     &  env & 26.7 & 25.9 &
$1.11\pm0.04$ & 47 & 4.6 & 1600 & $\ldots$\\

09452$+$1330  & VLT & st  & env    & 25.0$^g$ & 25.6 &  $1.03\pm
0.02$  &  150   &  45.8 & 1770 &  pc\\ 

09521$-$7508  & NTT & st  & $\ldots$ & $ <$26.6$^f$
& $<$25.3 & $\ldots$  &  $\ldots$  &  $\ldots$ &  $\ldots$ & $\ldots$ \\

10323$-$4611  & NTT & st & env & 25.9 & 25.0 &
$1.20\pm0.04$   & 12 & 4.1  & 550 & pc   \\

12384$-$4536  & NTT & st &  env & 25.8 & 25.3 &
$1.14\pm0.03$ & 38 & 3.6 & 2320 & $\ldots$ \\

14591$-$4438  & NTT & $\ldots$     & env  & 25.5 & 24.8 &
$1.10\pm0.03$ & 124 & 2.7 & 420 & $\ldots$ \\

16029$-$3041  & NTT & $\ldots$     & env  & 24.6$^g$ & 23.8
& $1.07\pm 0.02$ & 41 & 8.7 & 4660 & $\ldots$\\

17319$-$6234  & NTT & $\ldots$    & env & 24.2$^g$ & 23.9 &
$1.05\pm 0.02$ & 65 & 8.5 & 2760 & $\ldots$ \\

18240$+$2326  & OHP &   $\ldots$   & env  & 25.3$^e$ & 25.6  &
$1.14\pm 0.03$ & 46 & 5.0 & 1440 & $\ldots$\\

18467$-$4802  & VLT$^h$ & $\ldots$& env    & 23.7$^e$ & 24.5 
& $1.03 \pm 0.02$ & 113   & 5.0  & 2310 & $\ldots$ \\

19178$-$2620  & NTT & st & env & 24.5 & 24.0 &  $1.29\pm0.03$ & 156 
& 5.4 & 3130 & pc\\

20042$-$4241  & NTT & st  &$\ldots$     &  $<$27.6$^f$
  & $<$27.4 &  $\ldots$ &  $\ldots$ &  $\ldots$ & $\ldots$ & $\ldots$ \\ 

20077$-$0625  & DT  &  $\ldots$ & env   & 23.6$^e$ & 24.3 &
$1.28\pm 0.02$ &  58   & 14.7 & 3310  &  pc\\

23257$+$1038  & NTT &  $\ldots$ & env & 25.4$^e$ & 25.9 &  $1.01\pm 0.03$ &
 157 & 5.2 & 3780 &  pc\\

23320$+$4316  & OHP &  st & env & 26.8 & 26.2 & $1.13\pm 0.06$
& 140   & 7.0 & 1760 &  pc \\ 

\noalign{\smallskip}
\noalign{\smallskip}

\noalign{\smallskip}
\hline
\end{tabular}
\end{center}
{\small 
$^a$ Telescope used for photometry and imaging shown in Figs.~1--6 and
  Fig.~A.1\\ 
$^b$ Detections in $V$ or $B$: st\,=\,star, env\,=\,envelope \\ 
$^c$ Surface brightness in mag. arsec$^{-2}$ in $B$ or $V$ (designated in
  footnote) and effective $B_0$ (see text) \\
$^d$ Core: pc = polar core \\
$^e$ $V$-band measurement \\
$^f$ Nominal limit based on extended source because of bright central star \\
$^g$ $V$ surface brightness also measured: see section 3.3  \\
$^h$ VLT used for photometry, HST/ACS used for imaging \\
}

\end{table*}


\subsection{Additional observations}
For completeness, we include in the analysis the observations from our
earlier report (MH06) which are of comparable quality to the NTT
data. These observations are summarized in Table~3, and have been
processed in the same way.

The observations include short exposures with the Very Large
Telescope (VLT), longer exposures with the 1.54~m ESO Danish Telescope
(DT) and the 1.20~m telescope of the Observatoire de Haute-Provence
(OHP), and archival images made with the HST.  The VLT-FORS raw images
and calibration files were extracted from the ESO archive. The $V$-band
image of IRC+10216 was first shown by Leao et al.\ (\cite{leao06}). The HST
data are from programs 9463 and 10185, PI: R.~Sahai. Further details
of the observations are given in MH06.  Observations in MH06 but not
in Table~3 include several non-detections which are much poorer than
the NTT data, and AFGL~3068. In AFGL~3068, the illumination is highly
directional and the envelope shows a strong spiral pattern, which both
limit the ability to measure the overall shape.


   \begin{figure*}
   \setlength\fboxsep{0pt}
   \setlength\fboxrule{0.5pt}

  \vspace{0.3cm}
   \centerline{
\fbox{\includegraphics[width=0.27\linewidth]{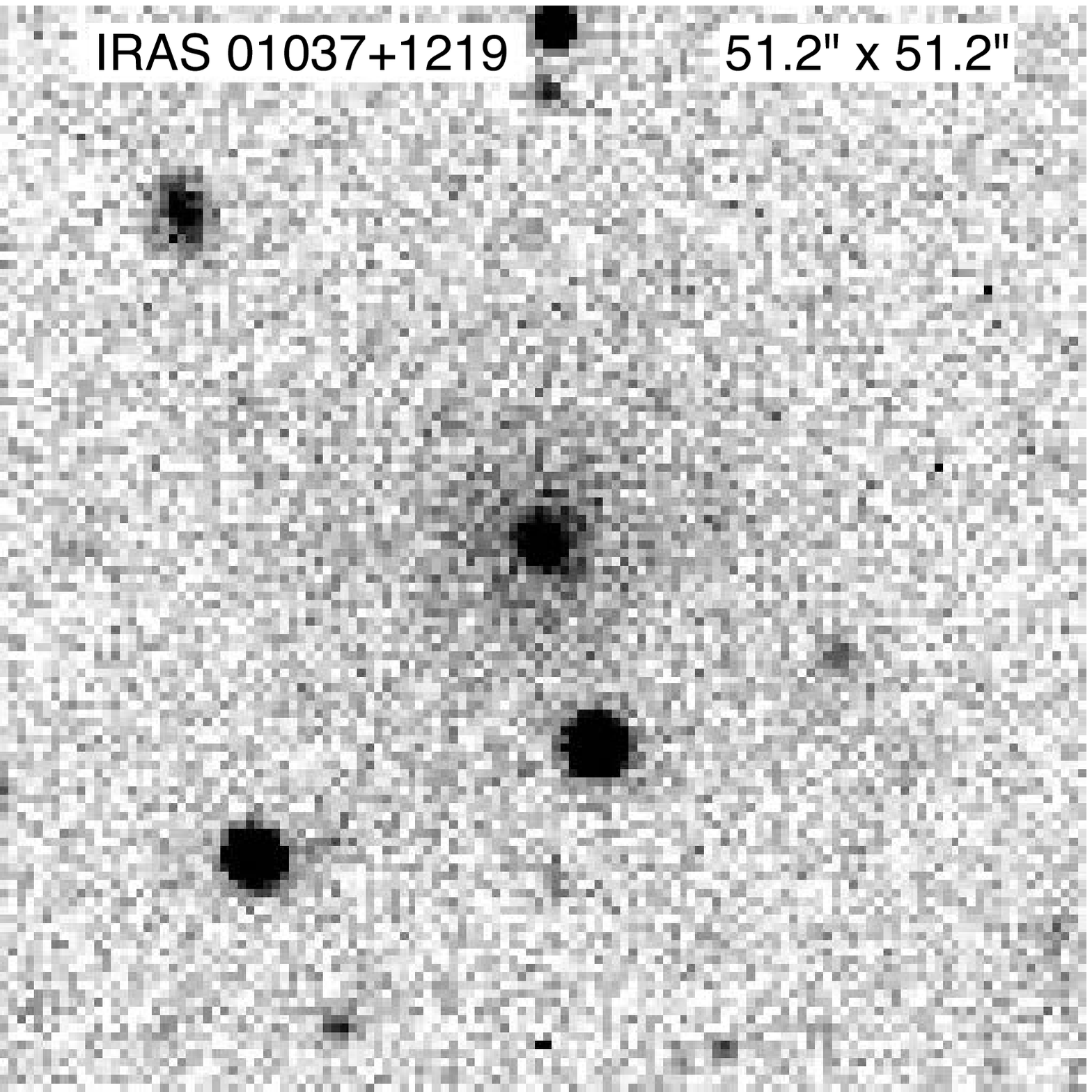}}\hspace{0.3cm}       
\fbox{\includegraphics[width=0.27\linewidth]{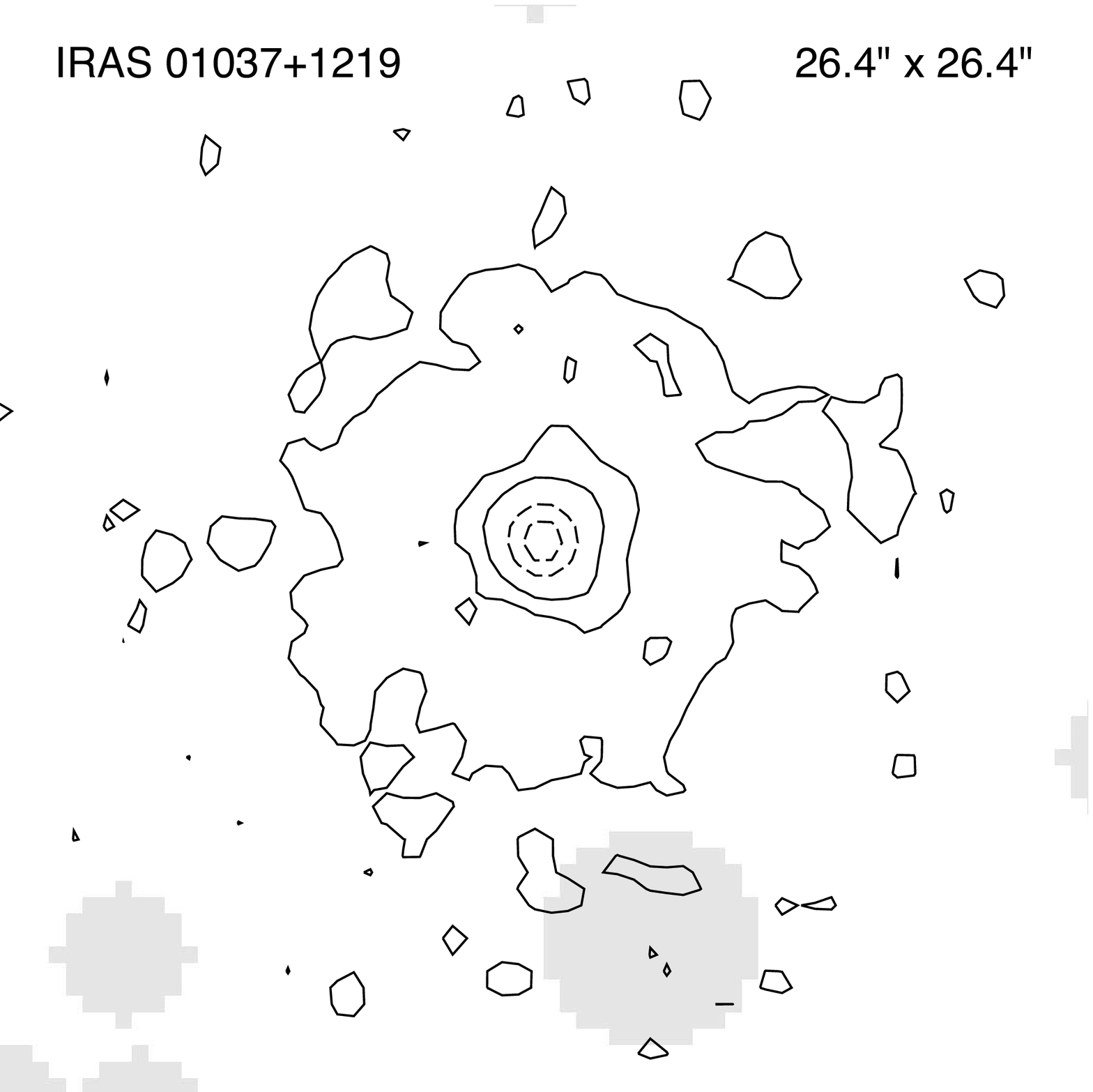}}\hspace{0.3cm}
\includegraphics[origin=c,angle=-90,width=0.27\linewidth,clip=true]{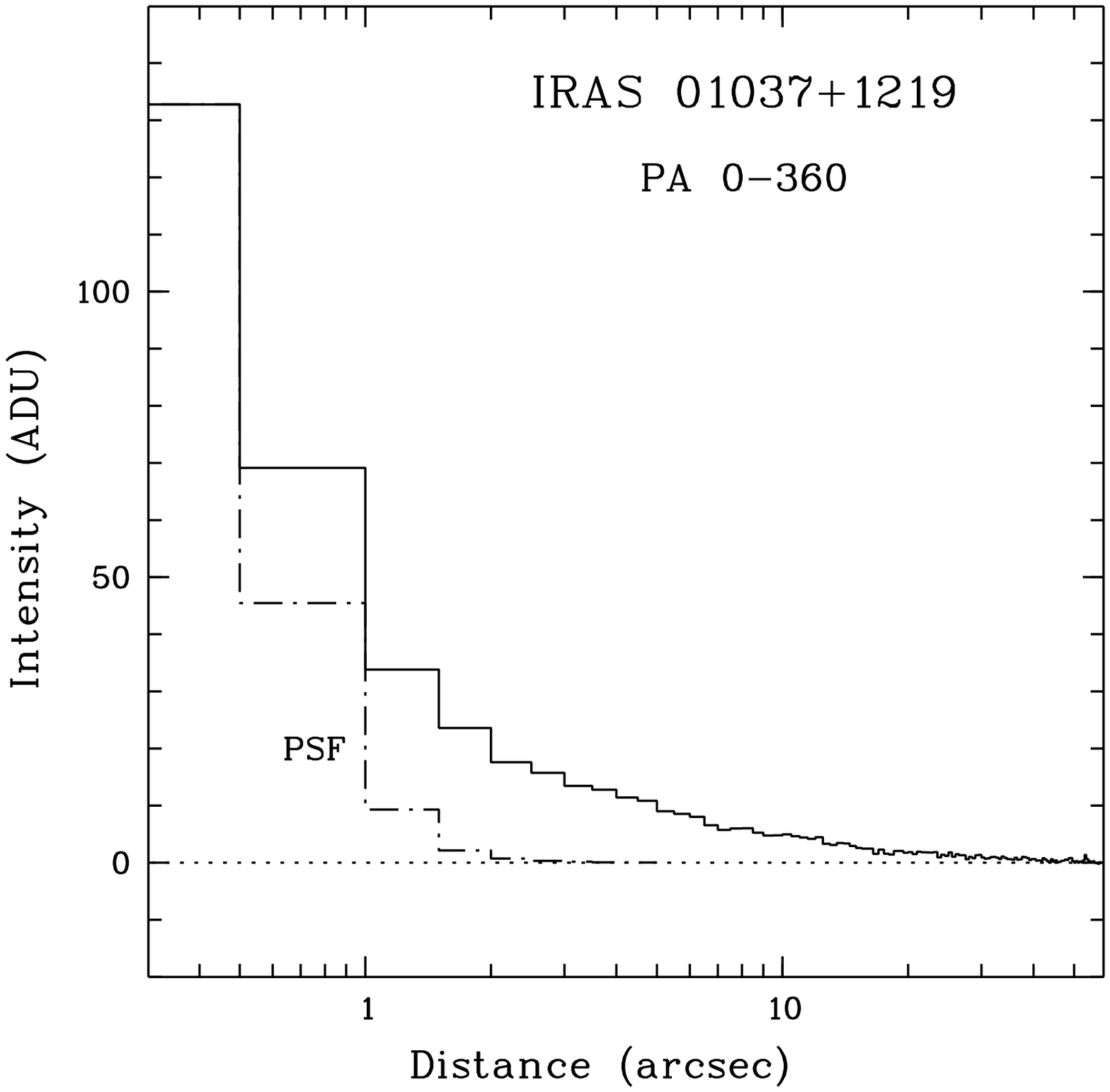}}       

  \vspace{0.6cm}
  \centerline{
\fbox{\includegraphics[width=0.27\linewidth]{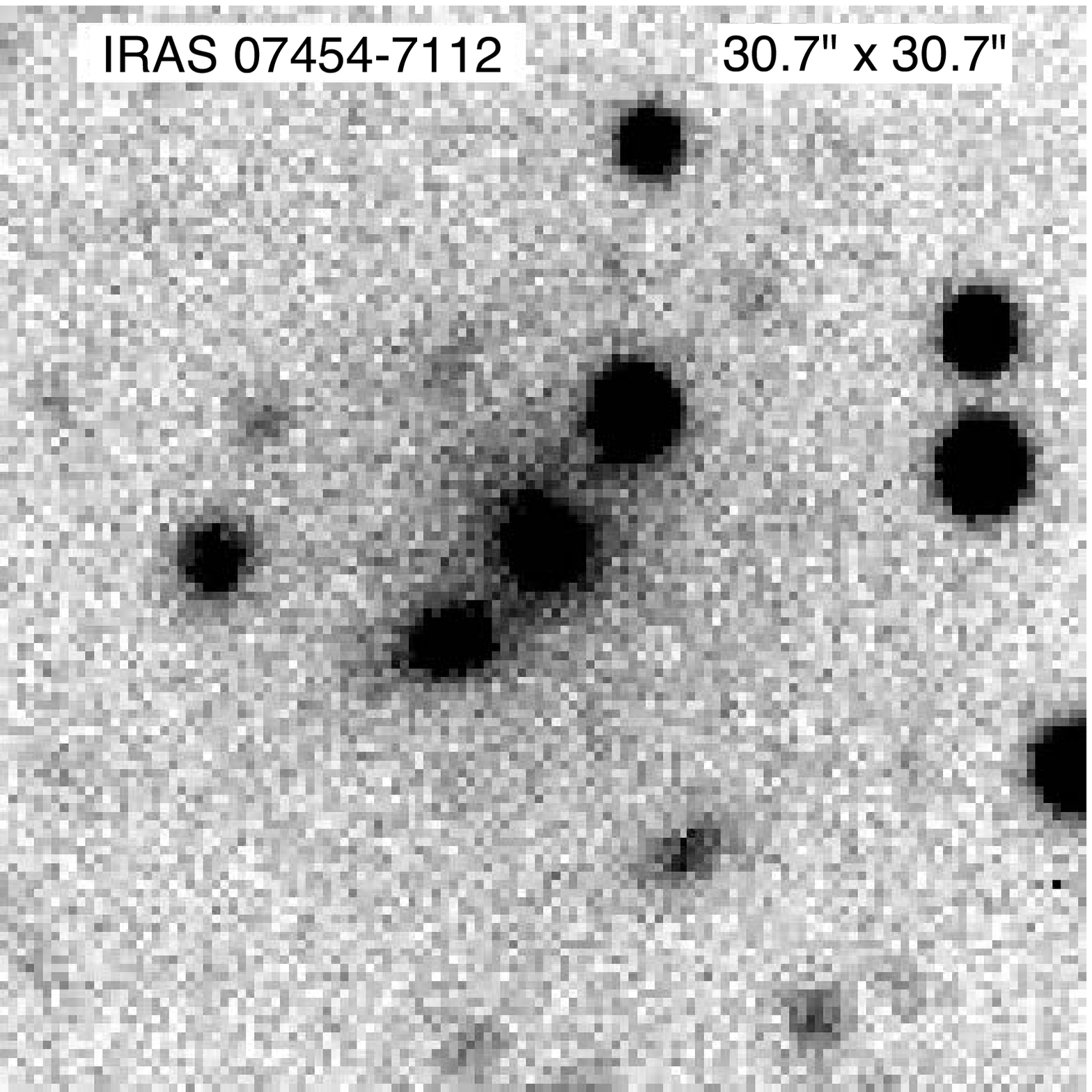}}\hspace{0.3cm}
\fbox{\includegraphics[width=0.27\linewidth]{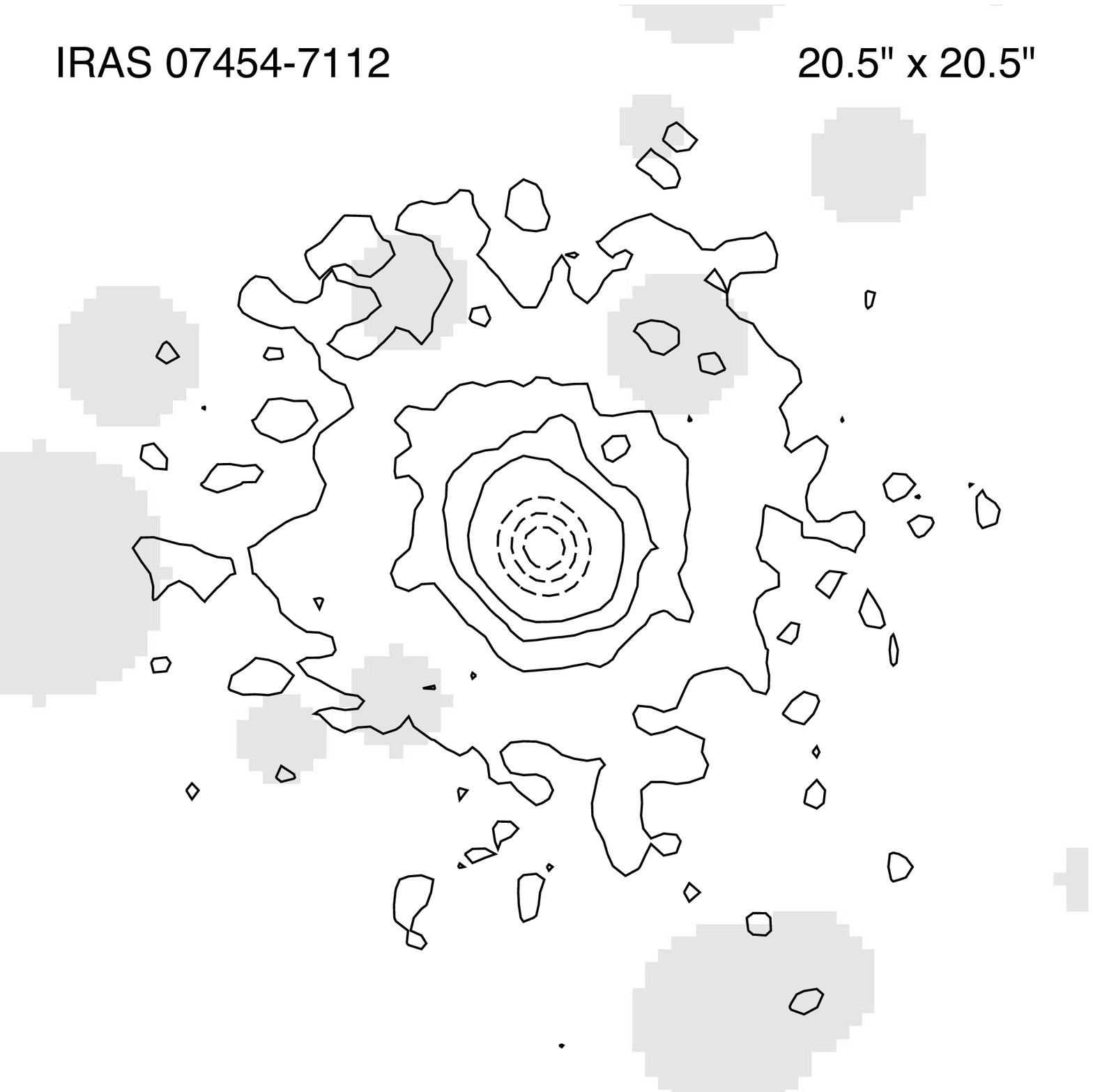}}\hspace{0.3cm}
\includegraphics[origin=c,angle=-90,width=0.27\linewidth,clip=true]{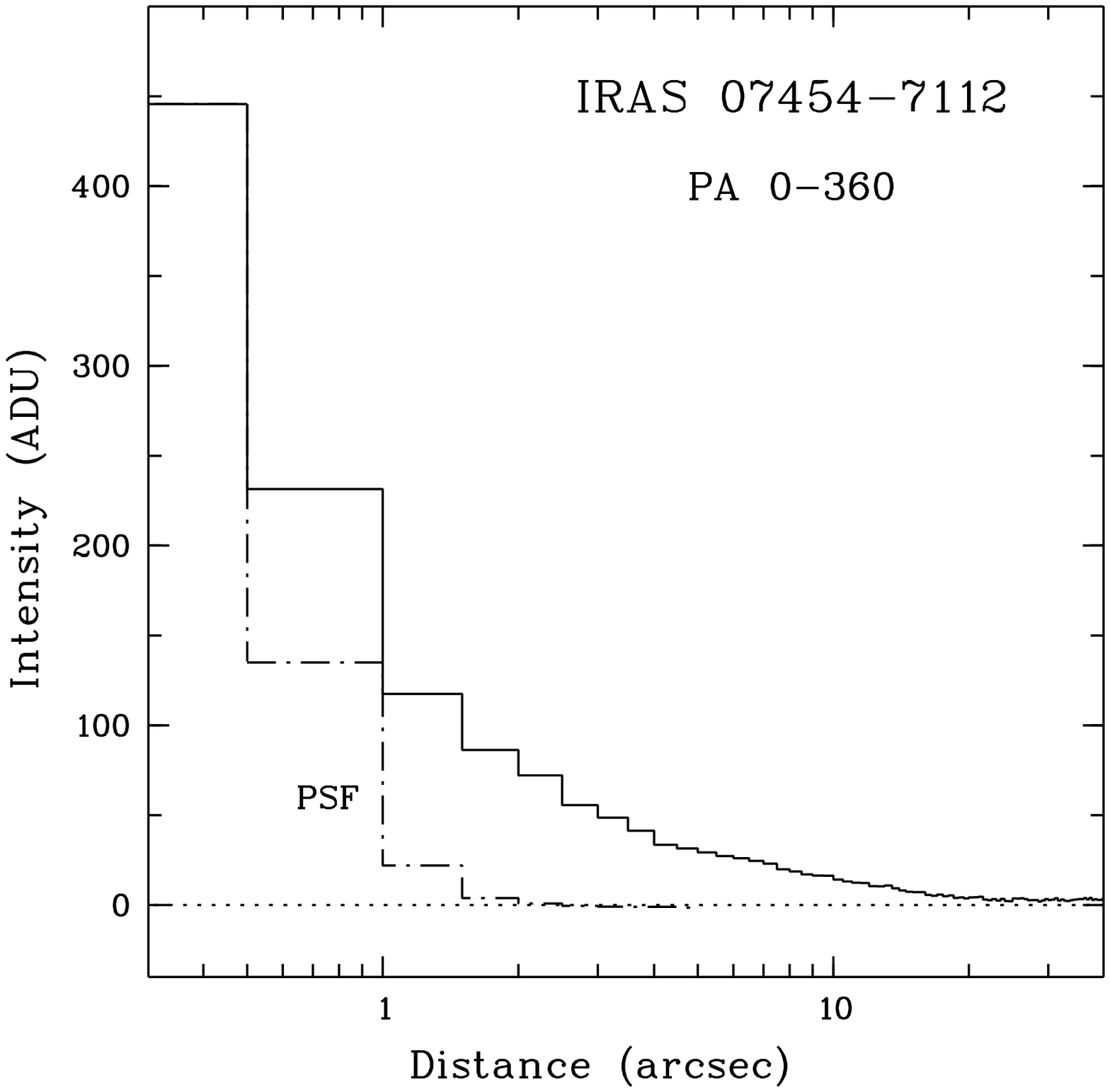}}

  \vspace{0.6cm}
  \centerline{
\fbox{\includegraphics[width=0.27\linewidth]{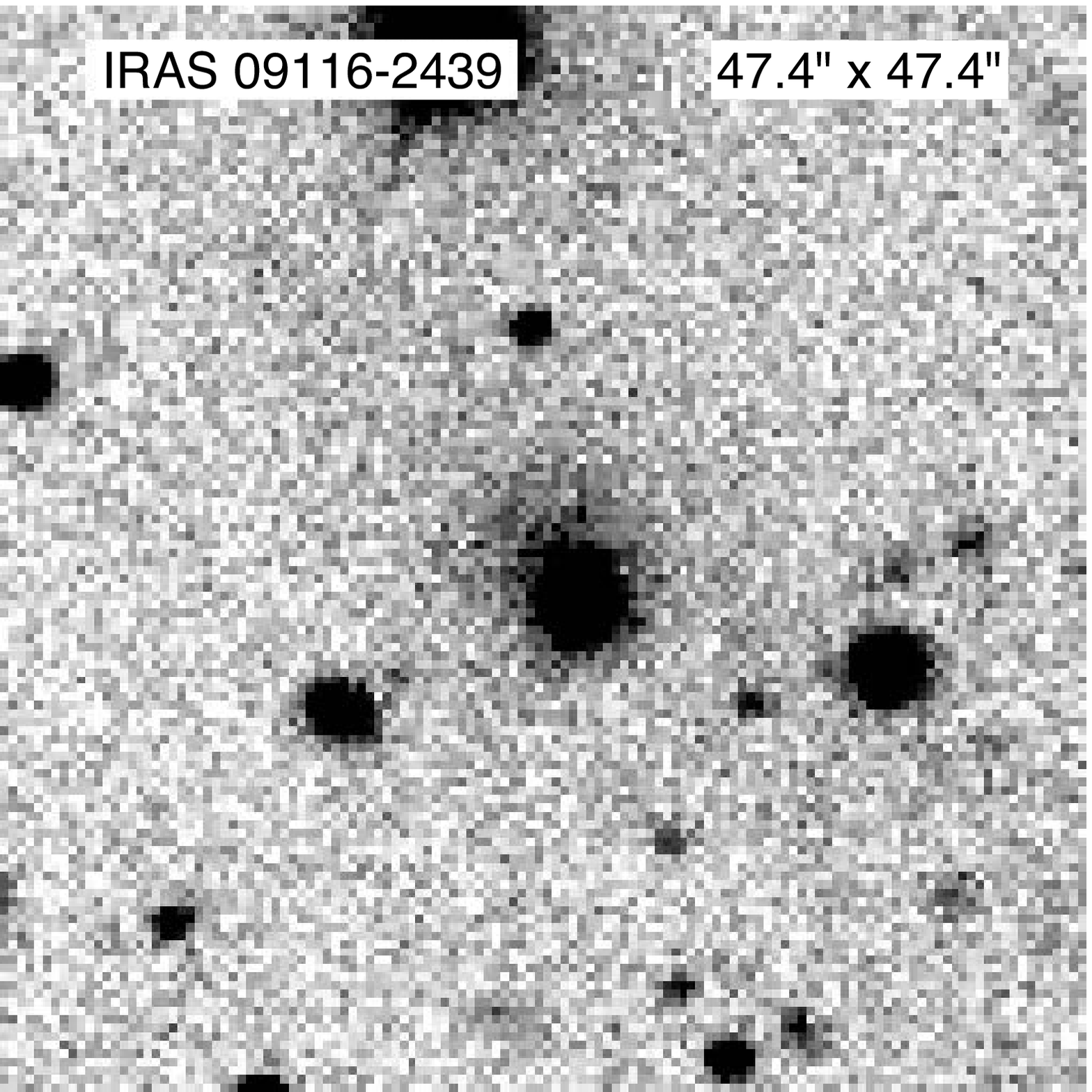}}\hspace{0.3cm}
\fbox{\includegraphics[width=0.27\linewidth]{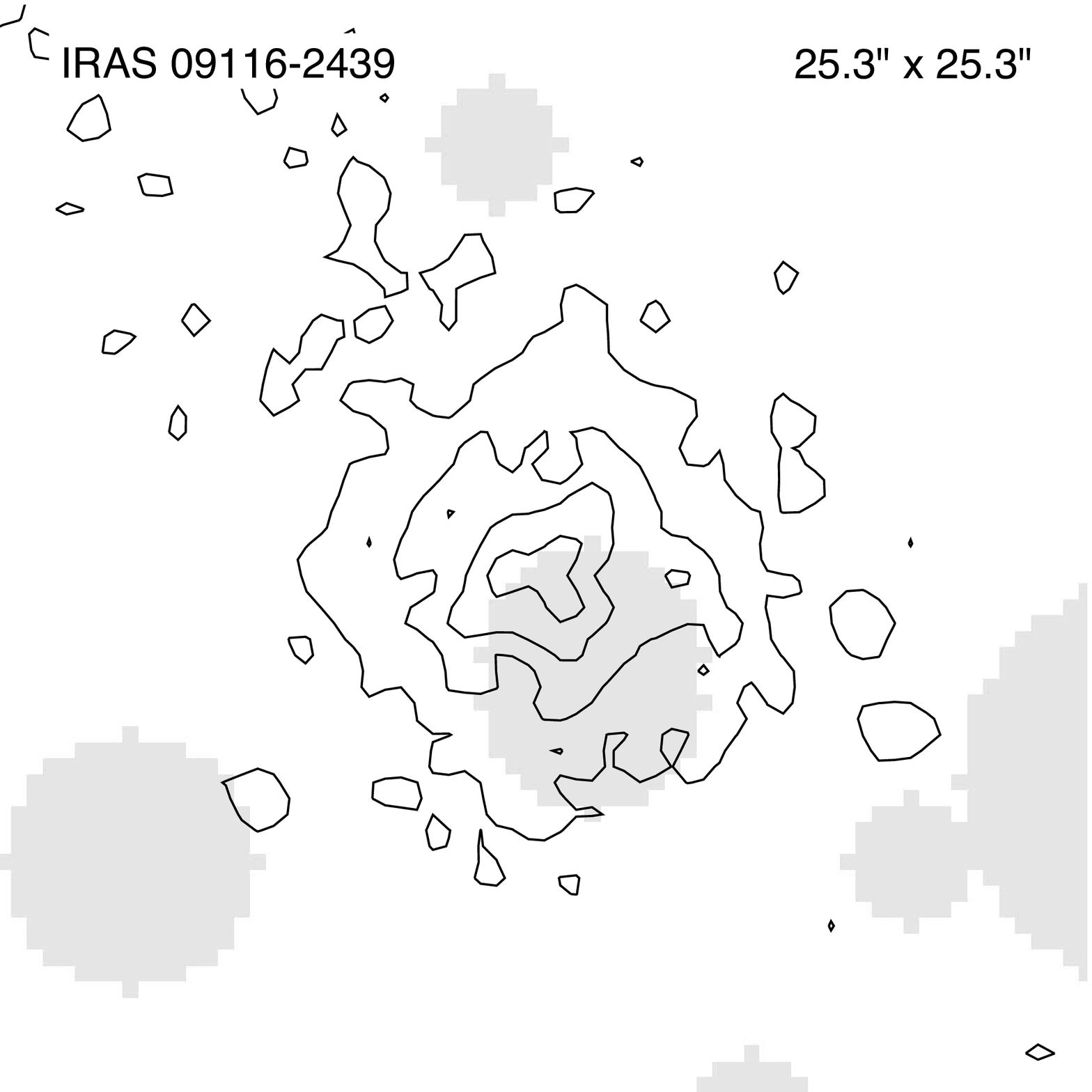}}\hspace{0.3cm}
\includegraphics[origin=c,angle=-90,width=0.27\linewidth,clip=true]{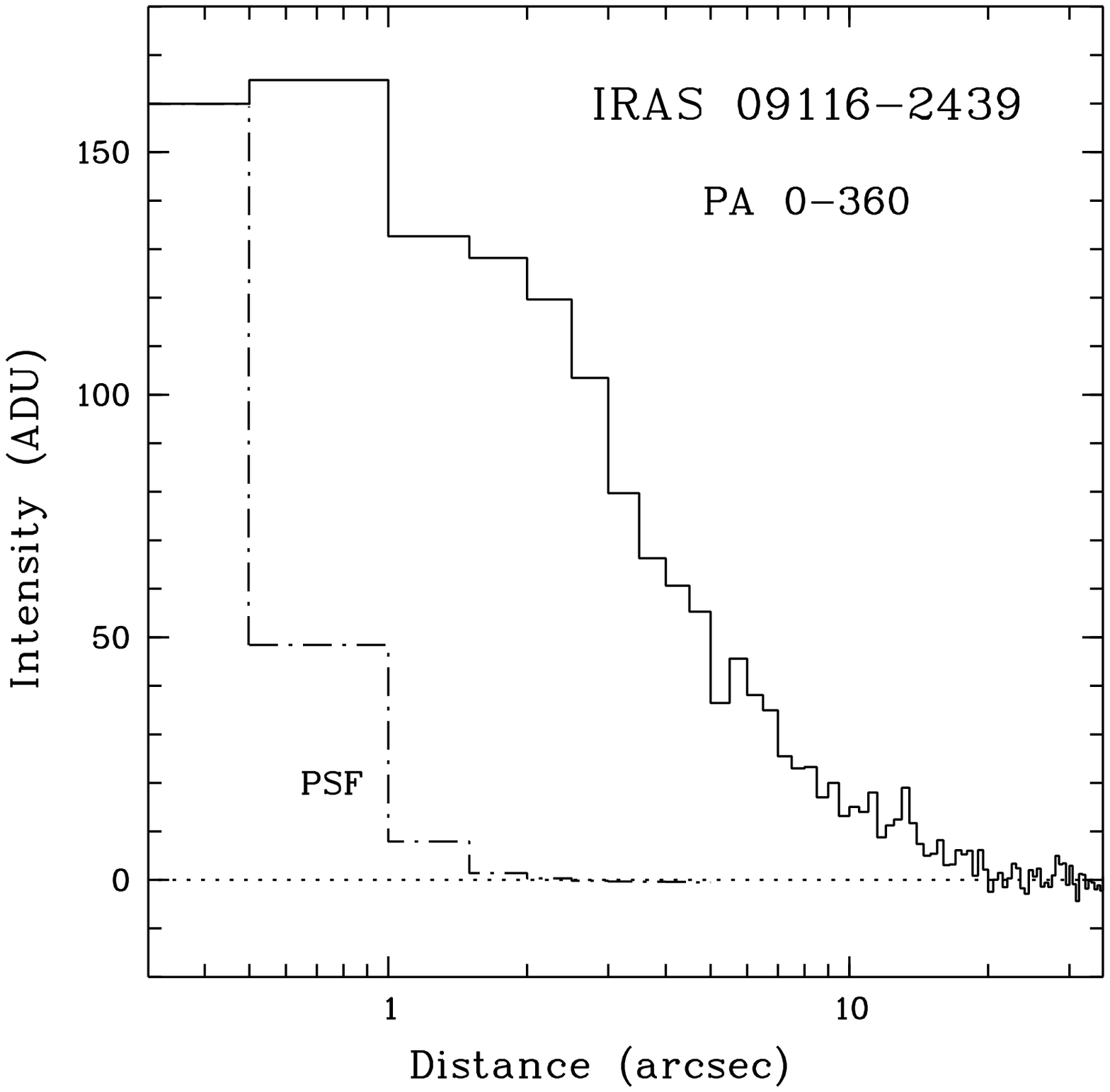}}

  \vspace{0.6cm}
   \centerline{
\fbox{\includegraphics[width=0.27\linewidth]{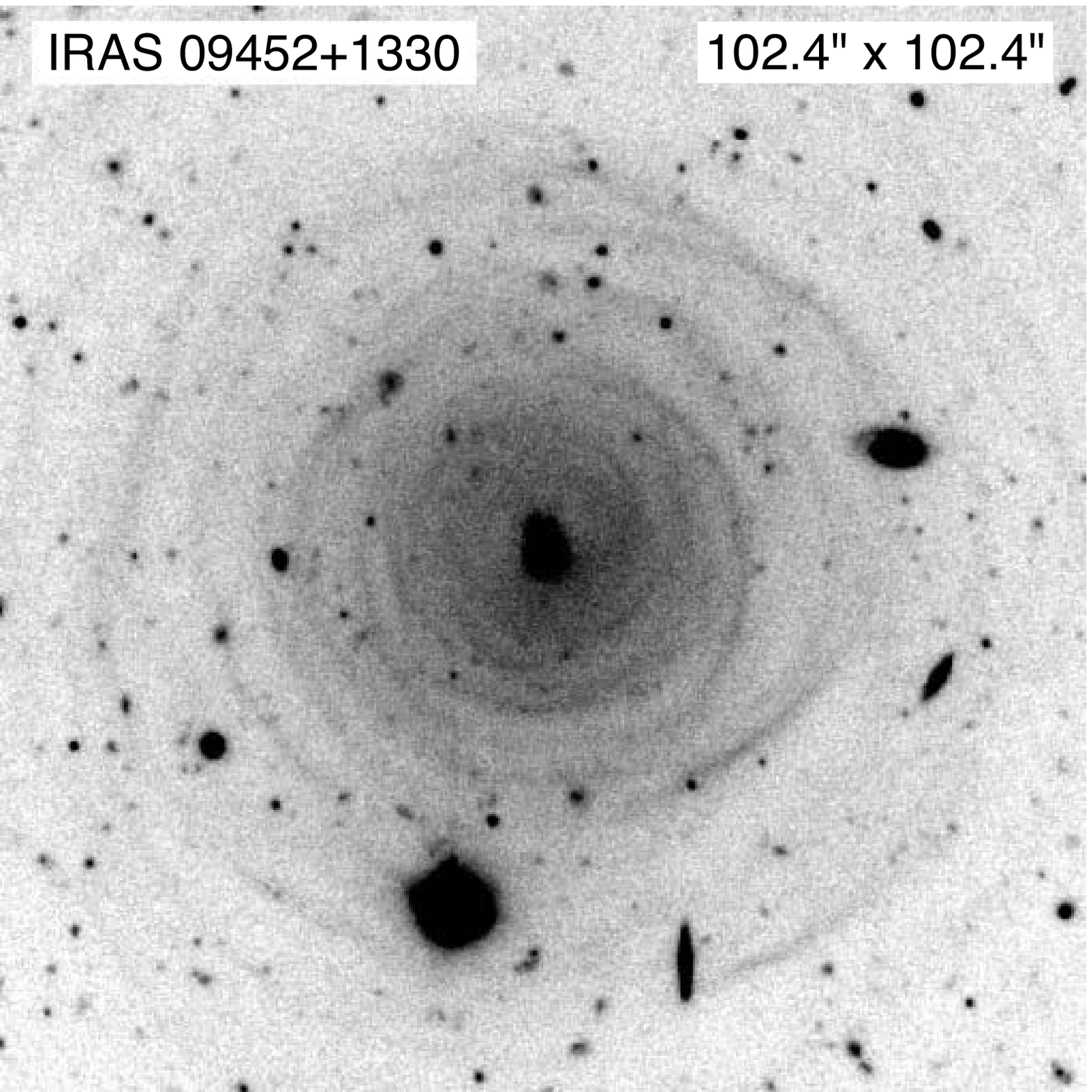}}\hspace{0.3cm}  
\fbox{\includegraphics[width=0.27\linewidth]{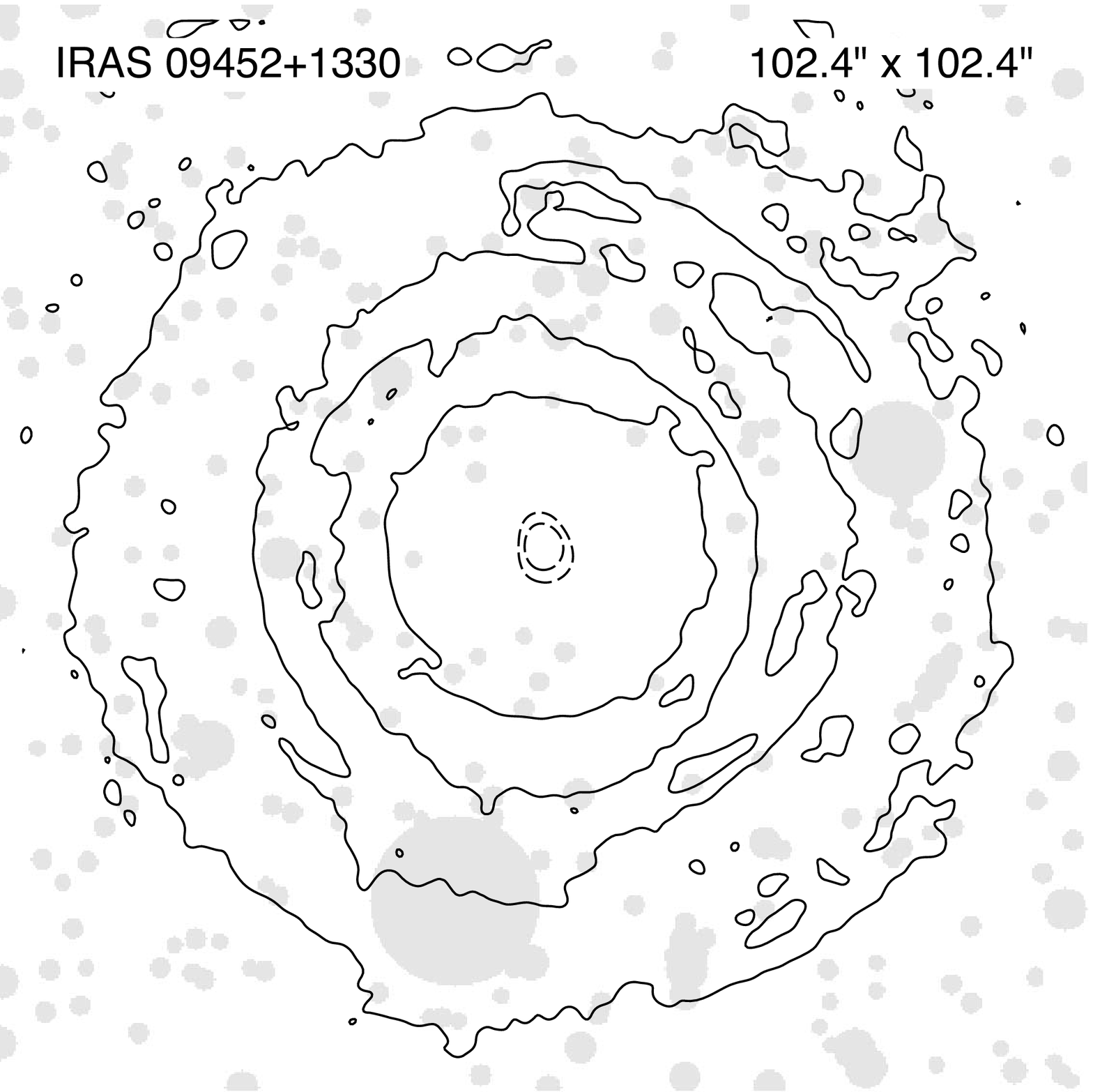}}\hspace{0.3cm}
\includegraphics[origin=c,angle=-90,width=0.27\linewidth,clip=true]{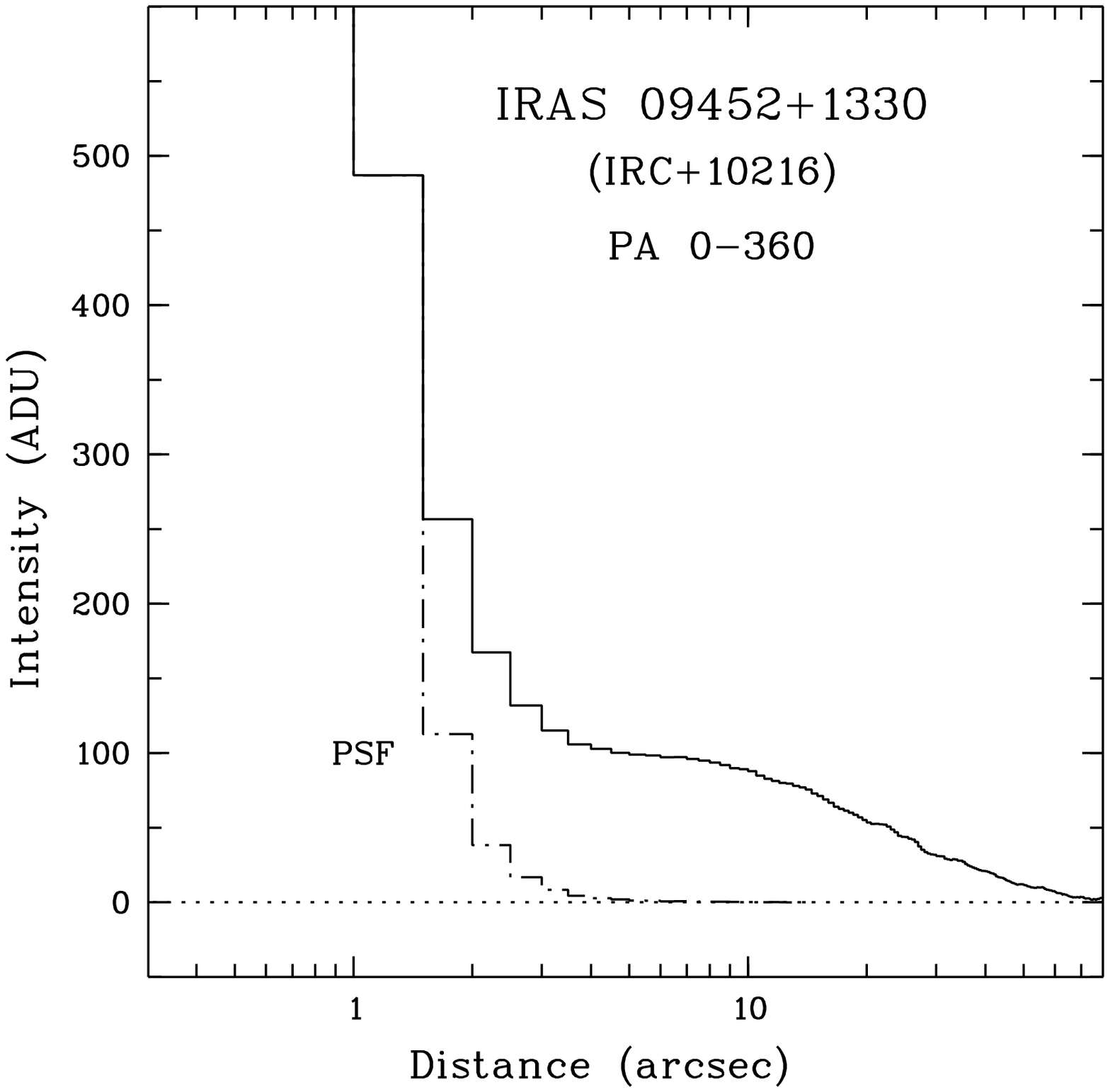}} 
  
\vspace{0.3cm}
\caption{Images and radial profiles of the AGB envelopes. Each
row shows from left to right: the image in grayscale; a close-up
contour map of the envelope, with field stars removed and their
cores masked; and the radial profile of the envelope with the PSF. The field
size is indicated in the images. The filters and contours,
specified by the lowest contour (and contour interval) in units
of the peak surface brightness of the envelope, are as follows
(from top to bottom):
IRAS 01037$+$1219, $V$-band, contours 0.33 (0.43);   
IRAS 07454$-$7112,  $B$-band, contours 0.31 (0.23);
IRAS 09116$-$2439,  $B$-band, contours 0.28 (0.26); 
IRAS 09452$+$1330, $V$-band, contours 0.32 (0.16).
The images are square, with north to the top and east
to the left. }

 \end{figure*}


  \begin{figure*}
   \setlength\fboxsep{0pt}
   \setlength\fboxrule{0.5pt}

  \vspace{0.3cm}
  \centerline{
\fbox{\includegraphics[width=0.27\linewidth]{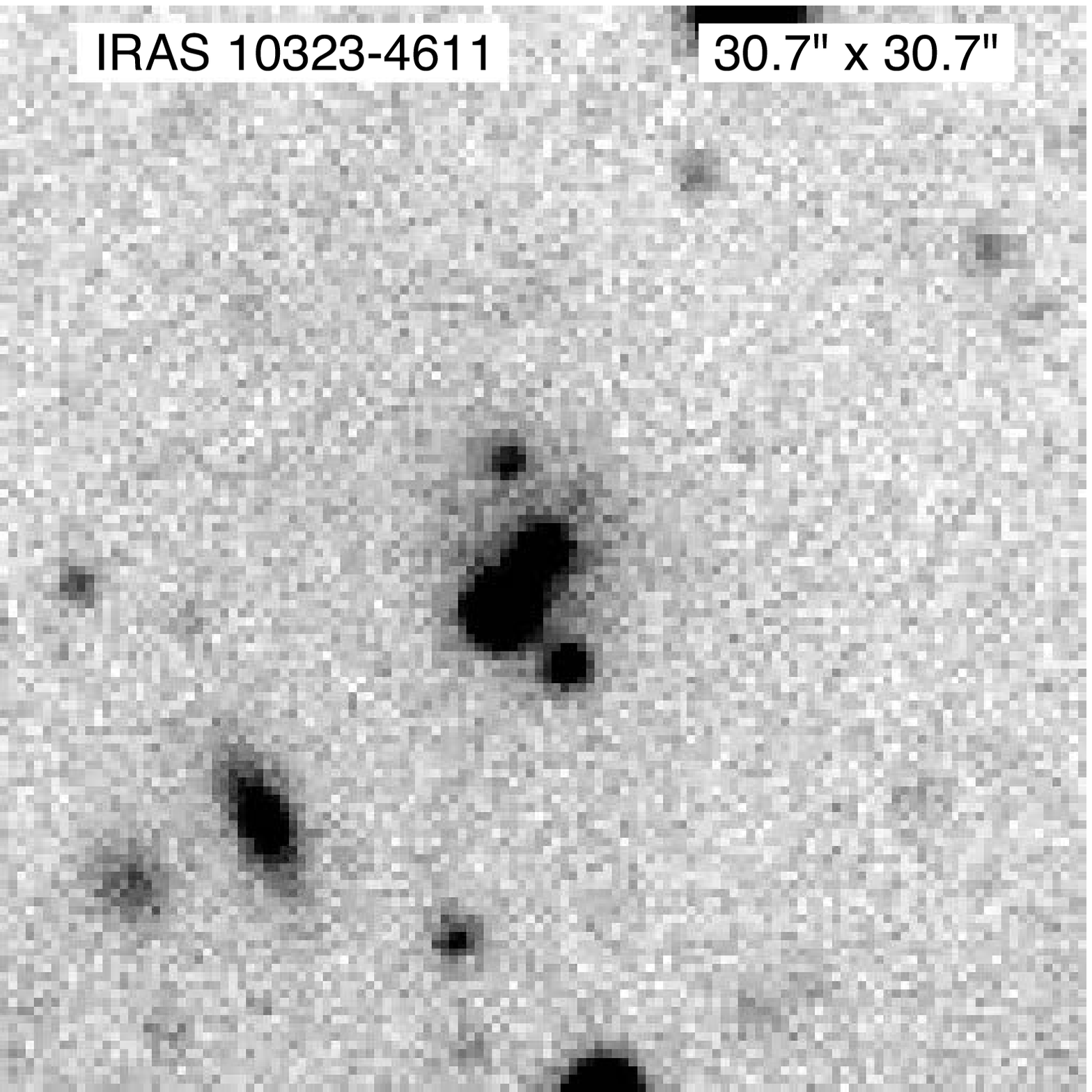}}\hspace{0.3cm}
\fbox{\includegraphics[width=0.27\linewidth]{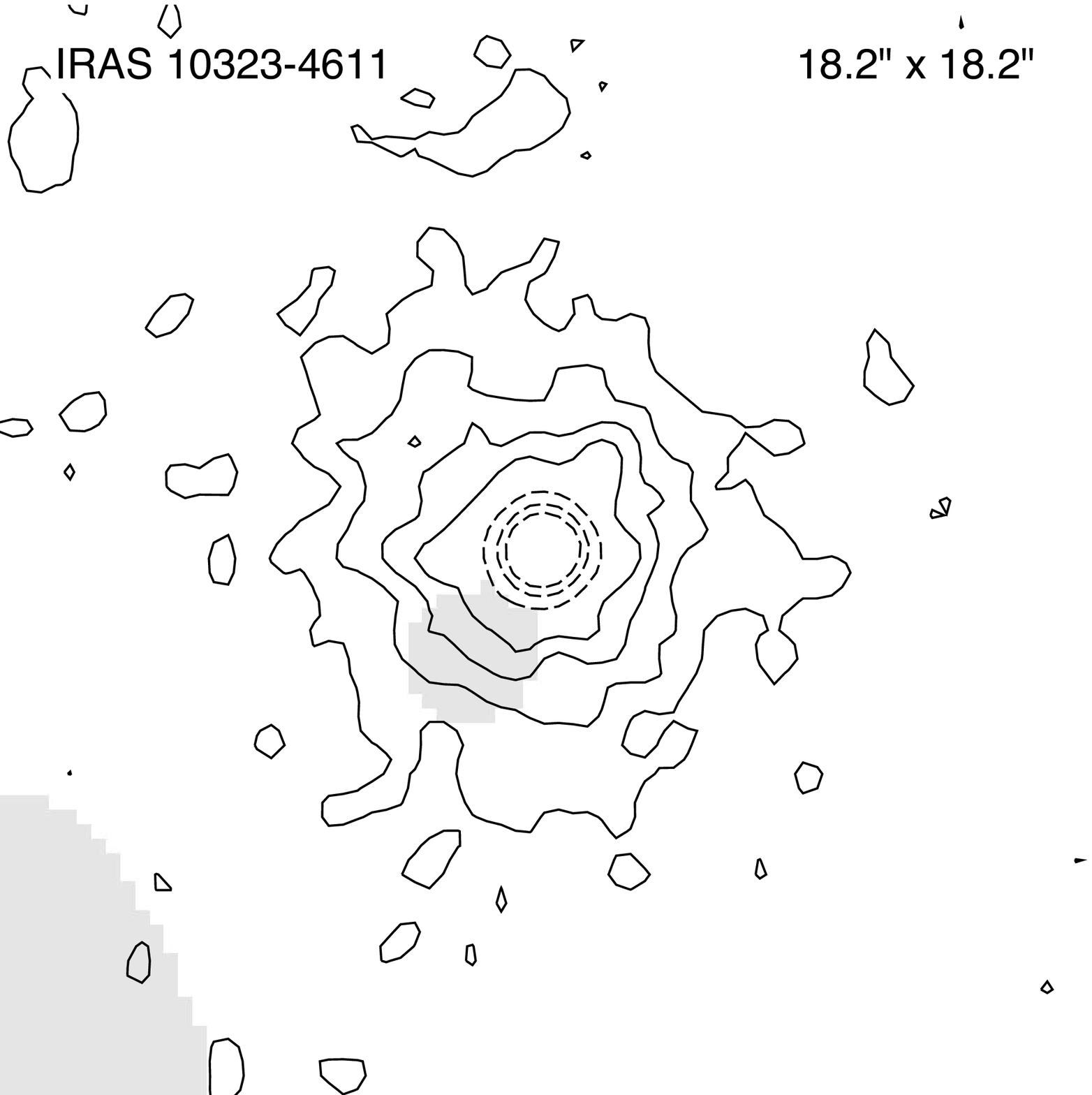}}\hspace{0.3cm}    
\includegraphics[origin=c,angle=-90,width=0.27\linewidth,clip=true]{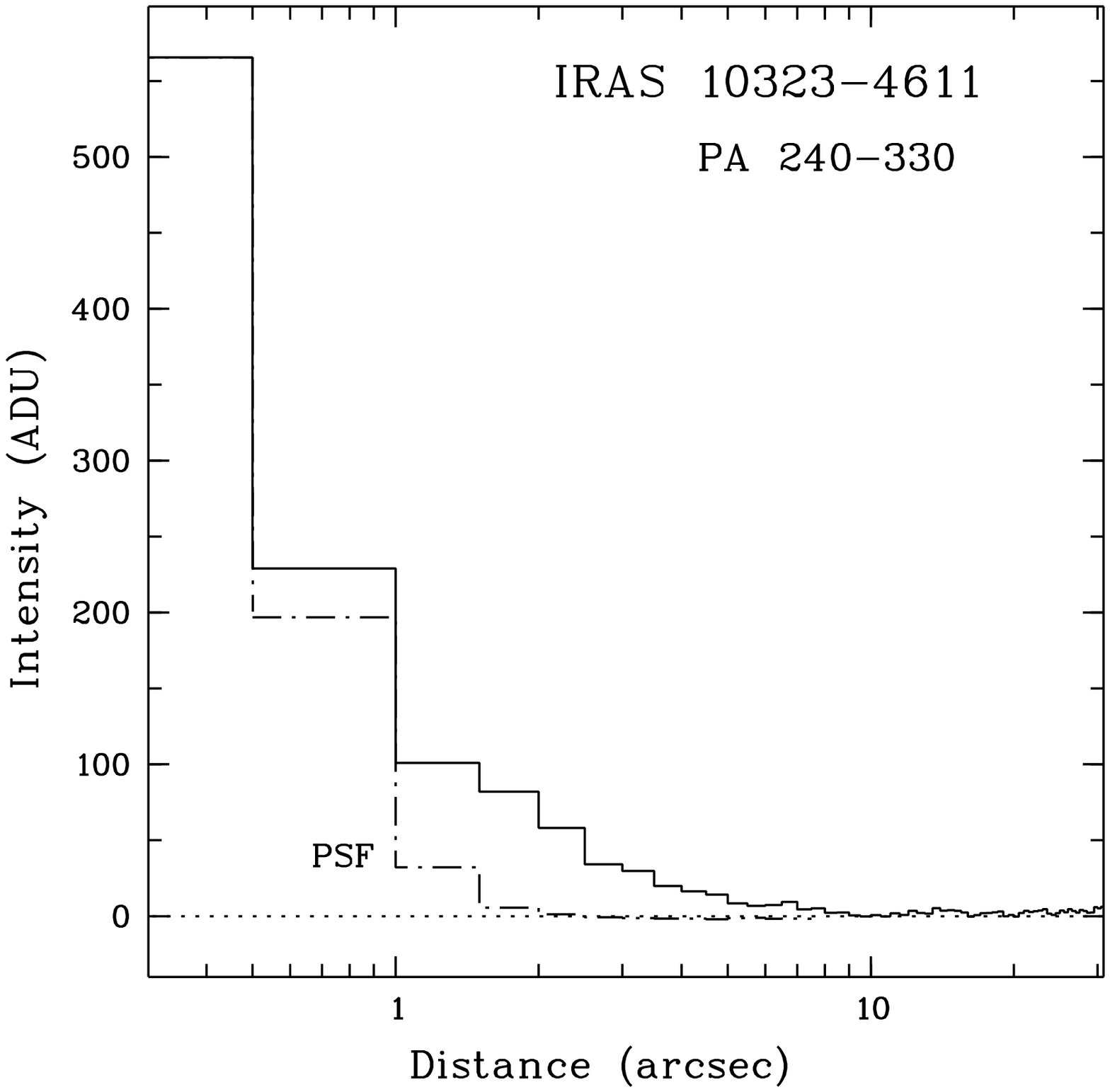}}     

  \vspace{0.6cm}
  \centerline{
\fbox{\includegraphics[width=0.27\linewidth]{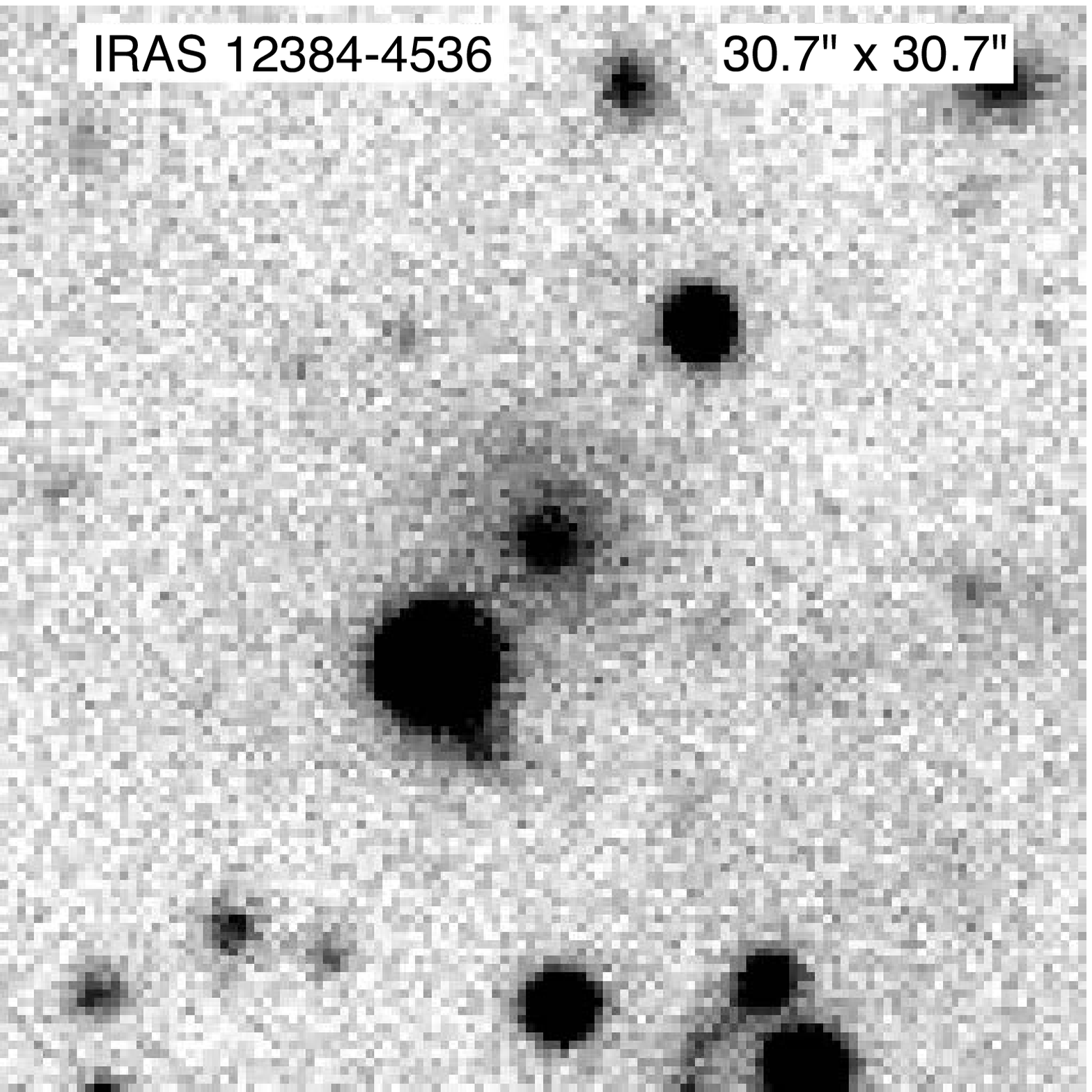}}\hspace{0.3cm}
\fbox{\includegraphics[width=0.27\linewidth]{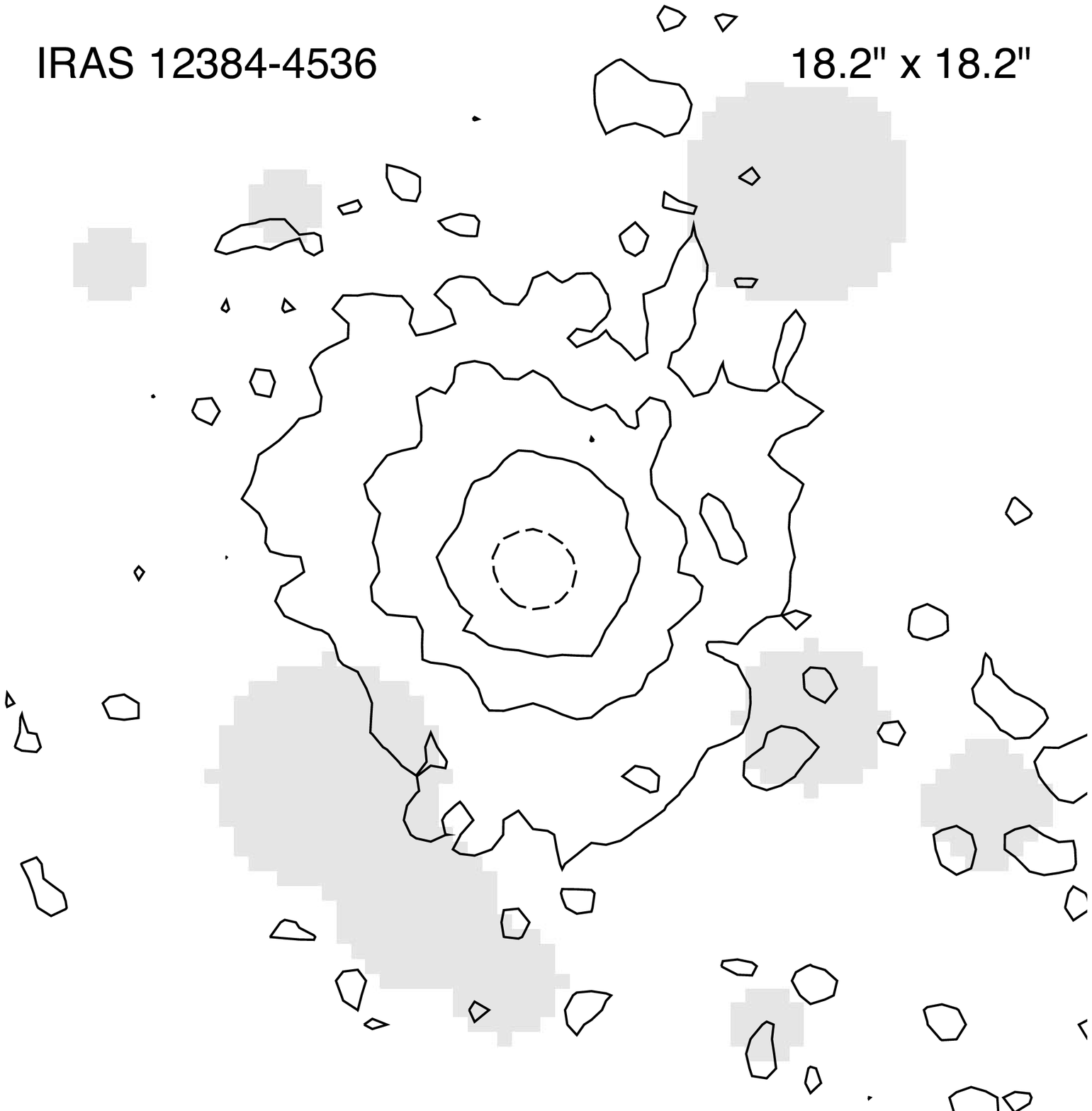}}\hspace{0.3cm}
\includegraphics[origin=c,angle=-90,width=0.27\linewidth,clip=true]{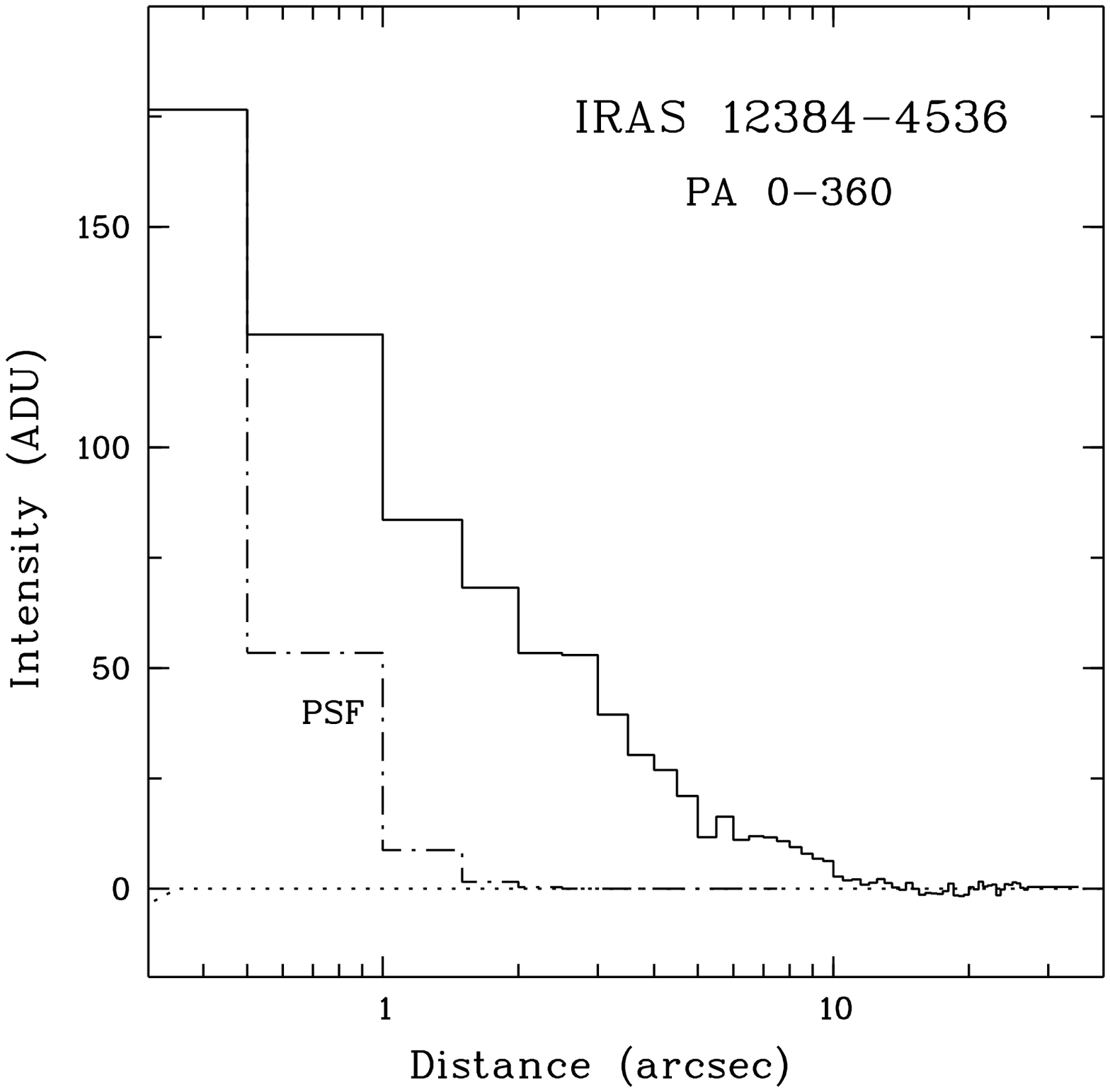}}

 \vspace{0.6cm}
   \centerline{
\fbox{\includegraphics[width=0.27\linewidth]{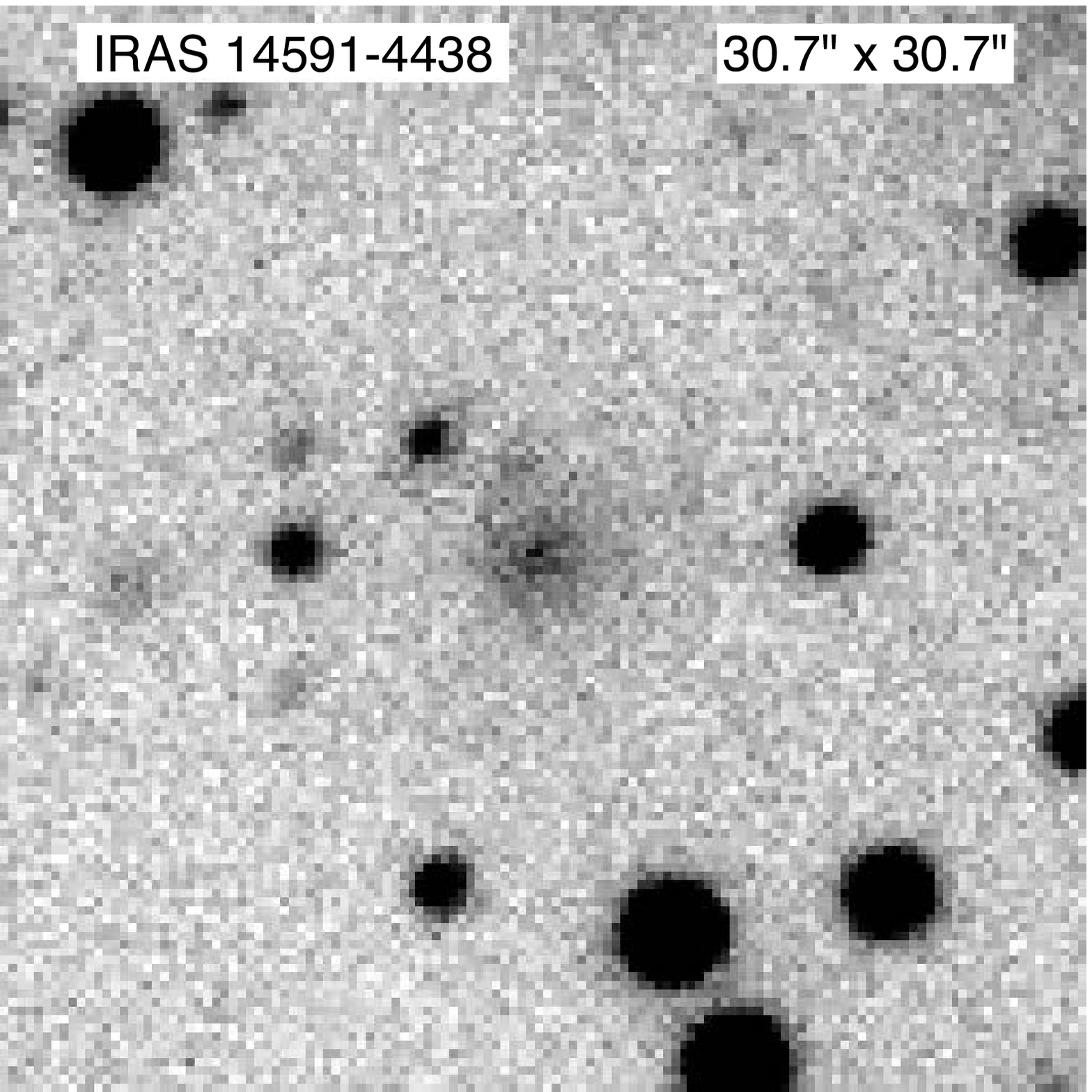}}\hspace{0.3cm}
\fbox{\includegraphics[width=0.27\linewidth]{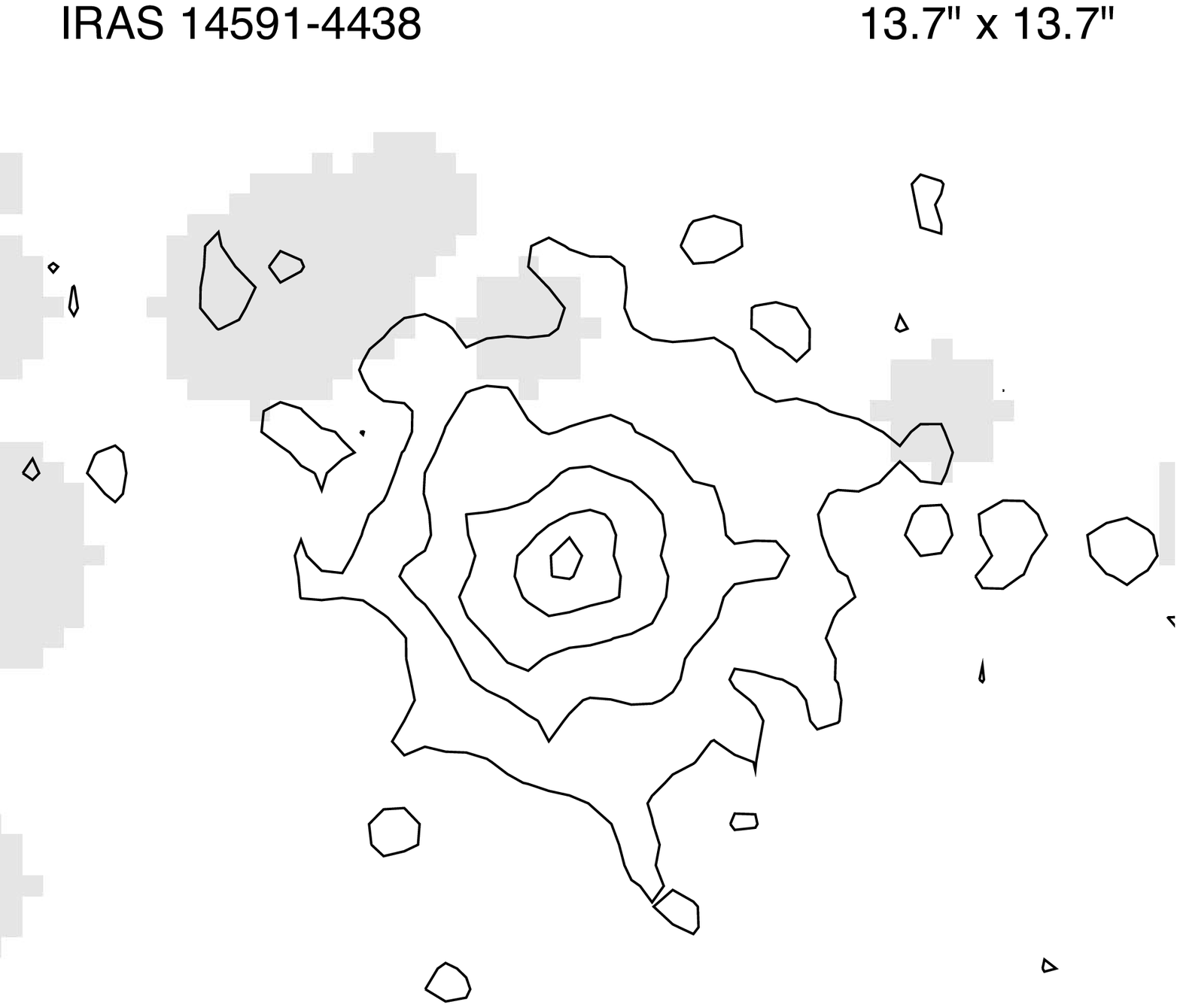}}\hspace{0.3cm}
\includegraphics[origin=c,angle=-90,width=0.27\linewidth,clip=true]{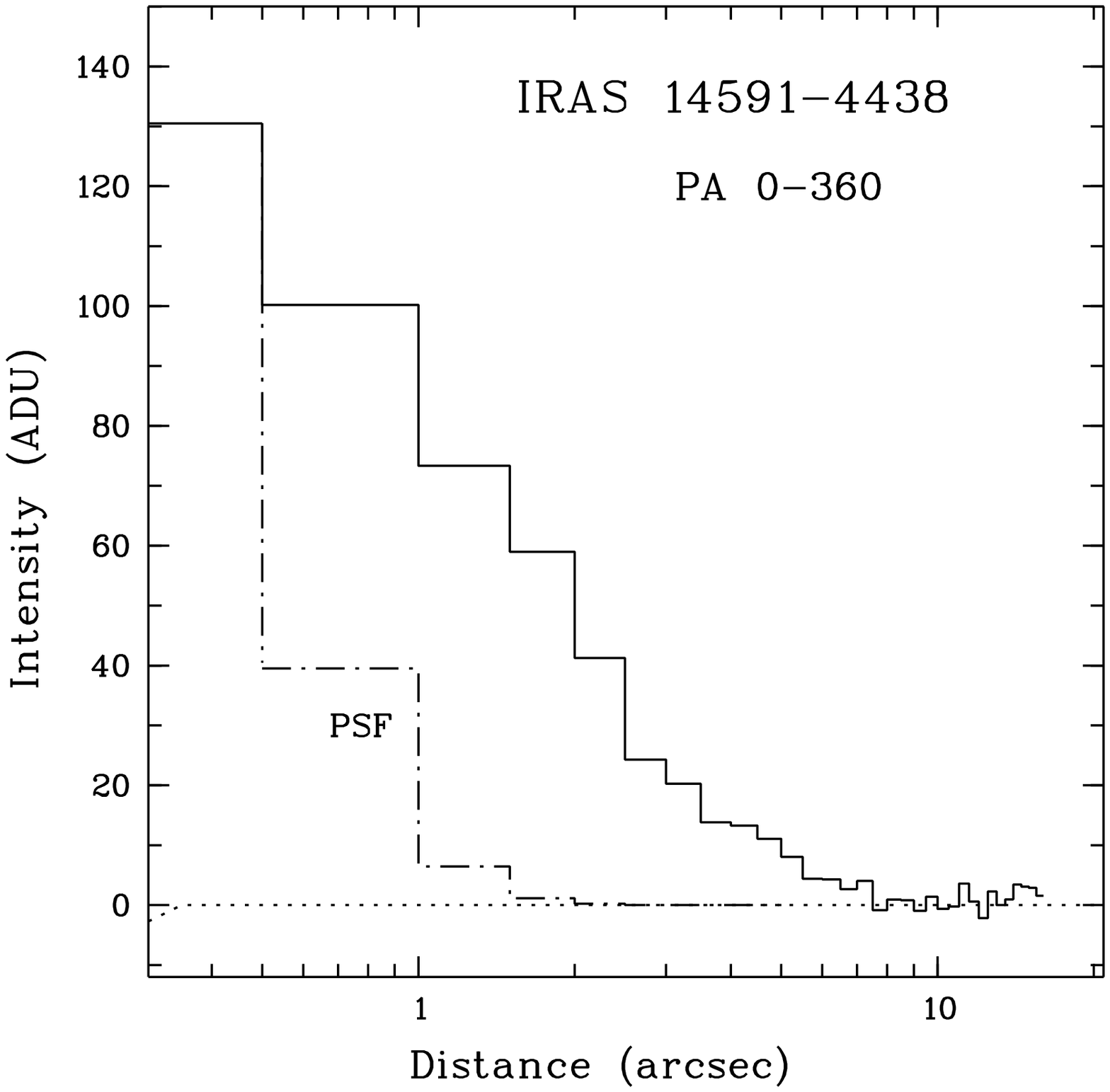} }

 \vspace{0.6cm}
  \centerline{
\fbox{\includegraphics[width=0.27\linewidth]{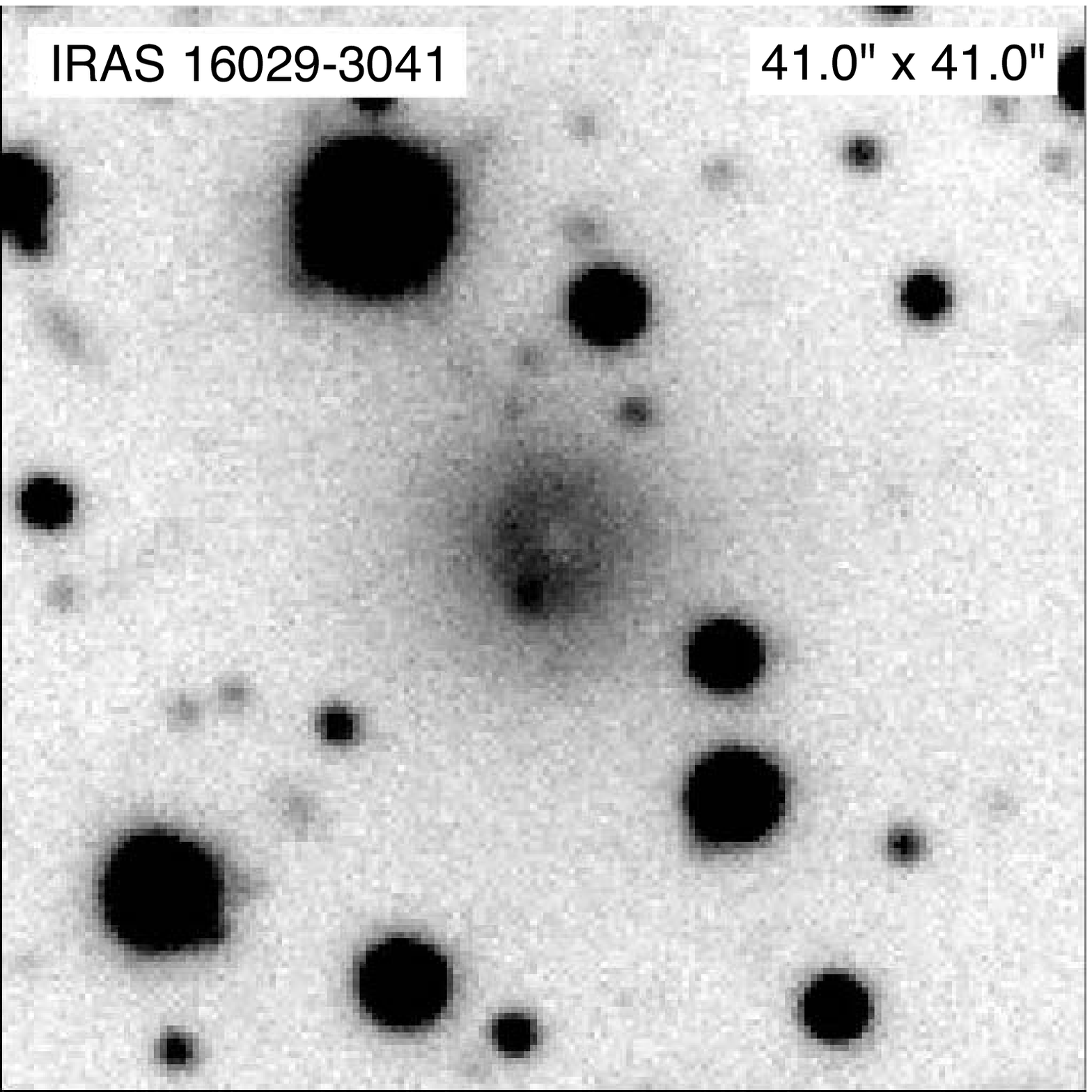}}\hspace{0.3cm}
\fbox{\includegraphics[width=0.27\linewidth]{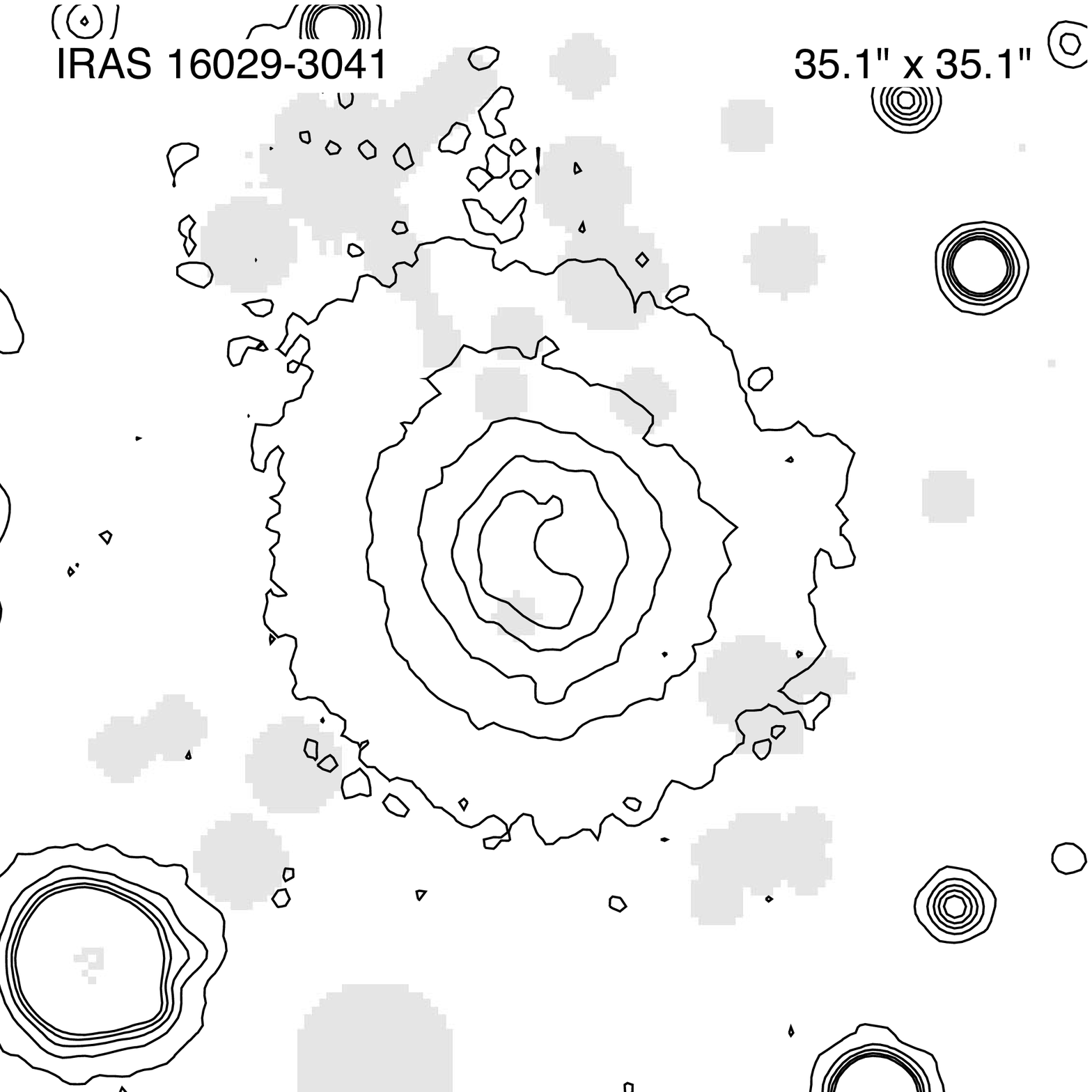}}\hspace{0.3cm}
\includegraphics[origin=c,angle=-90,width=0.27\linewidth,clip=true]{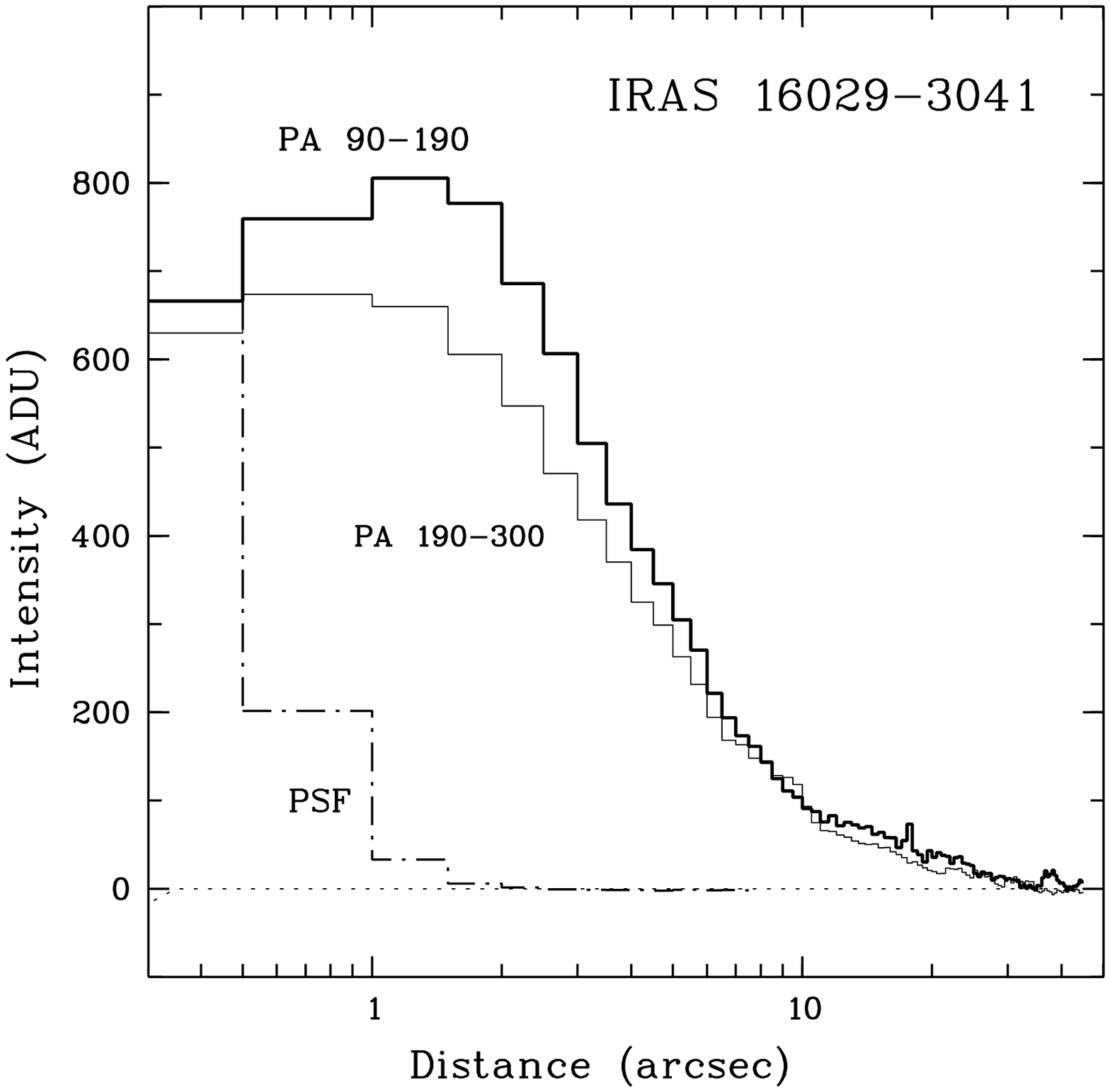} }

\vspace{0.3cm}
\caption{Continuation of Fig.~1. From top to bottom:
IRAS 10323$-$4611, $B$-band, contours 0.13 (0.25);
IRAS 12384$-$4536, $V$-band, contours 0.25 (0.25);
IRAS 14591$-$4438, $B$-band, contours 0.35 (0.33);
IRAS 16029$-$3041, $V$-band, contours 0.17 (0.17).
The images are square, with north to the top and east
to the left.}
 \end{figure*}


\begin{figure*}
   \setlength\fboxsep{0pt}
   \setlength\fboxrule{0.5pt}

  \vspace{0.3cm}
  \centerline{
\fbox{\includegraphics[width=0.27\linewidth]{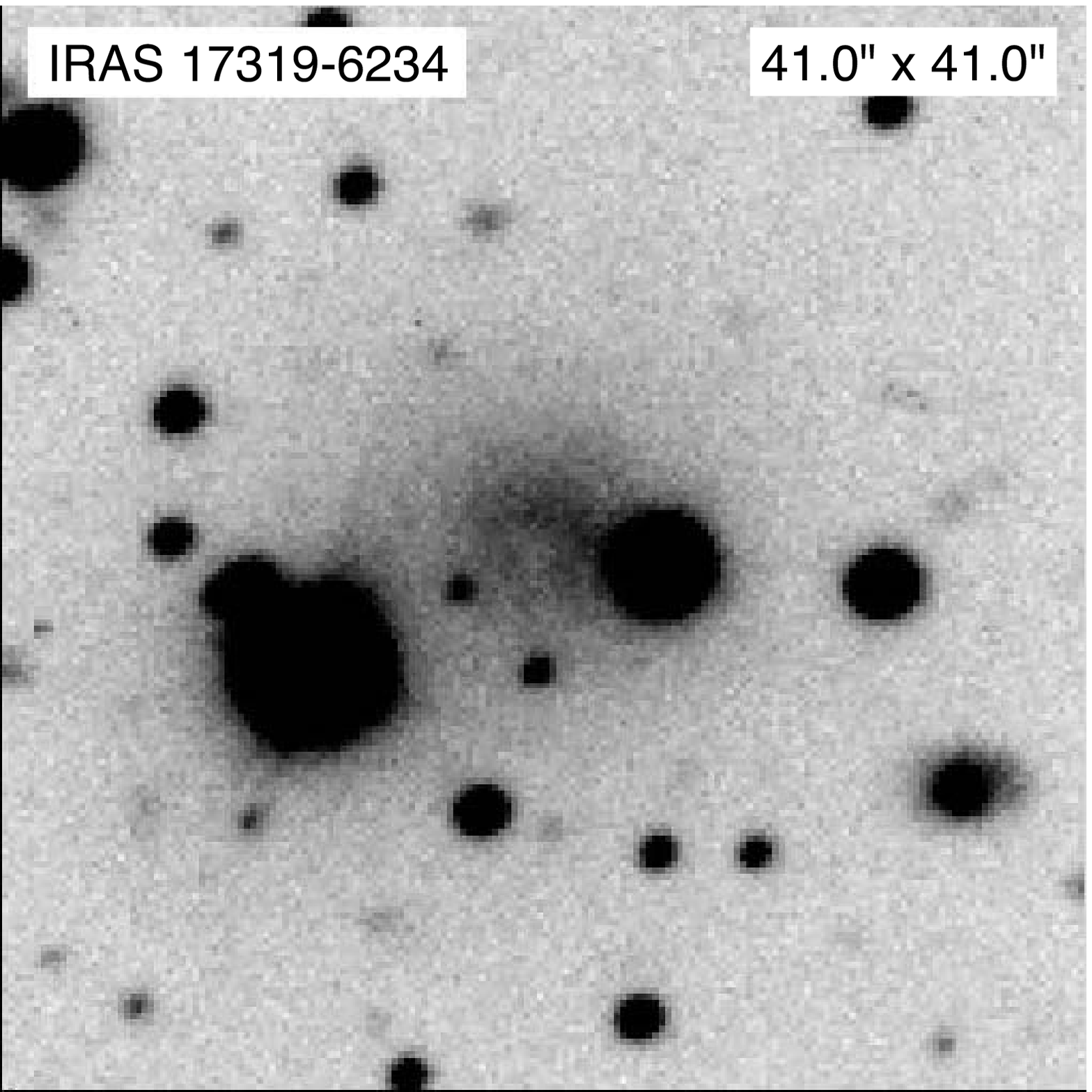}}\hspace{0.30cm}
\fbox{\includegraphics[width=0.27\linewidth]{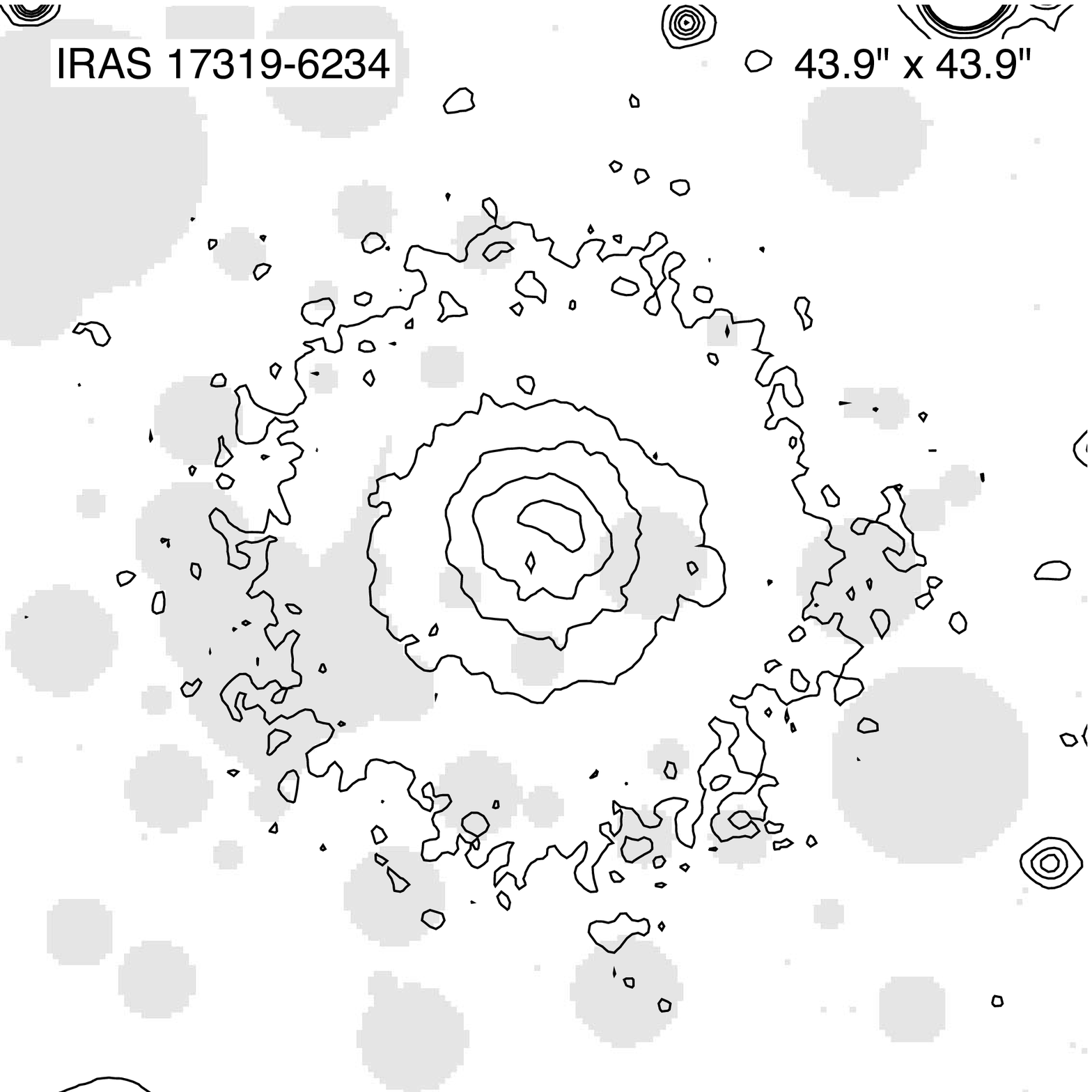}}\hspace{0.3cm}
\includegraphics[origin=c,angle=-90,width=0.27\linewidth,clip=true]{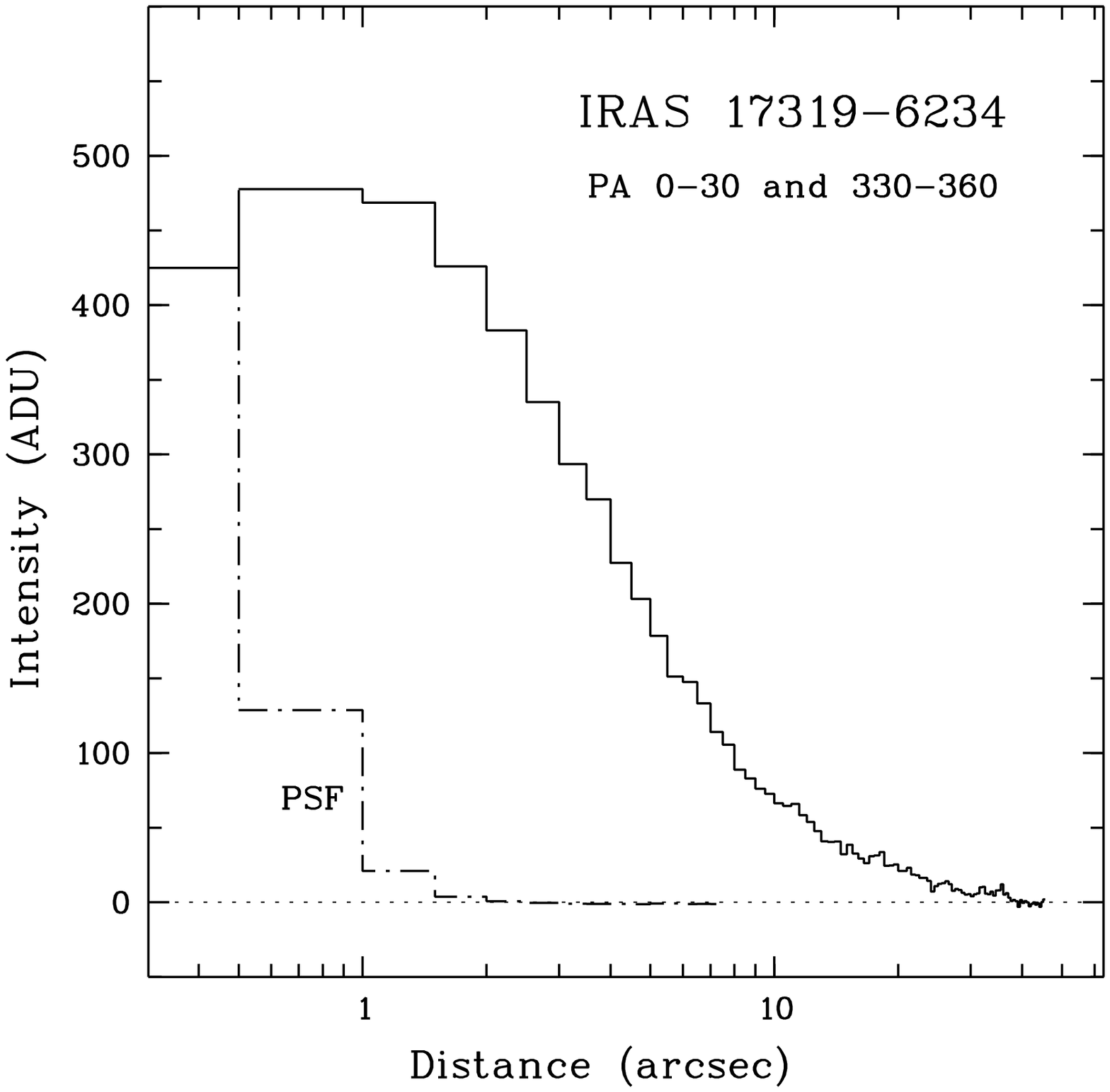}  }

 \vspace{0.6cm}
   \centerline{
\fbox{\includegraphics[width=0.27\linewidth]{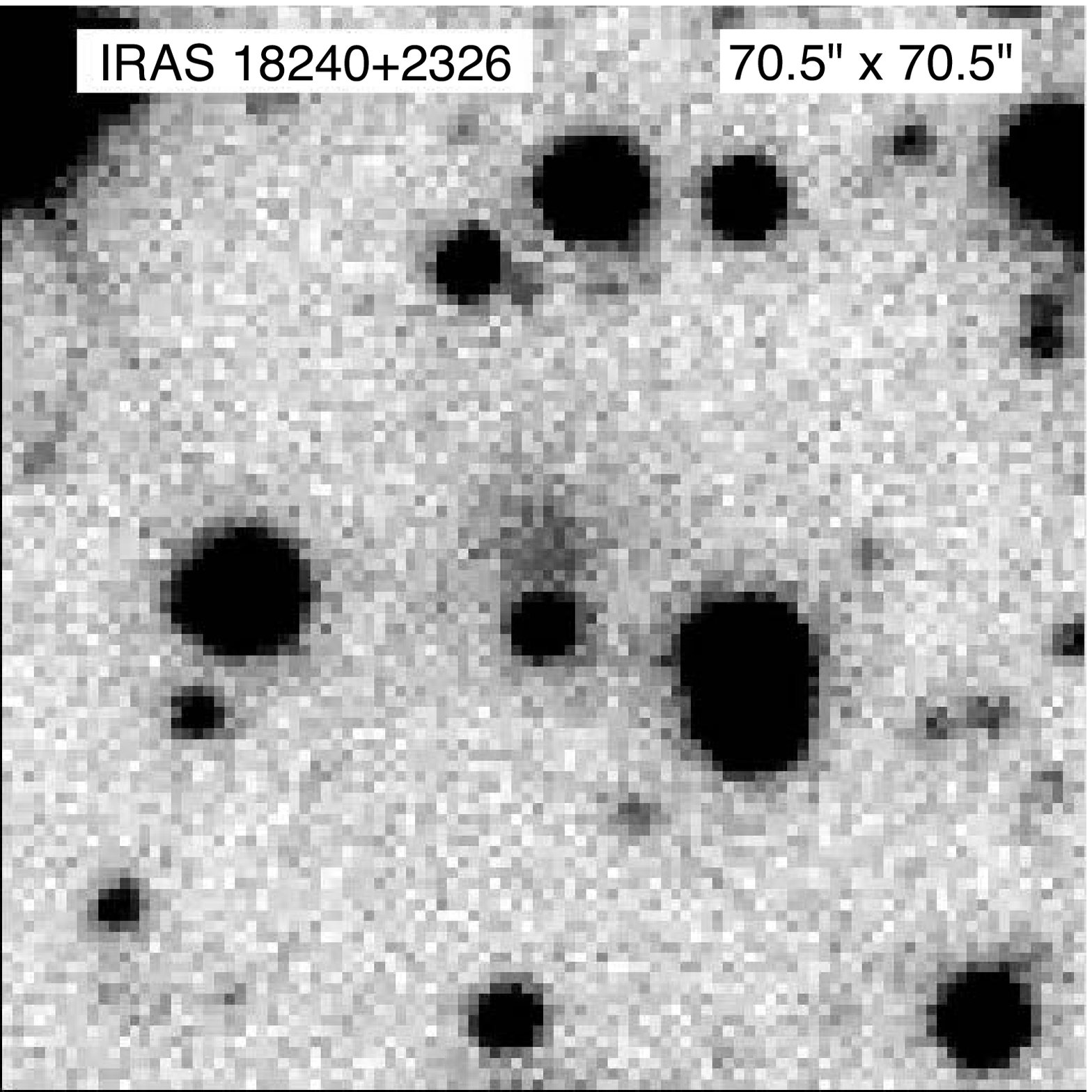}}\hspace{0.3cm}
\fbox{\includegraphics[width=0.27\linewidth]{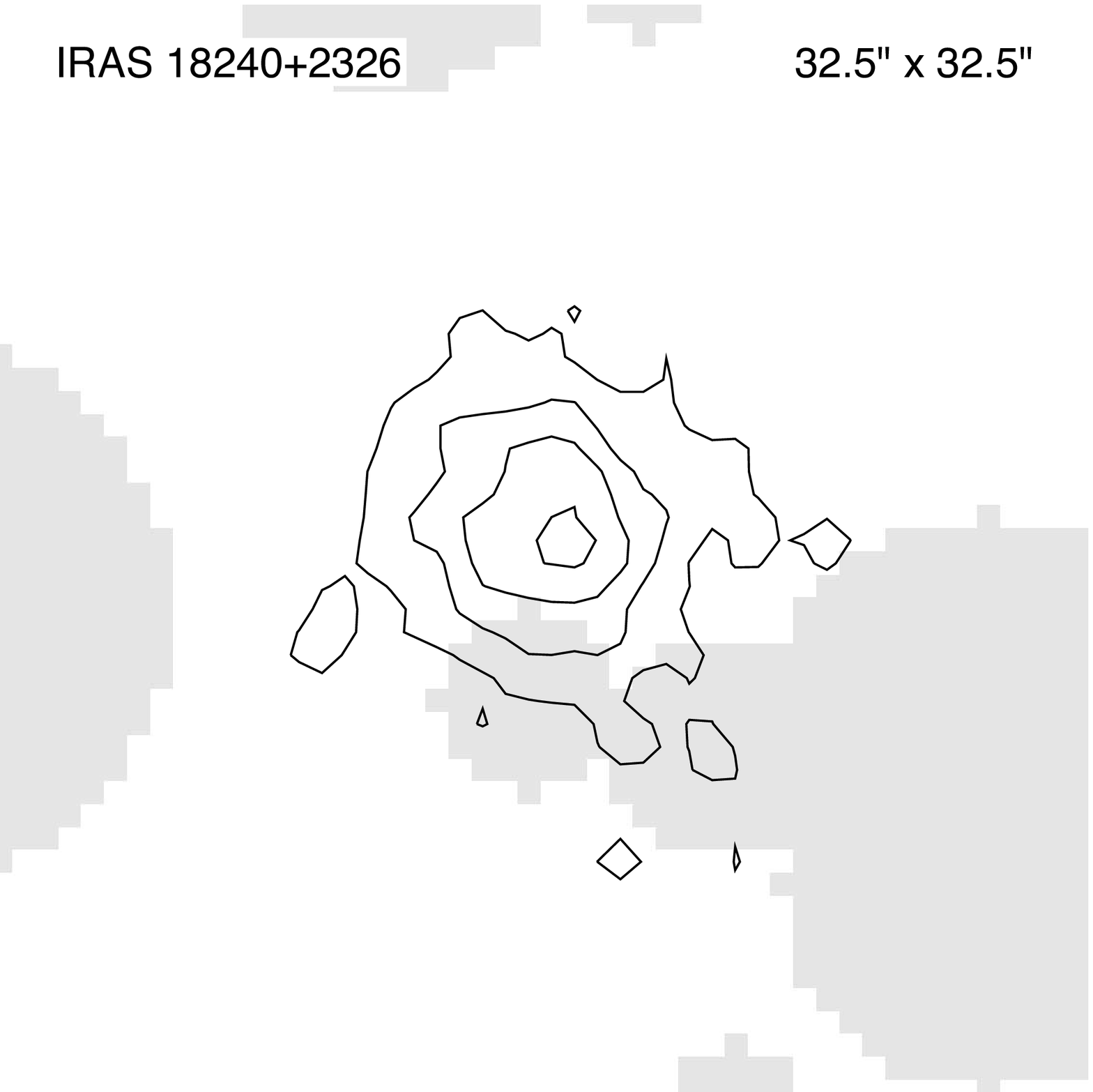}}\hspace{0.3cm}
\includegraphics[origin=c,angle=-90,width=0.27\linewidth,clip=true]{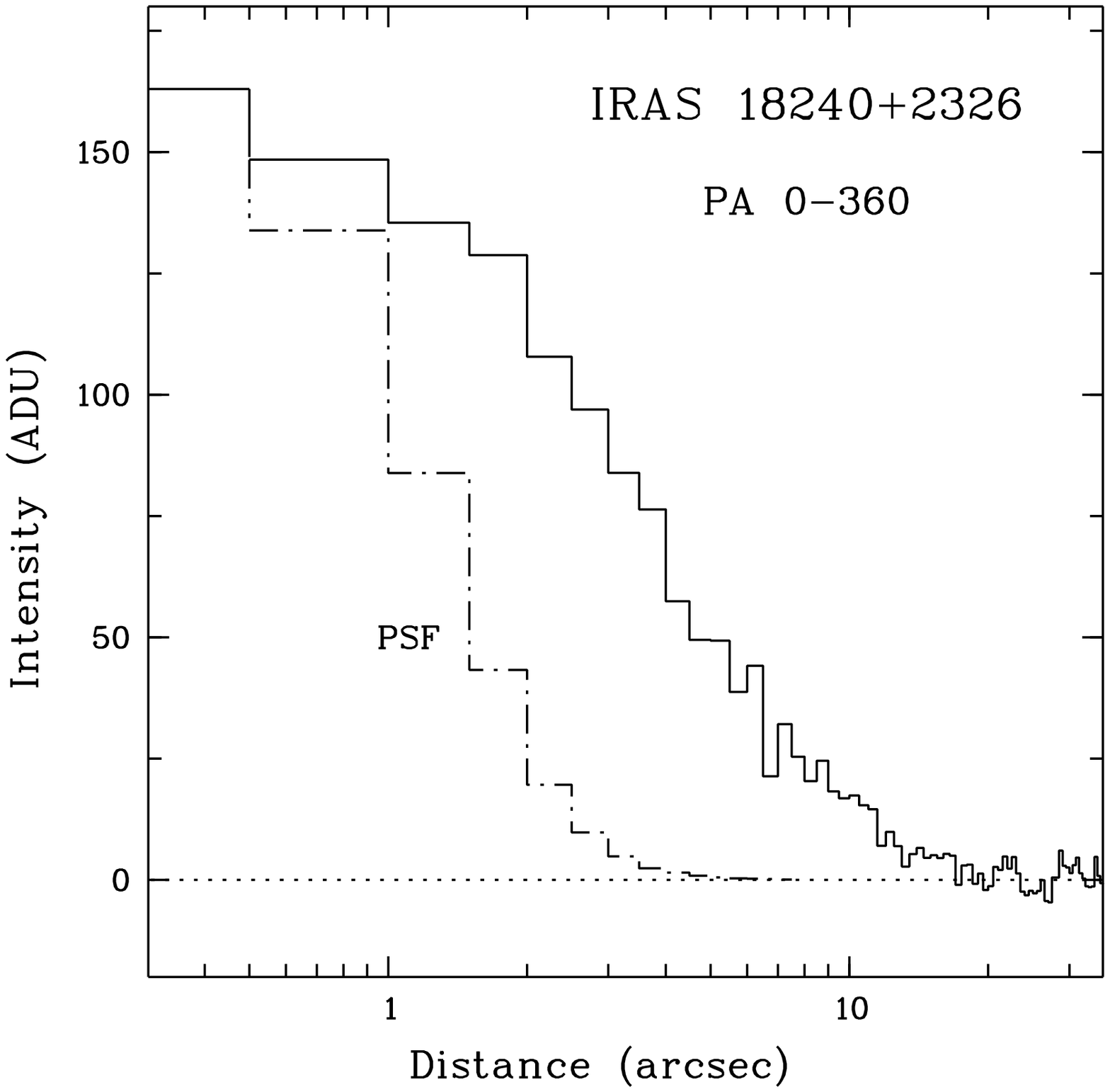} }

  \vspace{0.6cm}
  \centerline{
\fbox{\includegraphics[width=0.27\linewidth]{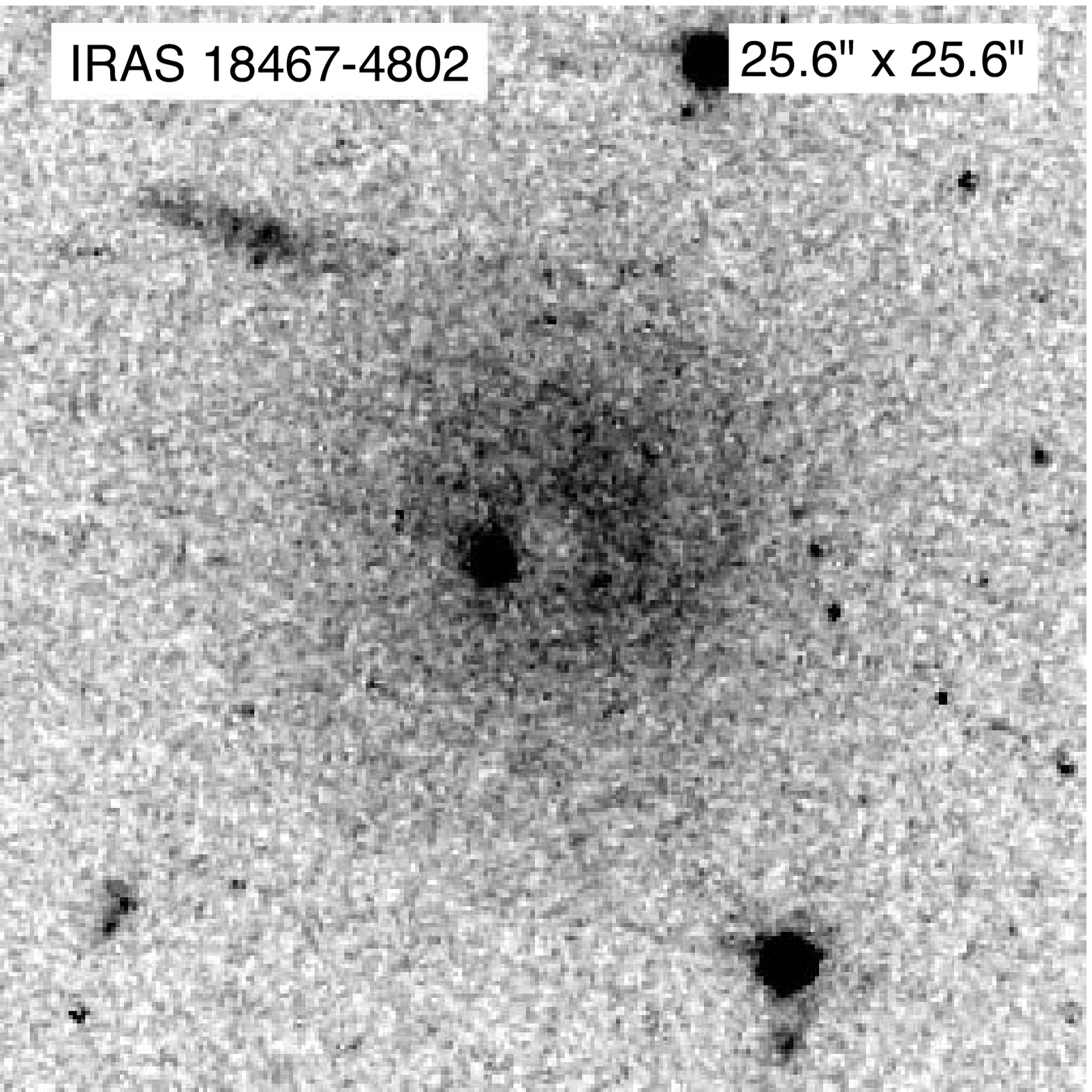}}\hspace{0.3cm}
\fbox{\includegraphics[width=0.27\linewidth]{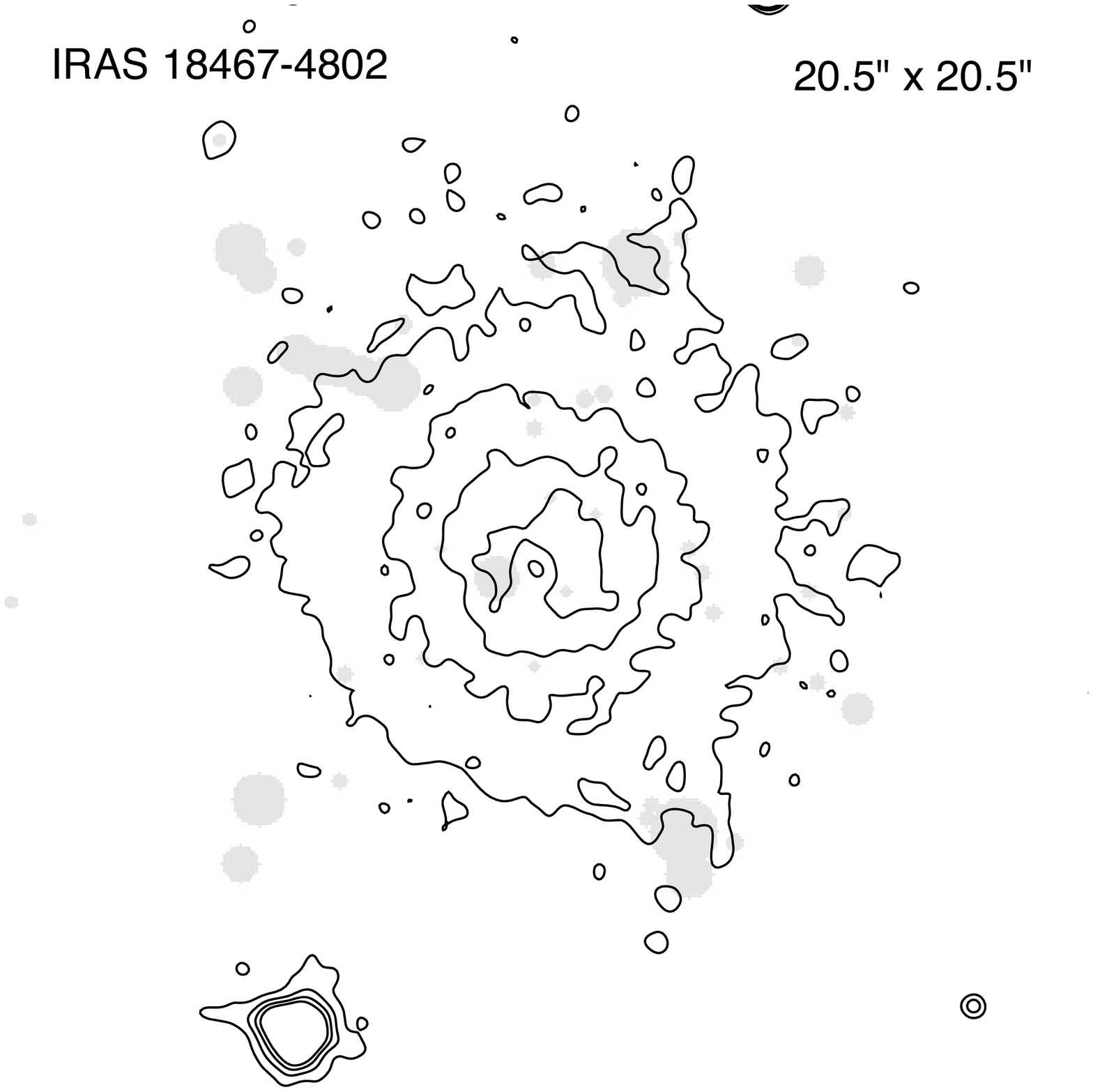}}\hspace{0.3cm}
\includegraphics[origin=c,angle=-90,width=0.27\linewidth,clip=true]{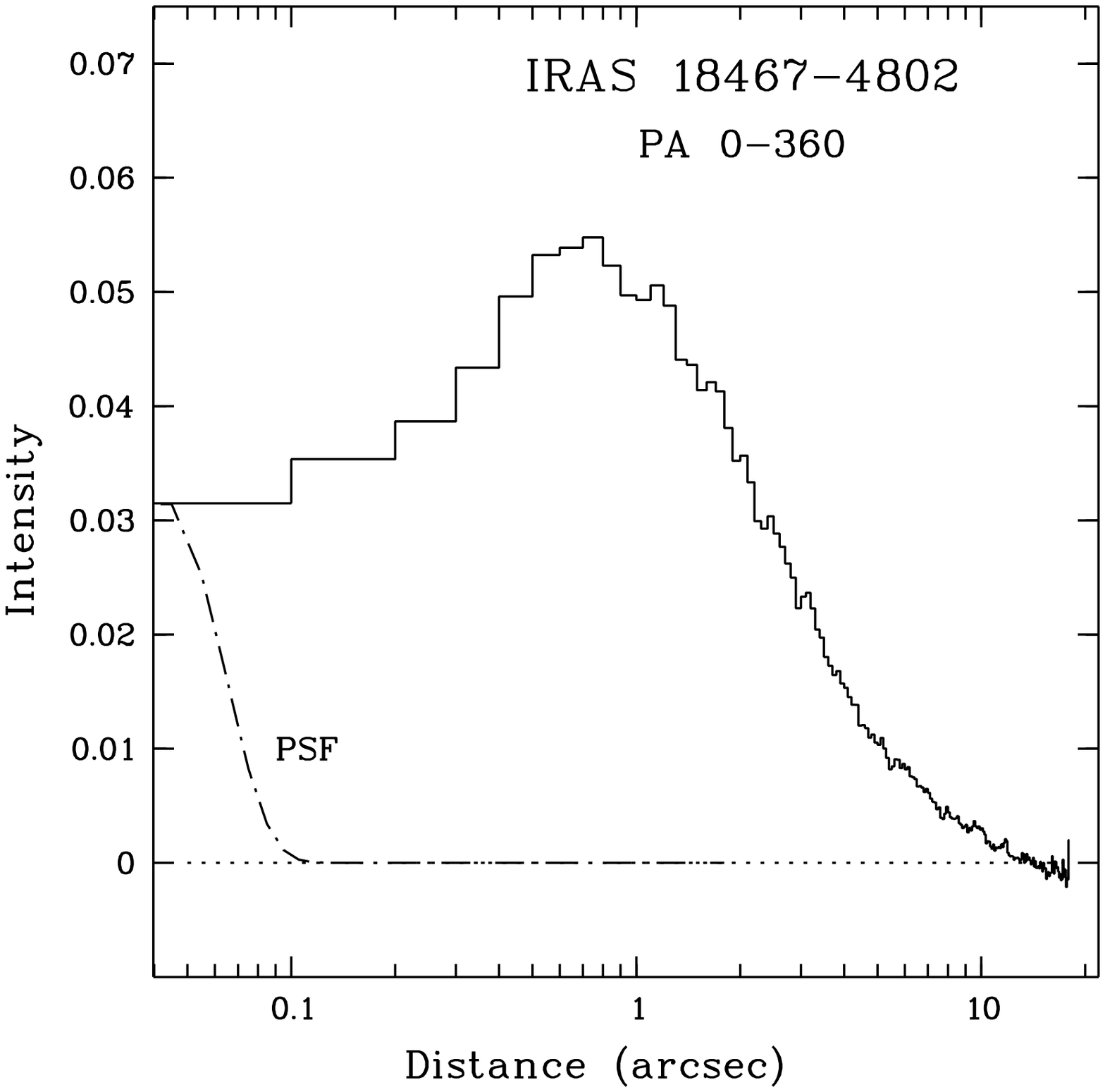} }

  \vspace{0.6cm}
  \centerline{
\fbox{\includegraphics[width=0.27\linewidth]{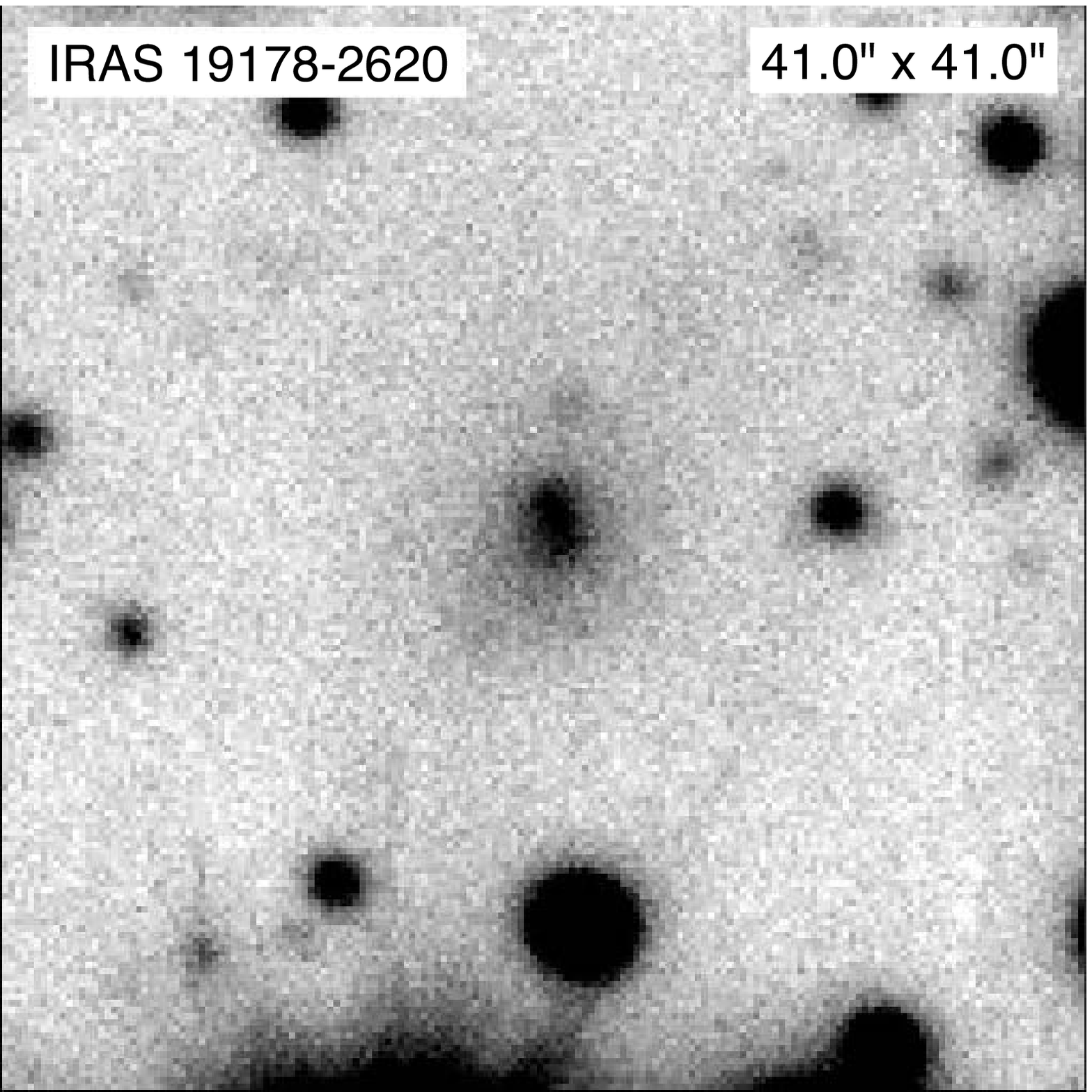}}\hspace{0.3cm}
\fbox{\includegraphics[width=0.27\linewidth]{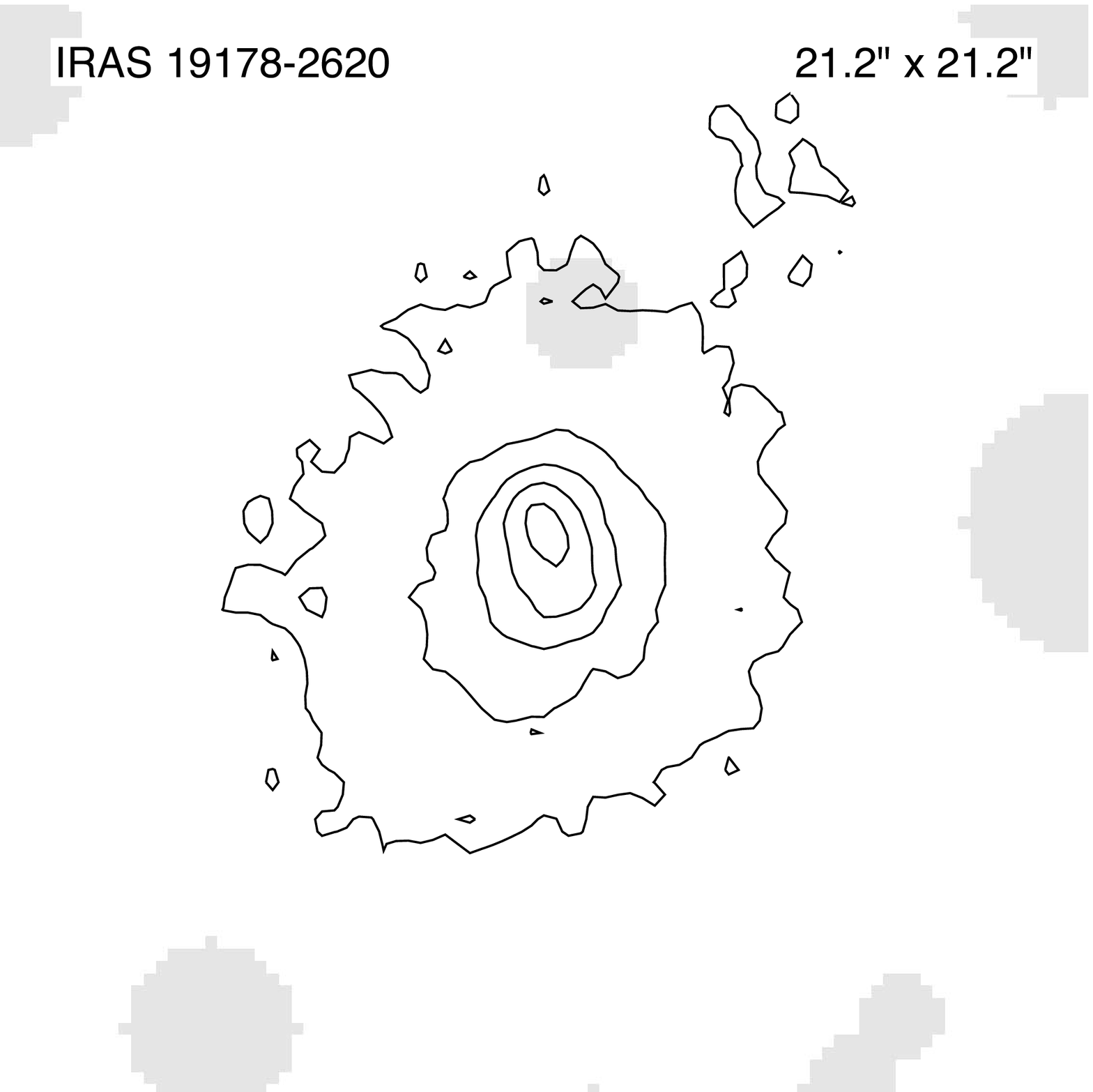}}\hspace{0.3cm}
\includegraphics[origin=c,angle=-90,width=0.27\linewidth,clip=true]{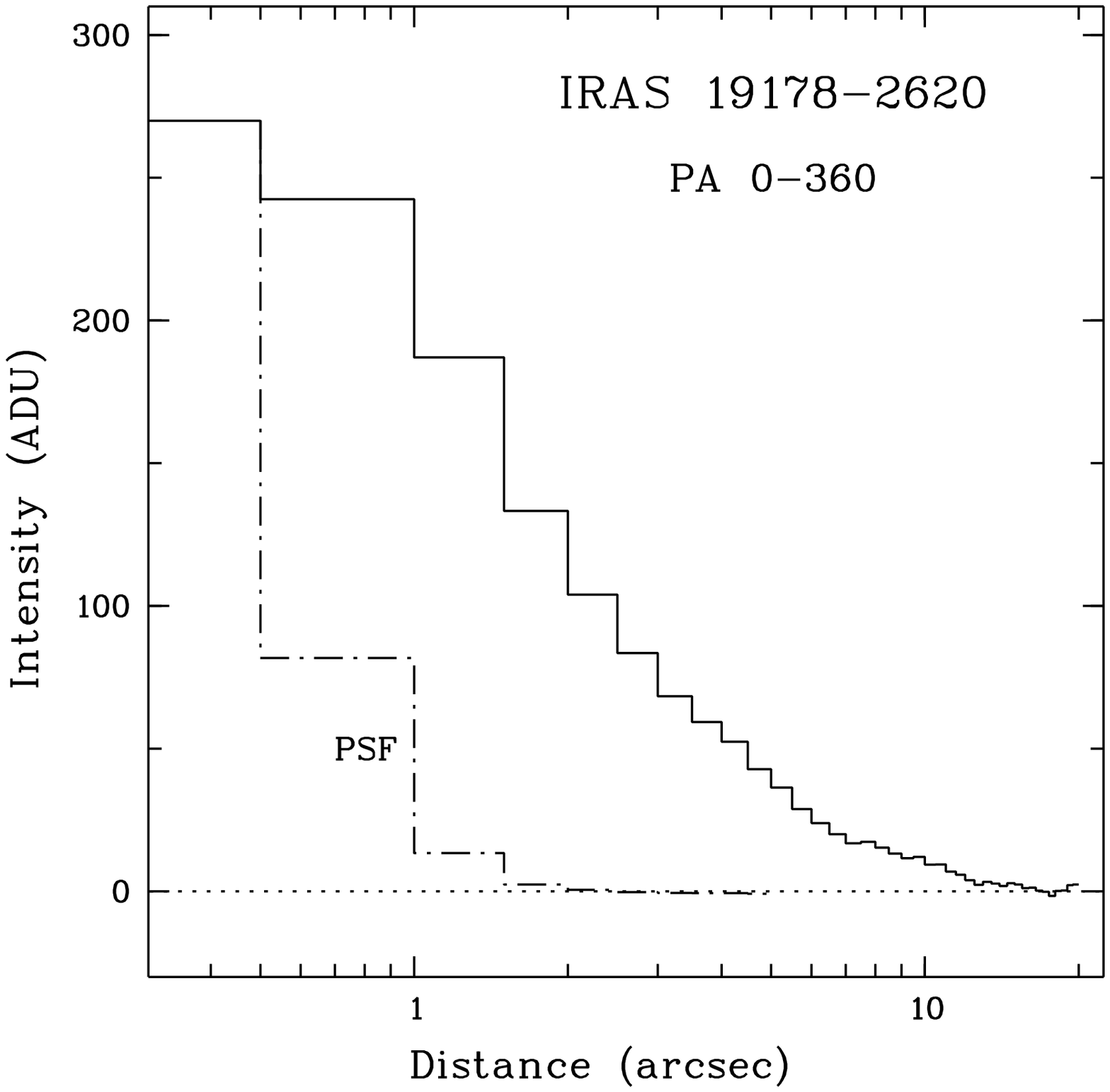}  }
\vspace{0.3cm}

  \caption{Continuation of Fig.~1. From top to bottom:
IRAS 17319$-$6234, $V$-band, contours 0.12 (0.19);
IRAS 18240$+$2326, $V$-band, contours 0.34 (0.23);
IRAS 18467$-$4802, $F606W$ (HST/ACS), contours 0.27 (0.25);
IRAS 19178$-$2620, $B$-band, contours 0.18 (0.22).
The images are square, with north to the top and east
to the left.}
 
\end{figure*}


\begin{figure*}
   \setlength\fboxsep{0pt}
   \setlength\fboxrule{0.5pt}

   \vspace{0.3cm}
   \centerline{
\fbox{\includegraphics[width=0.27\linewidth]{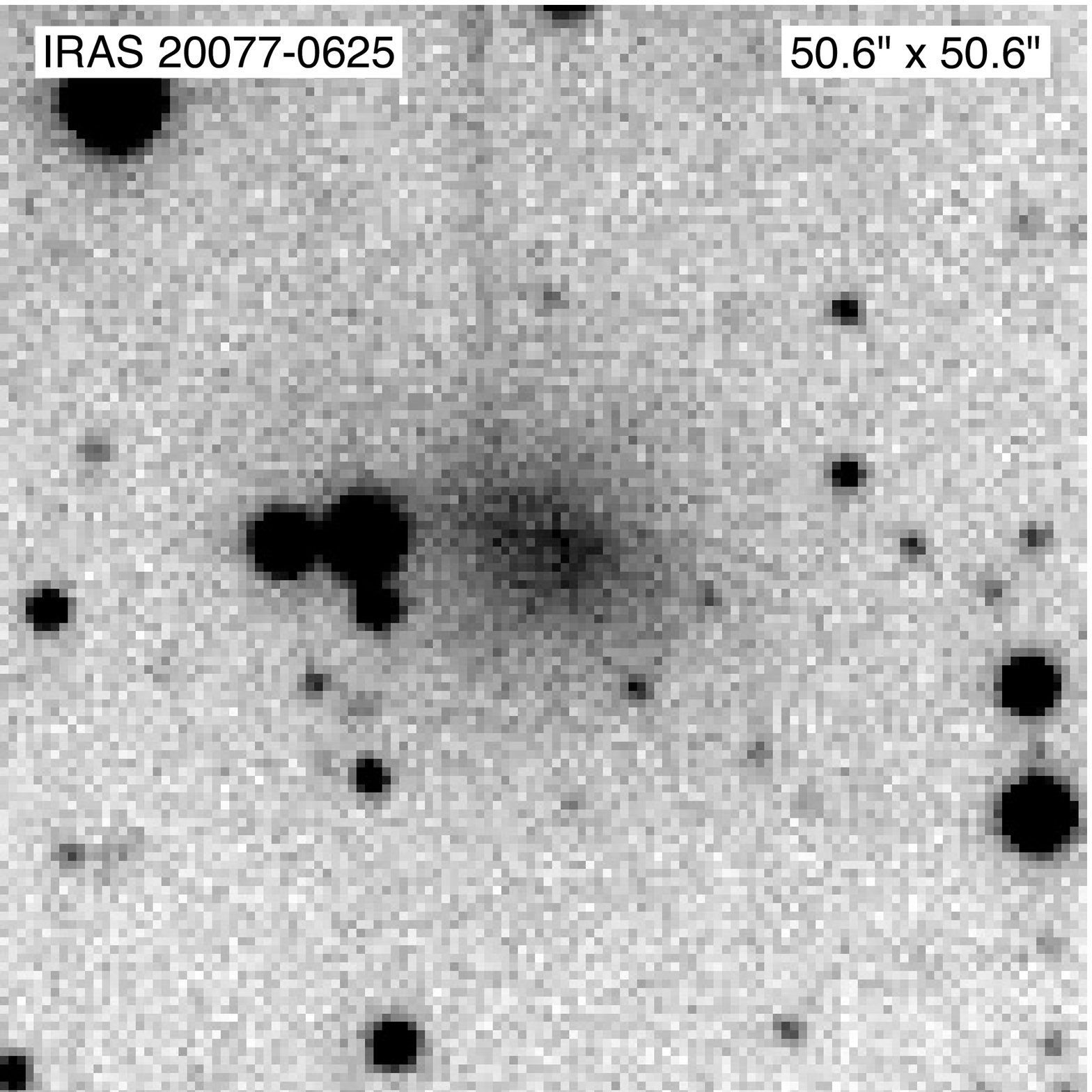}}\hspace{0.3cm}
\fbox{\includegraphics[width=0.27\linewidth]{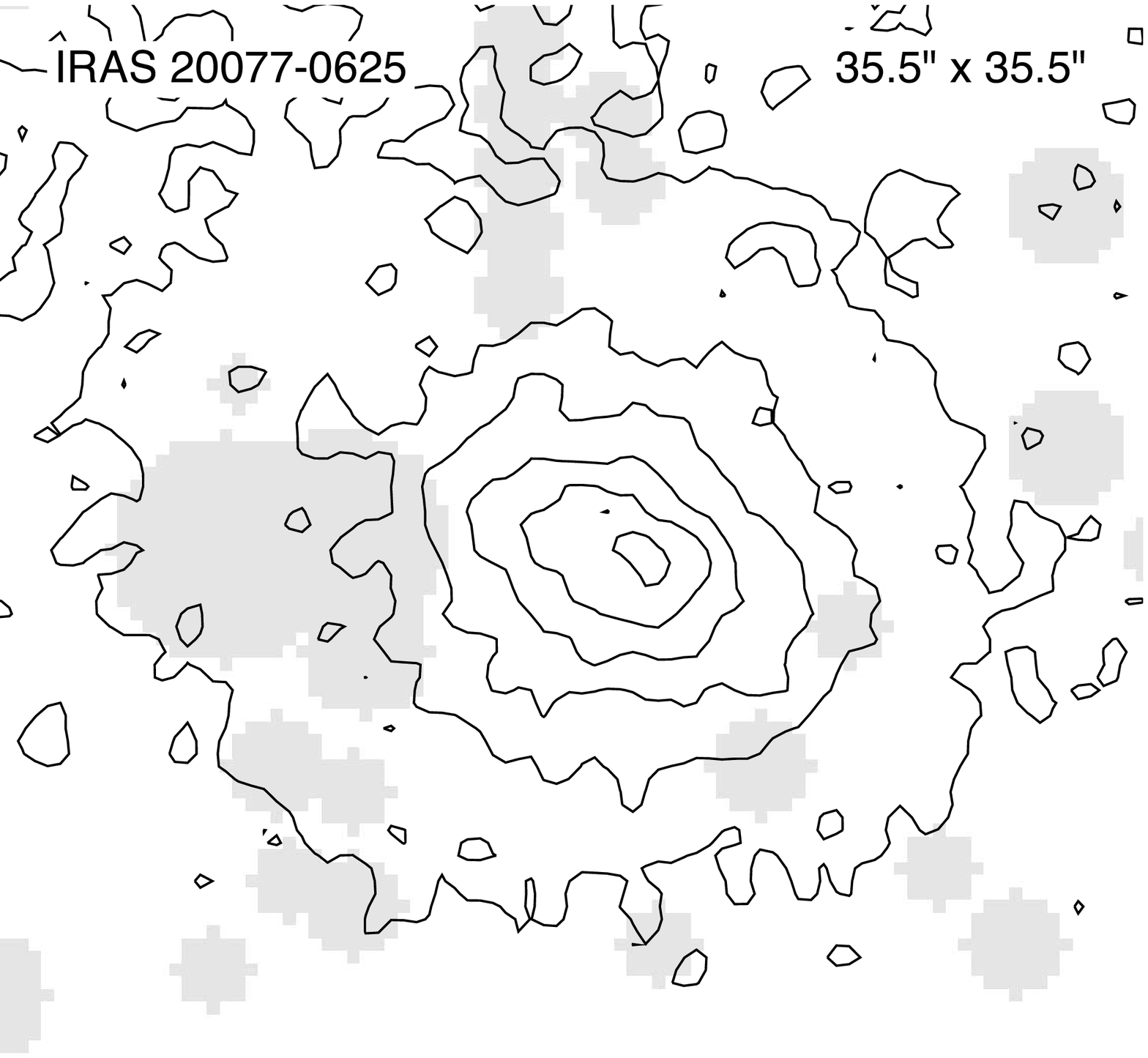}}\hspace{0.3cm}
\includegraphics[origin=c,angle=-90,width=0.27\linewidth,clip=true]{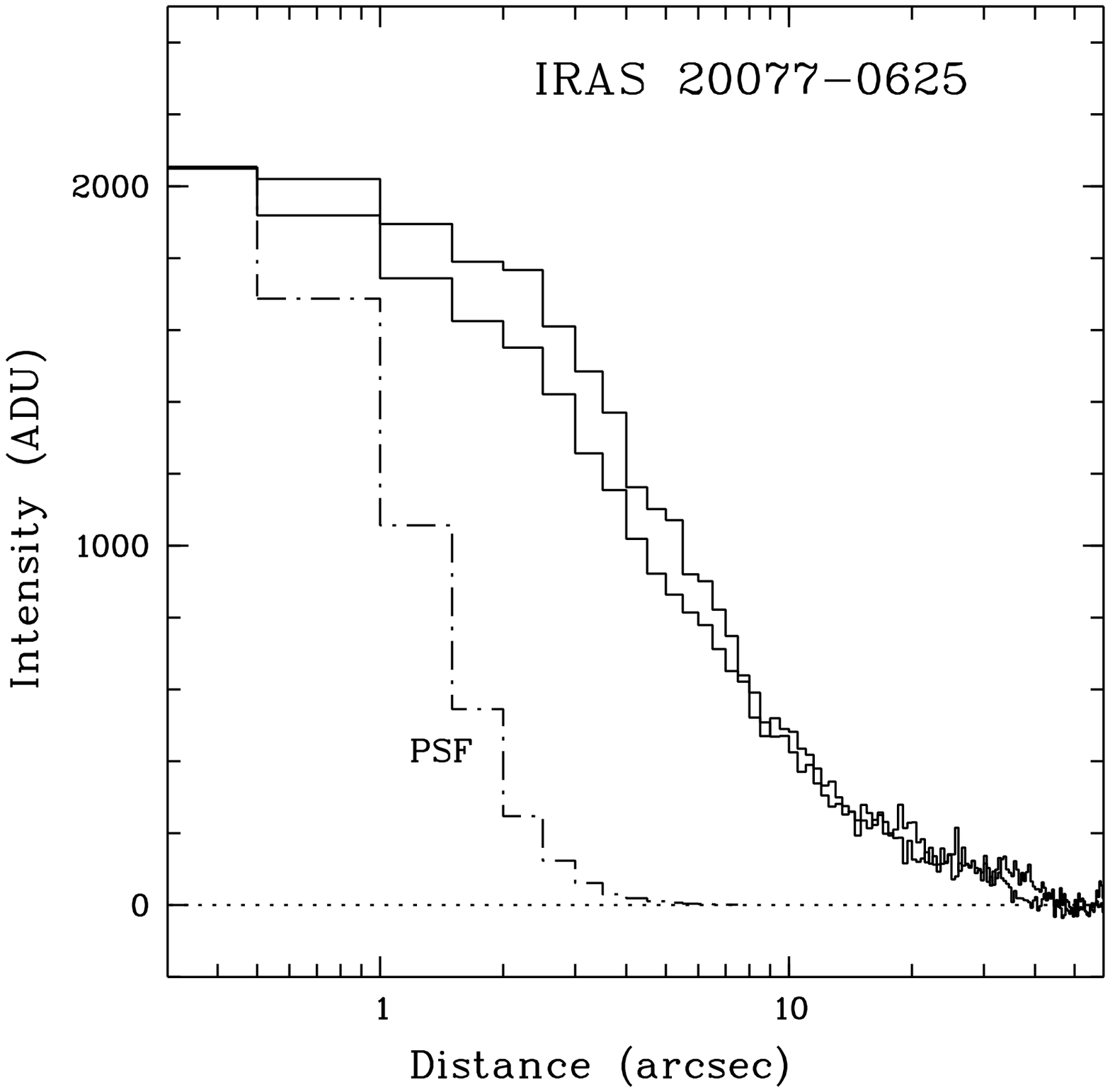} }

  \vspace{0.6cm}
  \centerline{
\fbox{\includegraphics[width=0.27\linewidth]{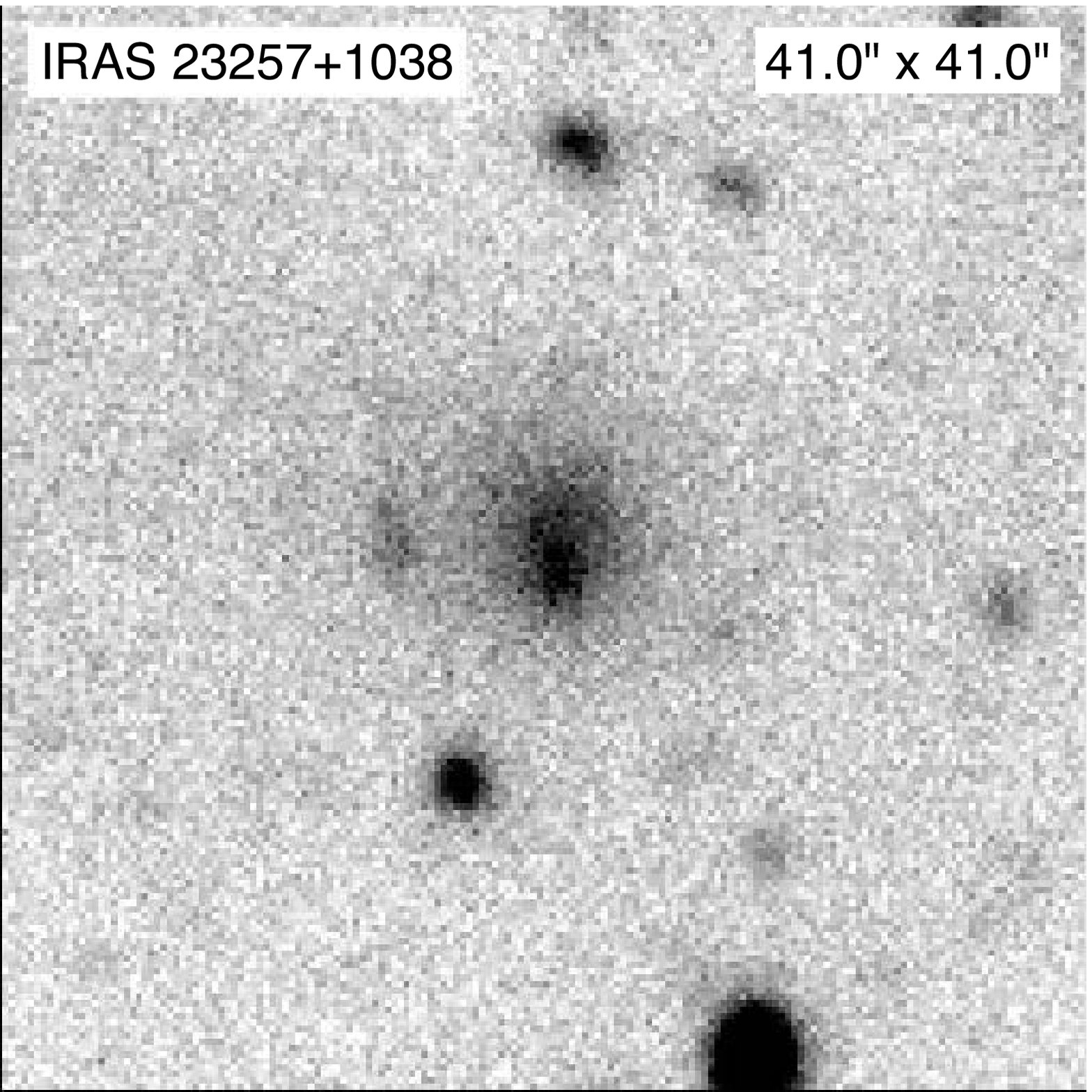}}\hspace{0.3cm}
\fbox{\includegraphics[width=0.27\linewidth]{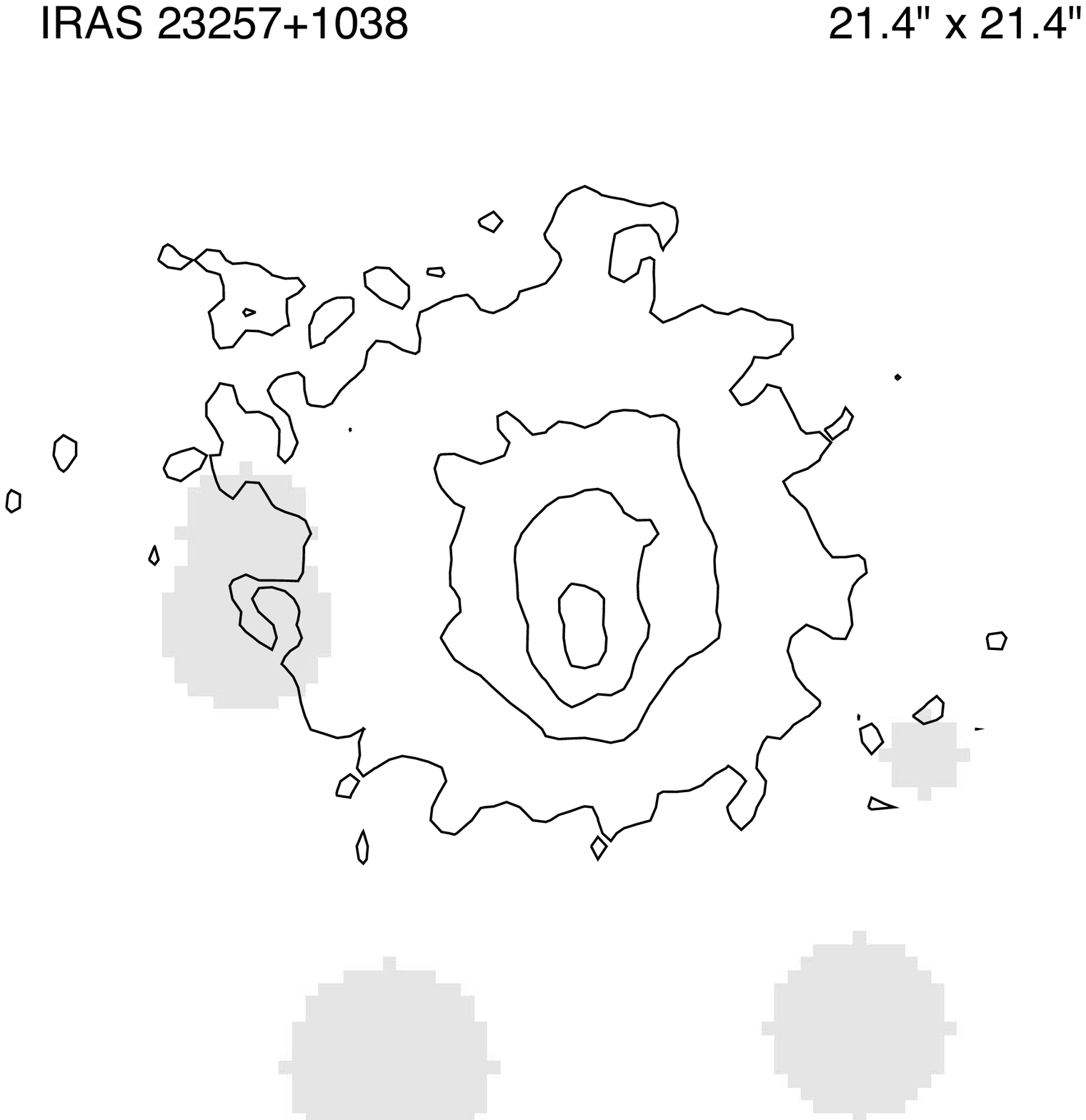}}\hspace{0.3cm}
\includegraphics[origin=c,angle=-90,width=0.27\linewidth,clip=true]{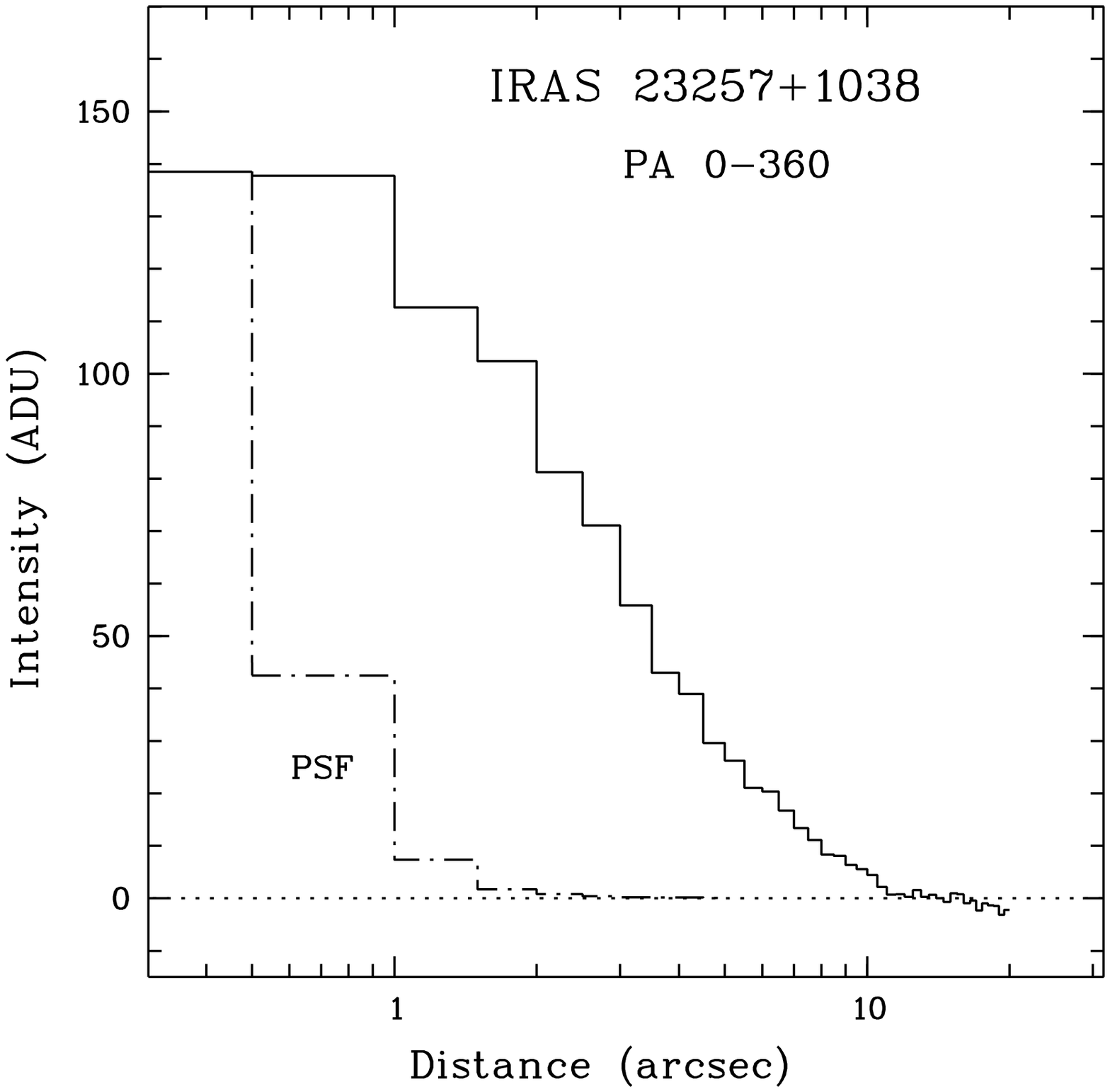} }

  \vspace{0.6cm}
  \centerline{
\fbox{\includegraphics[width=0.27\linewidth]{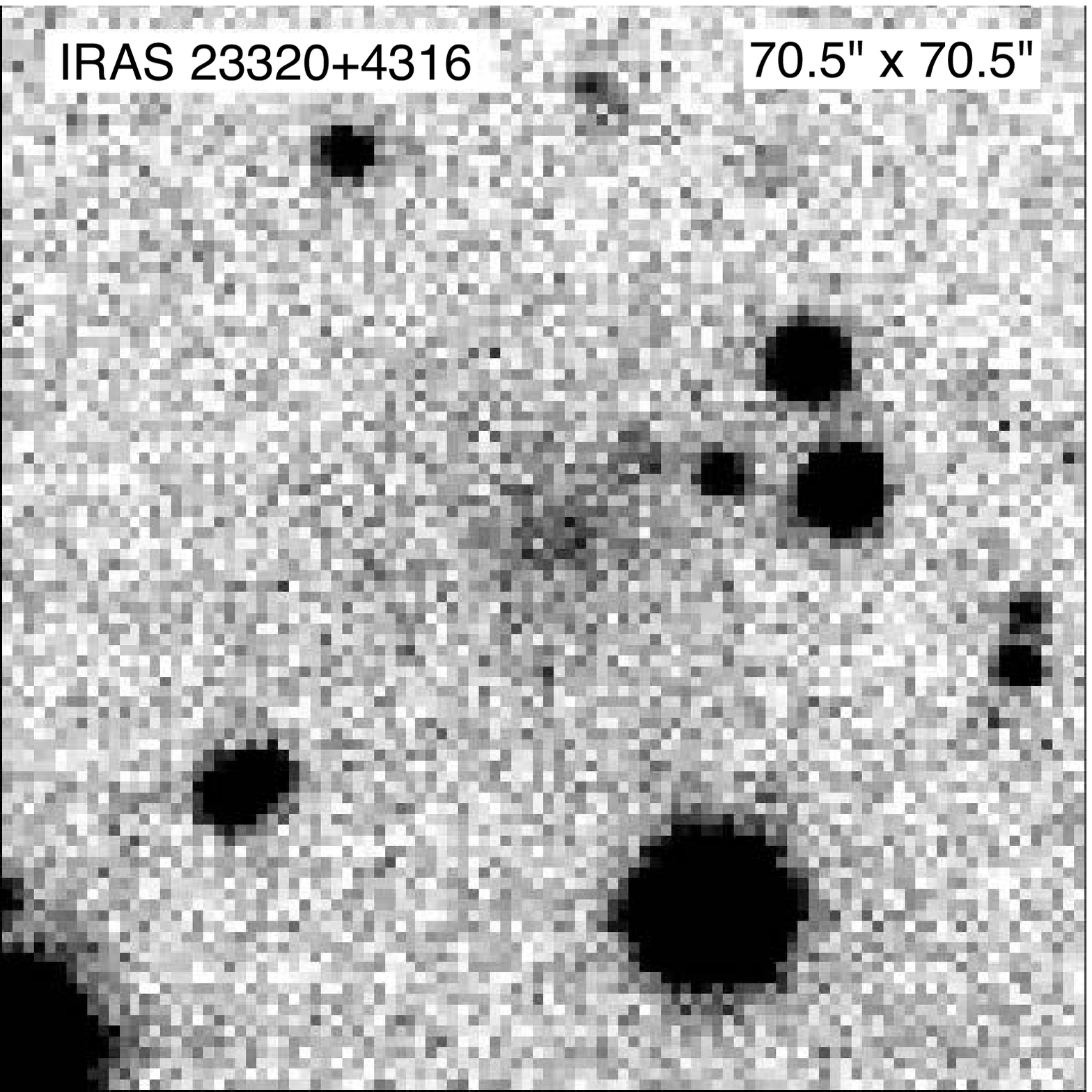}}\hspace{0.3cm}
\fbox{\includegraphics[width=0.27\linewidth]{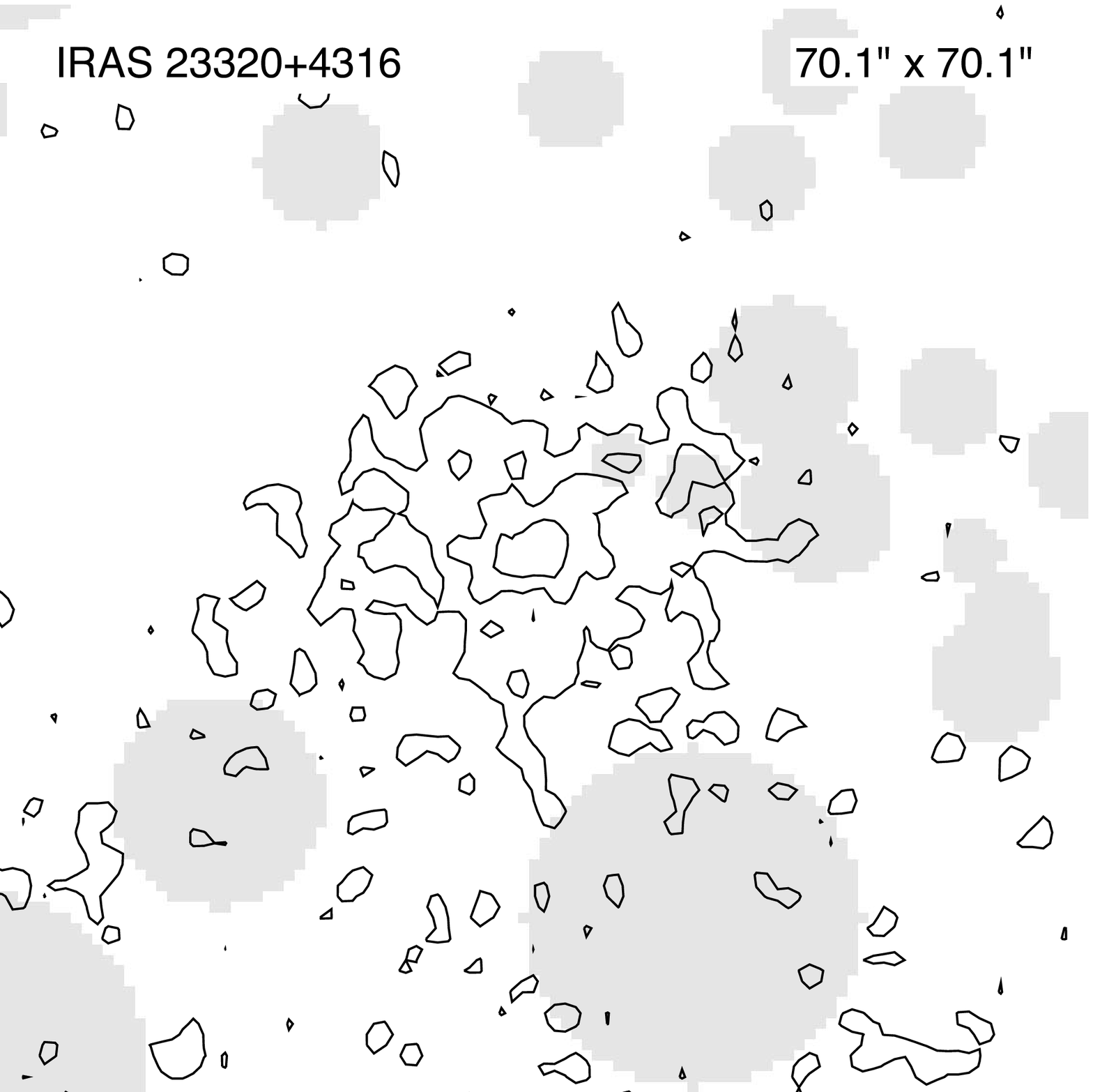}}\hspace{0.3cm}
\includegraphics[origin=c,angle=-90,width=0.27\linewidth,clip=true]{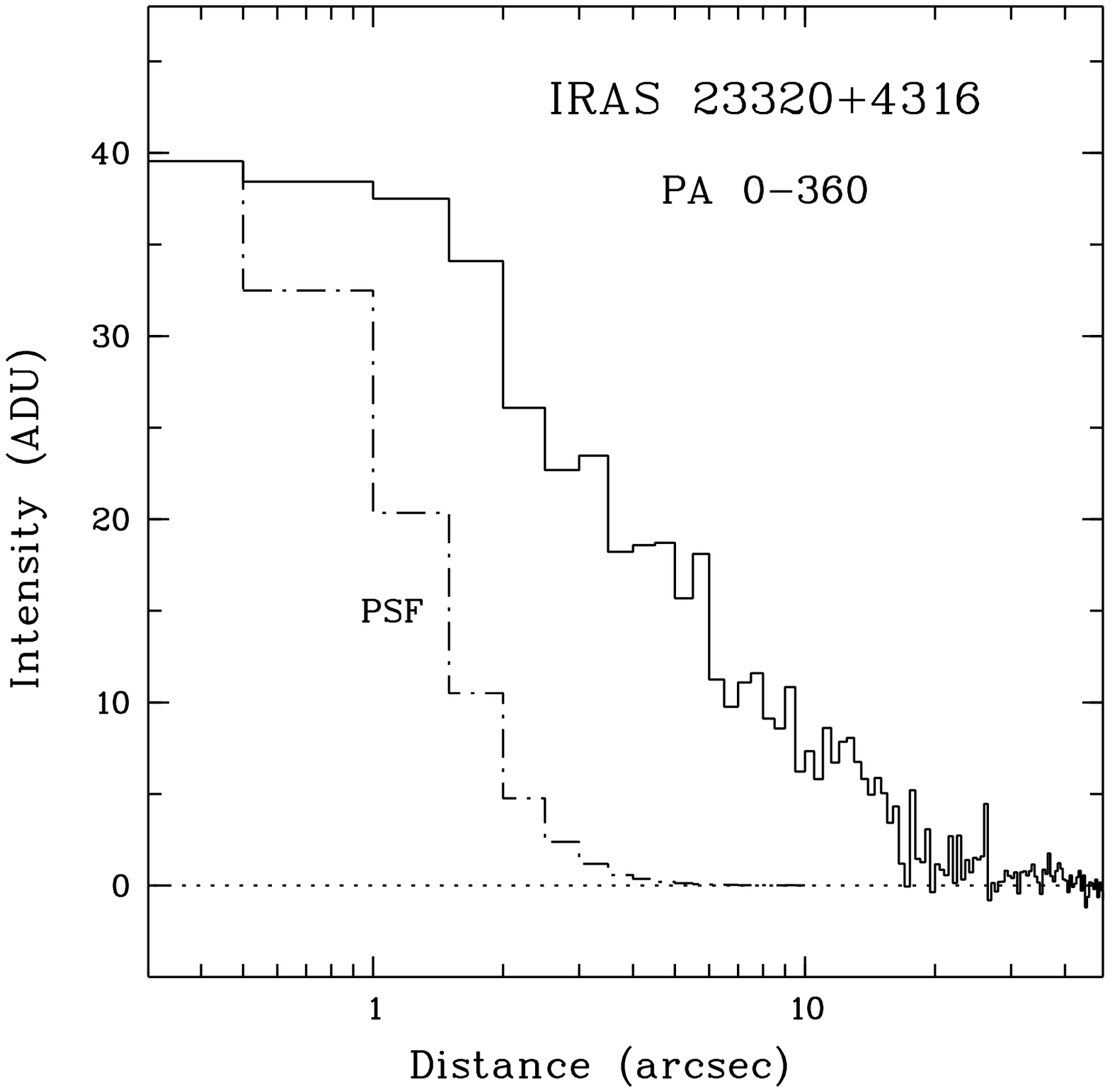}  }
\vspace{0.3cm}

  \caption{Continuation of Fig.~1. From top to bottom:
IRAS 20077$-$0625, $V$-band, contours 0.20 (0.18);
IRAS 23257$+$1038, $V$-band, contours 0.20 (0.31);
IRAS 23320$+$4316, $B$-band, contours 0.29 (0.29).
The images are square, with north to the top and east
to the left.
}
 \end{figure*}

\onlfig{5}{
\begin{figure*}
   \vspace{0.3cm}
   \centerline{
   \hspace{0.3cm}
  \includegraphics[width=0.3\linewidth]{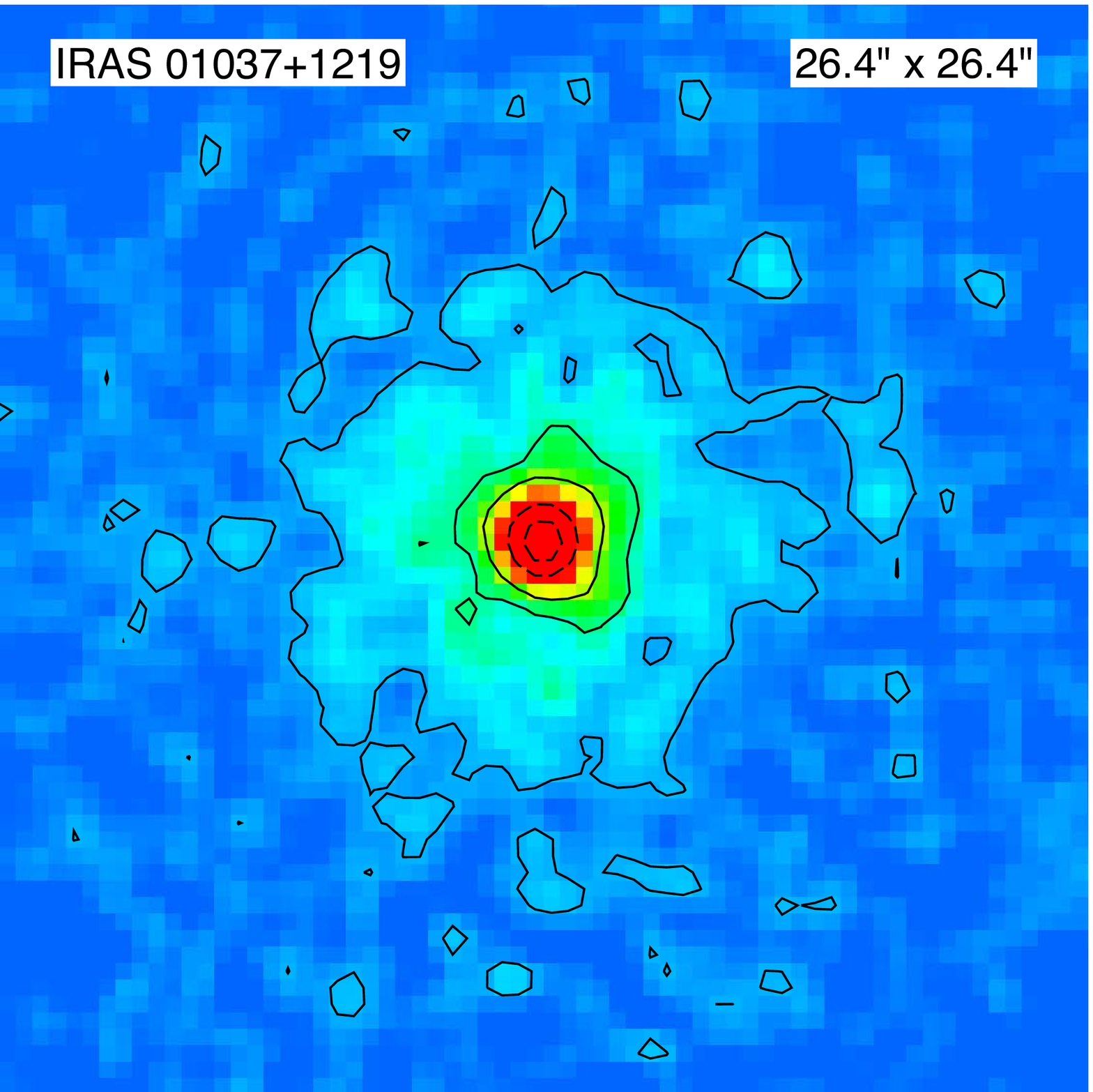}\hspace{0.3cm}
\includegraphics[width=0.3\linewidth]{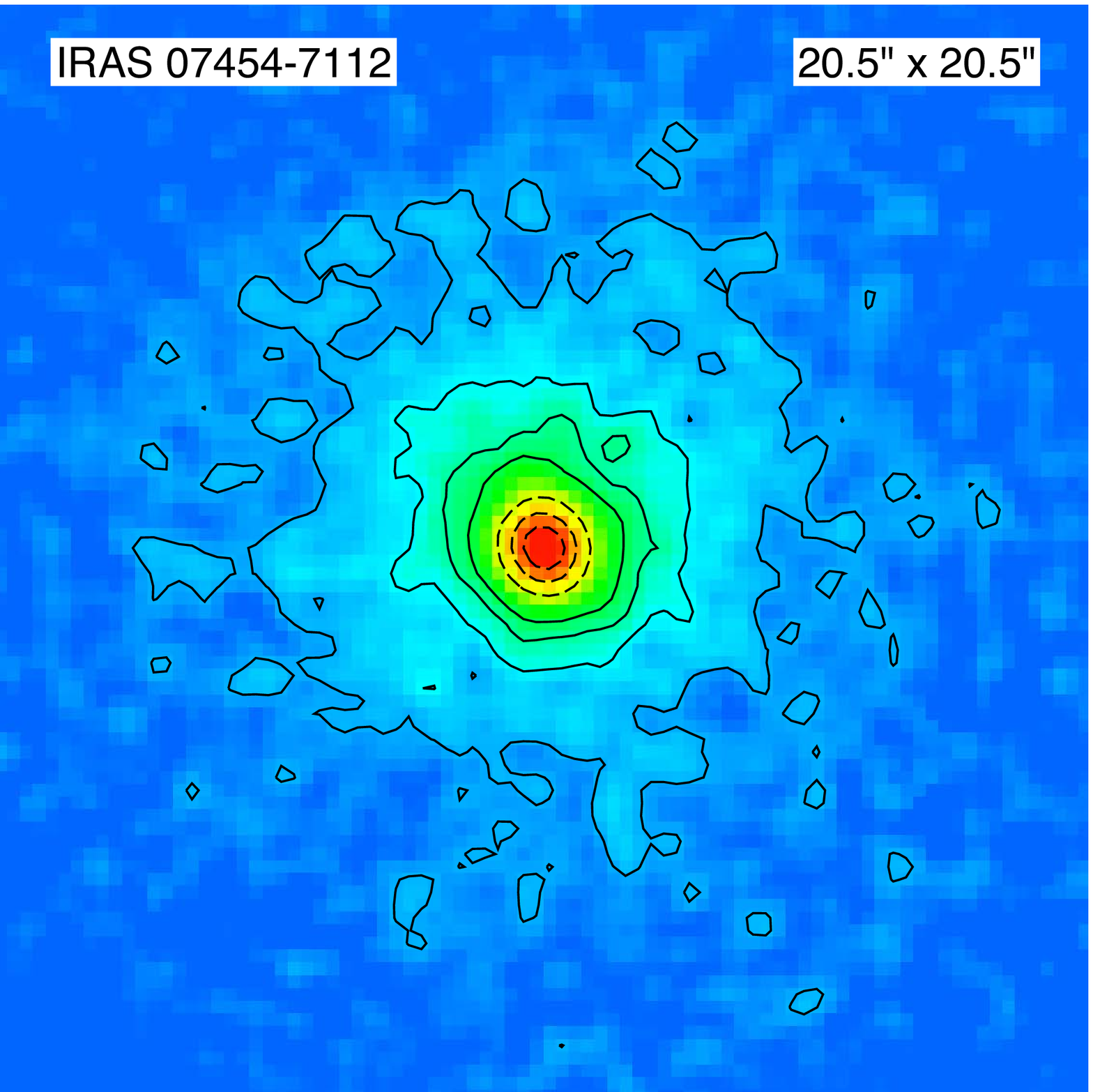}\hspace{0.3cm}
\includegraphics[width=0.3\linewidth]{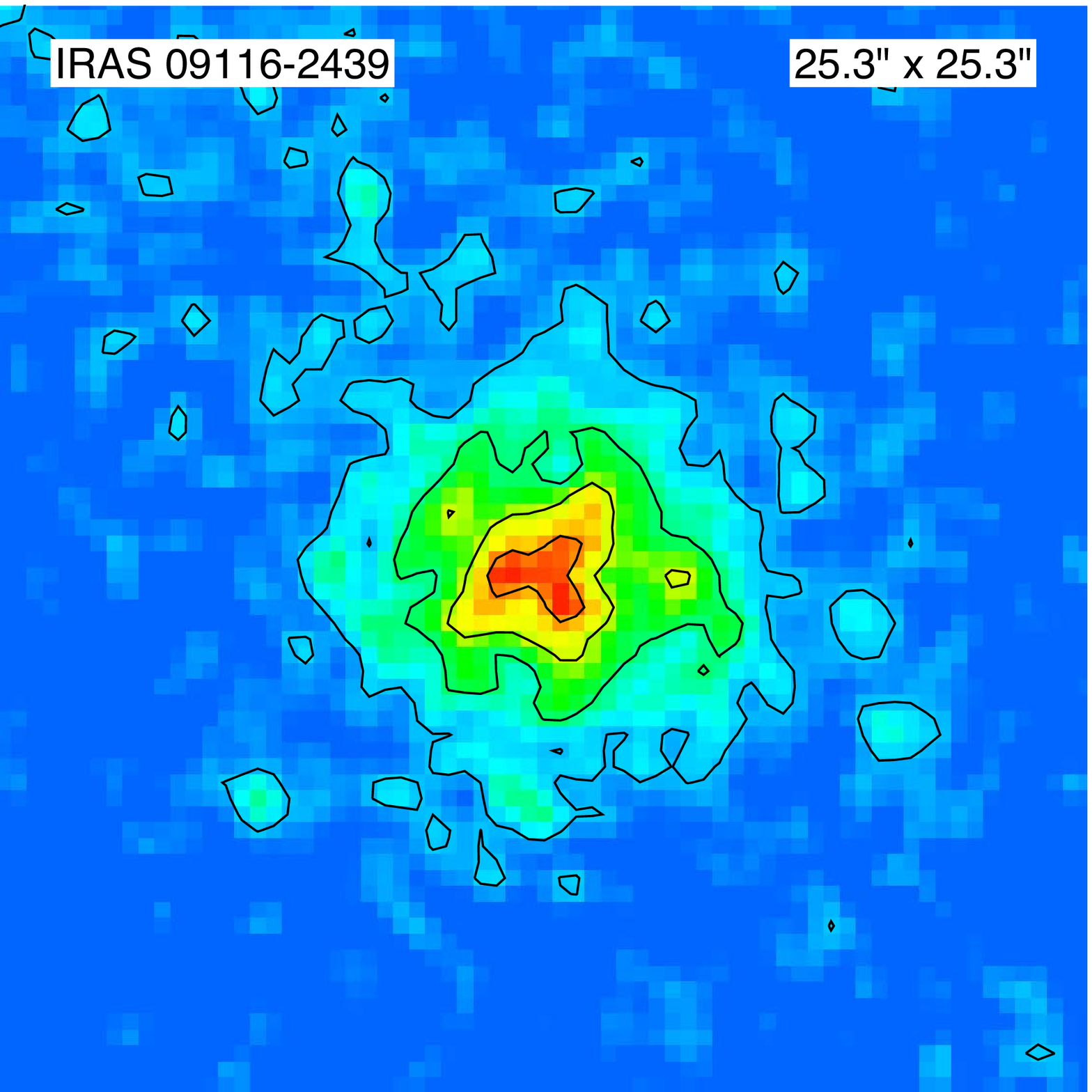}\hspace{0.3cm}  }
\vspace{0.6cm}
\centerline{
   \hspace{0.3cm}
  \includegraphics[width=0.3\linewidth]{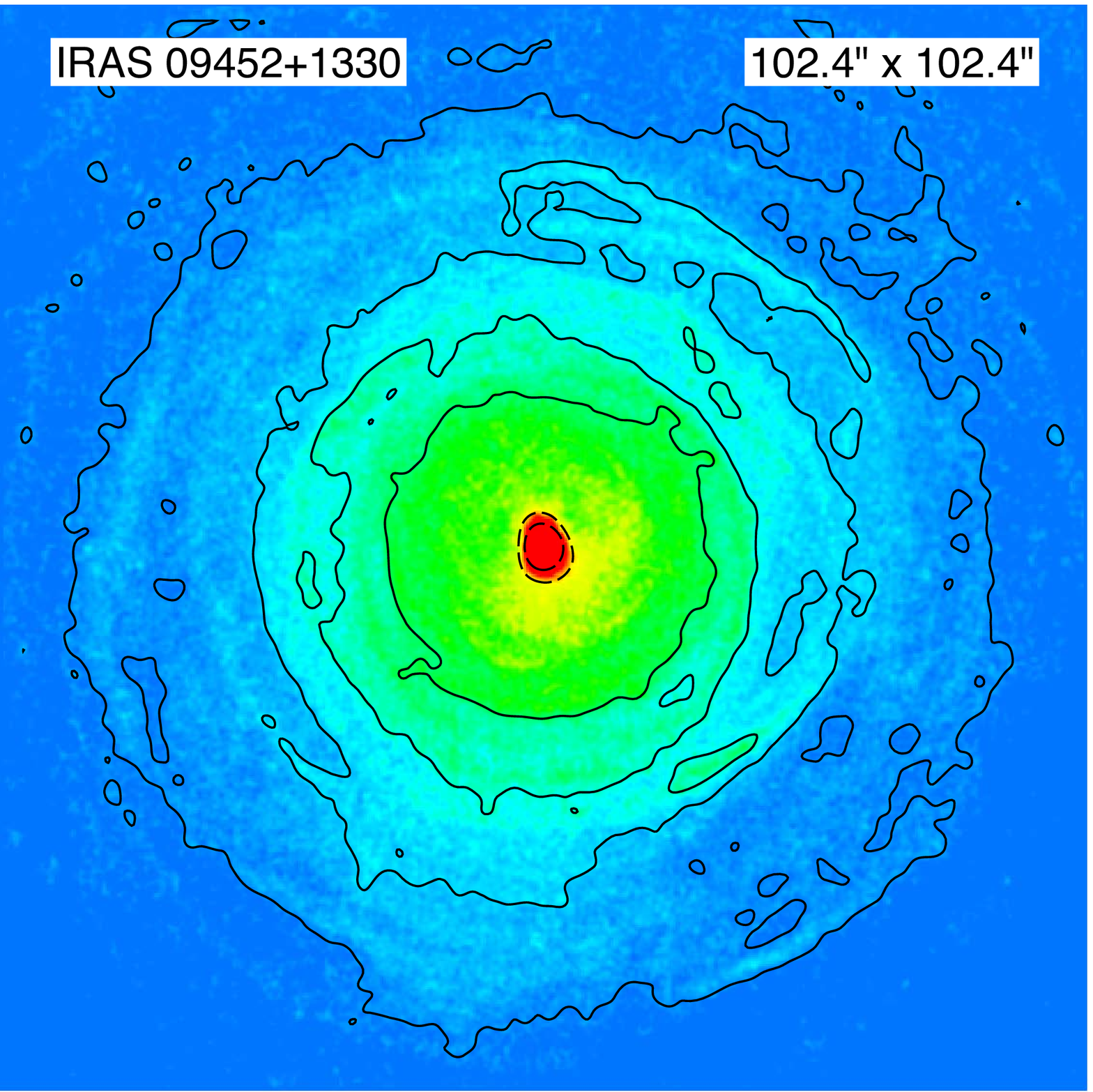}\hspace{0.3cm}
\includegraphics[width=0.3\linewidth]{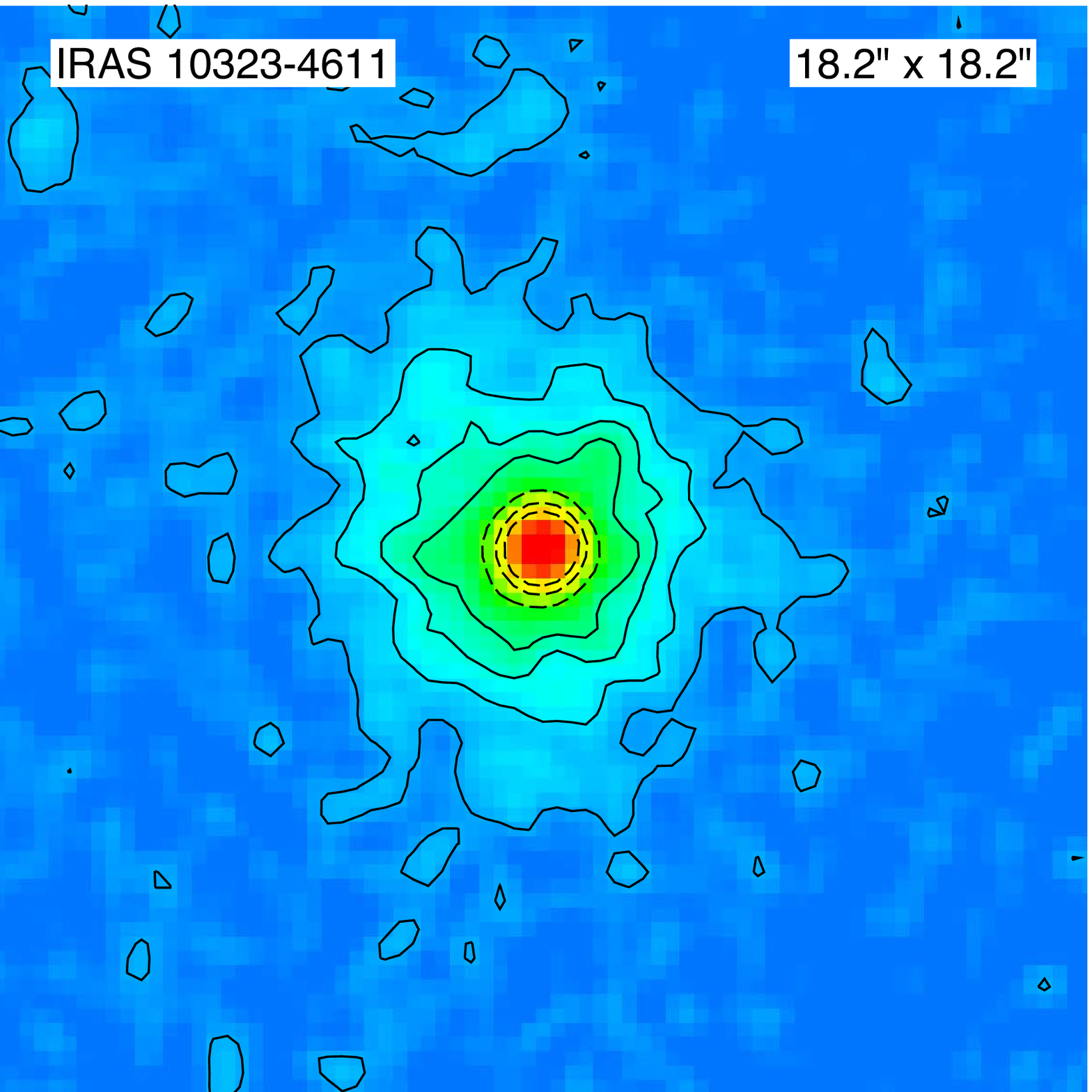}\hspace{0.3cm}
\includegraphics[width=0.3\linewidth]{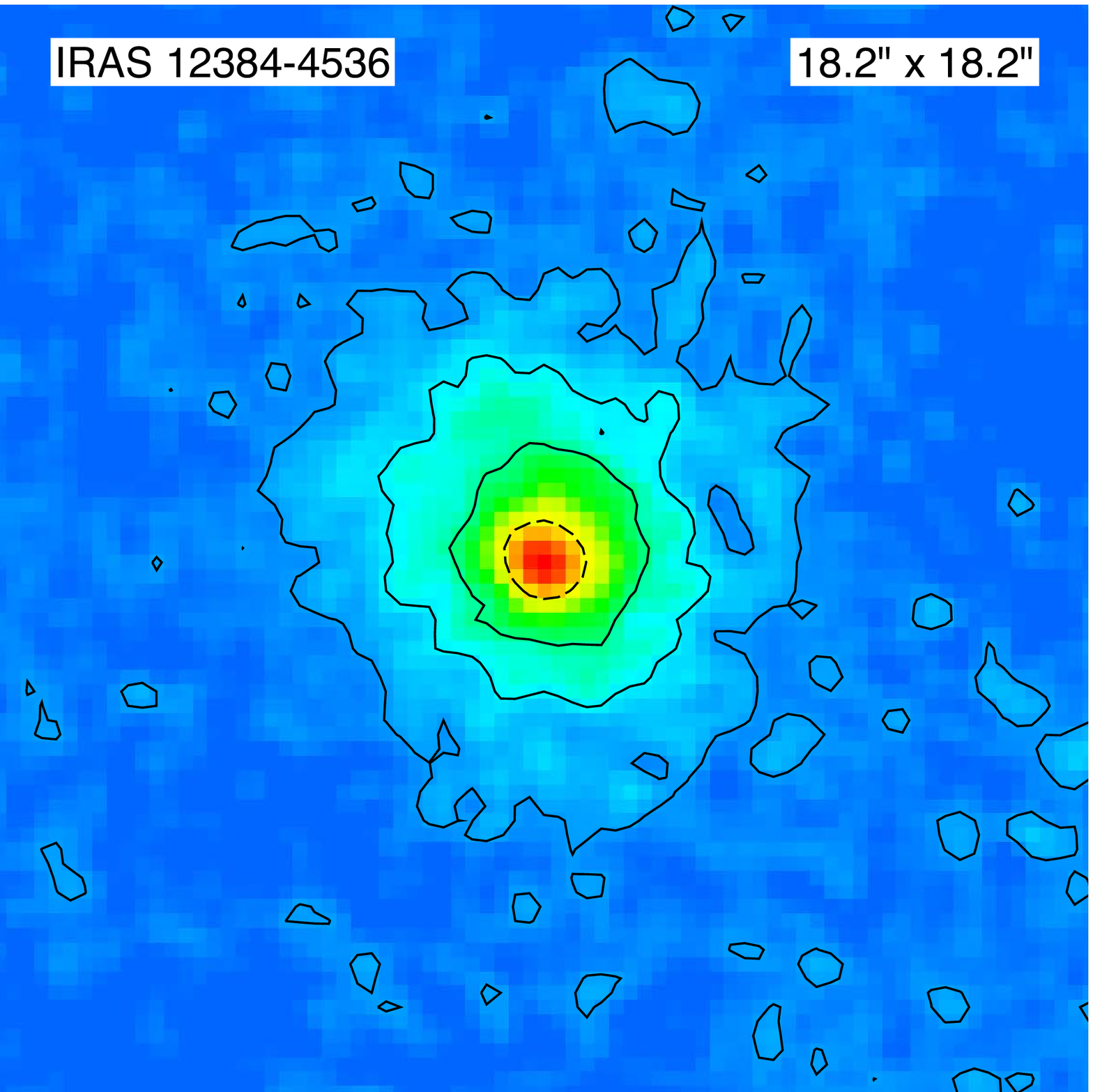}\hspace{0.3cm}  }
\vspace{0.6cm}
 \centerline{ 
  \hspace{0.3cm}
  \includegraphics[width=0.3\linewidth]{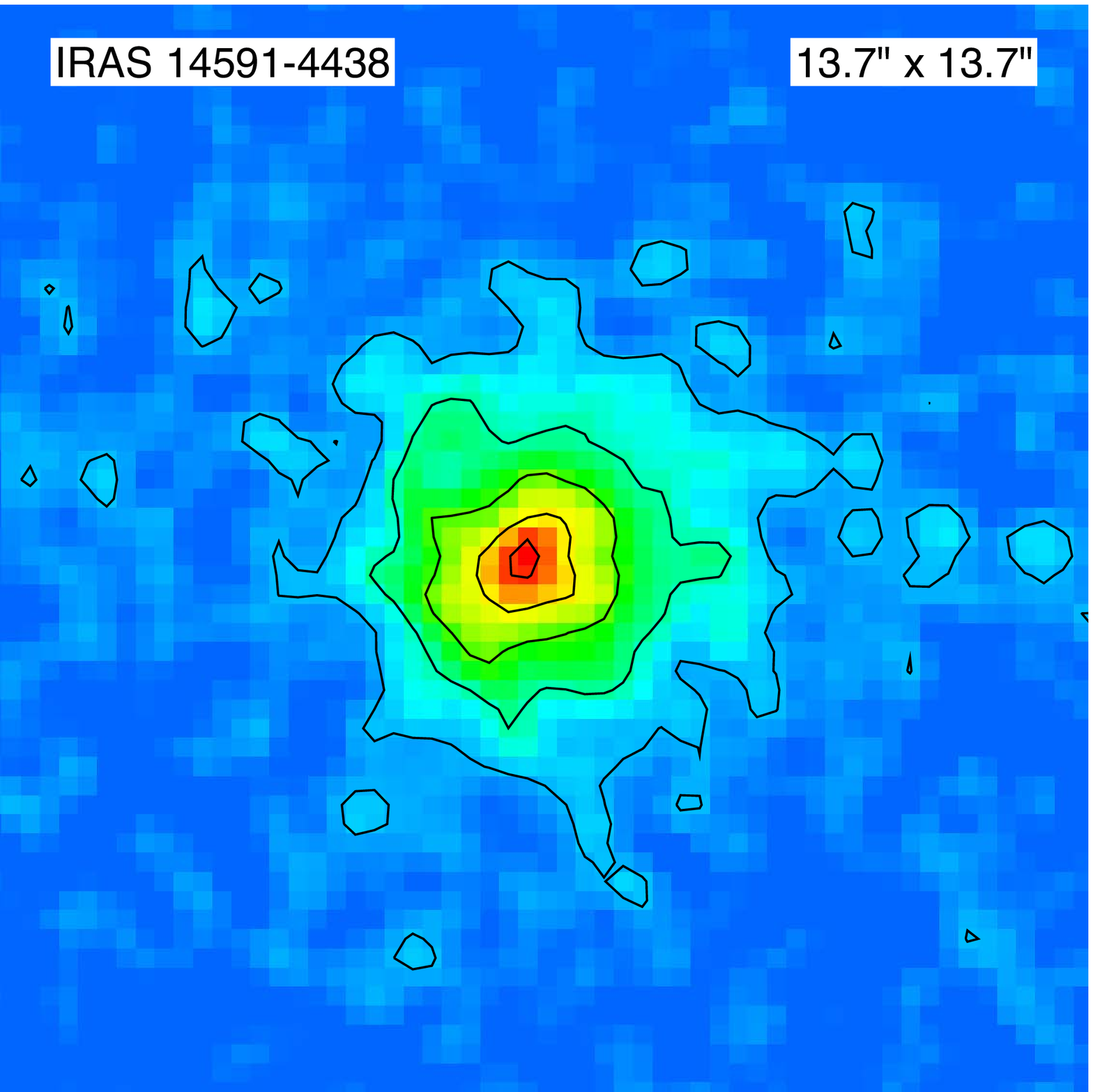}\hspace{0.3cm}
\includegraphics[width=0.3\linewidth]{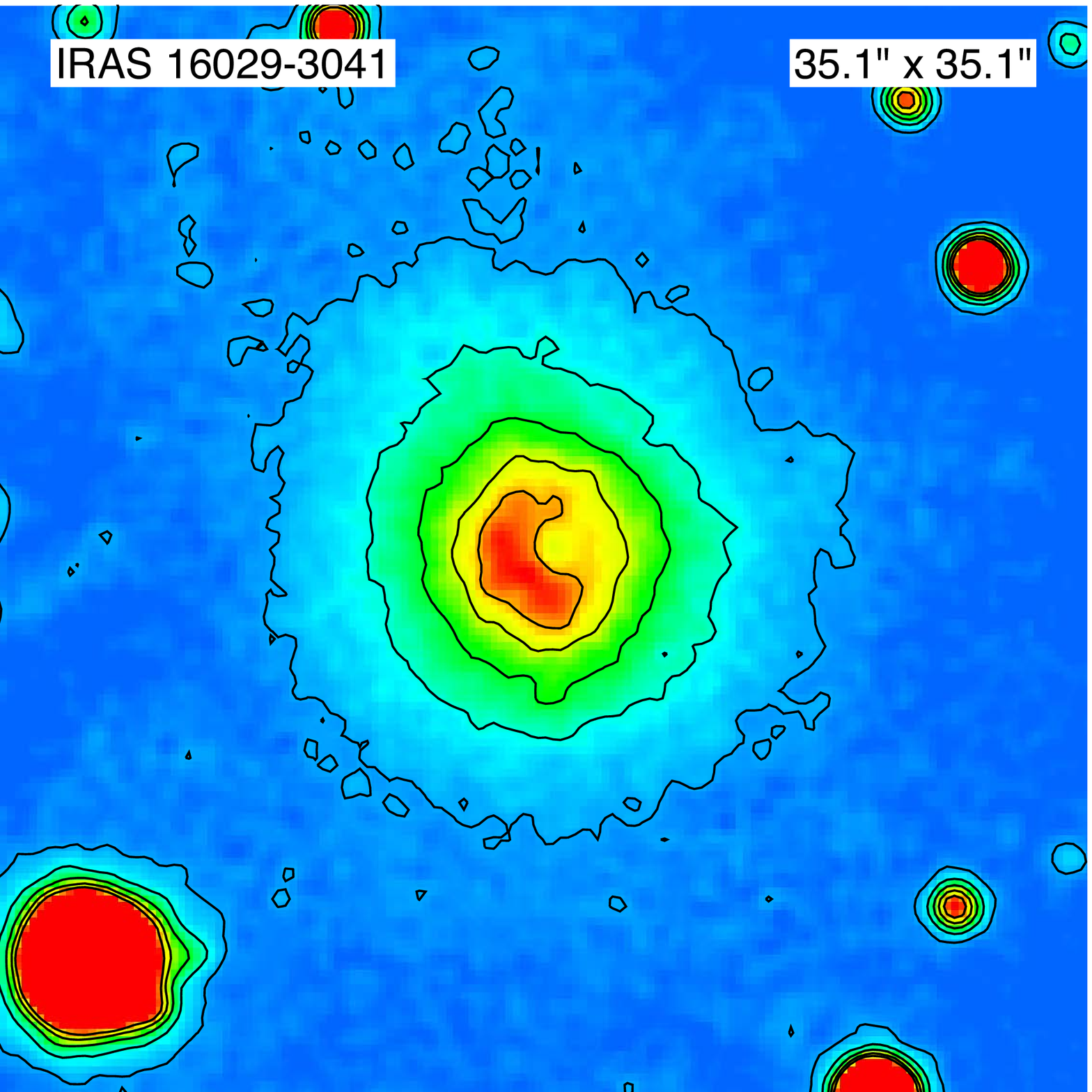}\hspace{0.3cm}
\includegraphics[width=0.3\linewidth]{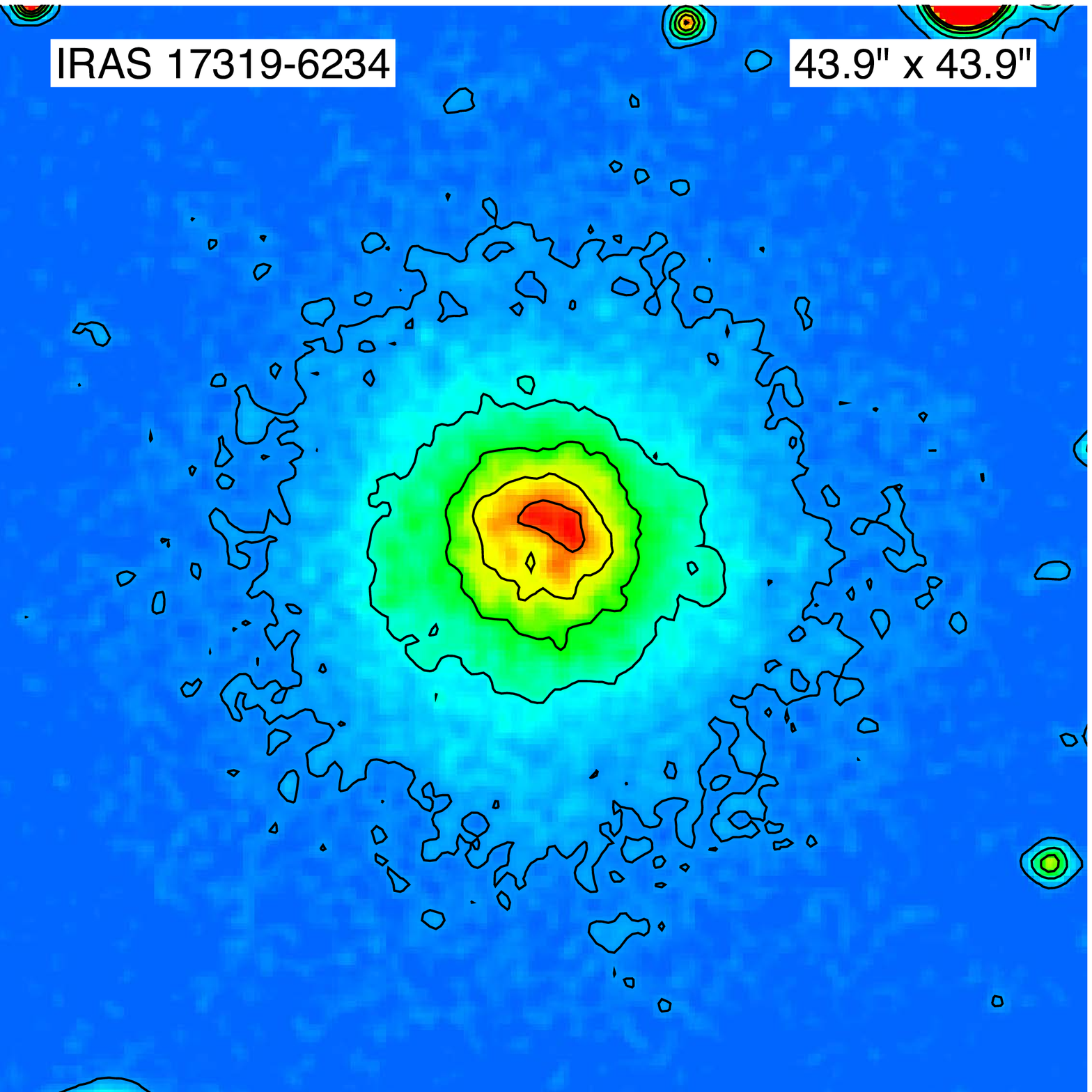}\hspace{0.3cm}  }
\vspace{0.3cm}

\caption{Images of the circumstellar envelopes (on-line only). The images are shown
in color with the field stars replaced, and with the contours of
Figs.1--4 superposed.
Top row, left to right: 
IRAS 01037$+$1219, $V$-band;   
IRAS 07454$-$7112, $B$-band;
IRAS 09116$-$2439, $B$-band. 
Middle row, left to right:
IRAS 09452$+$1330, $V$-band;
IRAS 10323$-$4611, $B$-band;
IRAS 12384$-$4536, $V$-band. 
Bottom row, left to right:
IRAS 14591$-$4438, $B$-band;
IRAS 16029$-$3041, $V$-band;
IRAS 17319$-$6234, $V$-band.
All images are square, with north to the top and east
to the left.
}
 \end{figure*}
} 

\onlfig{6}{
\begin{figure*}
\vspace{0.3cm}
 \centerline{ 
  \hspace{0.3cm}
  \includegraphics[width=0.3\linewidth]{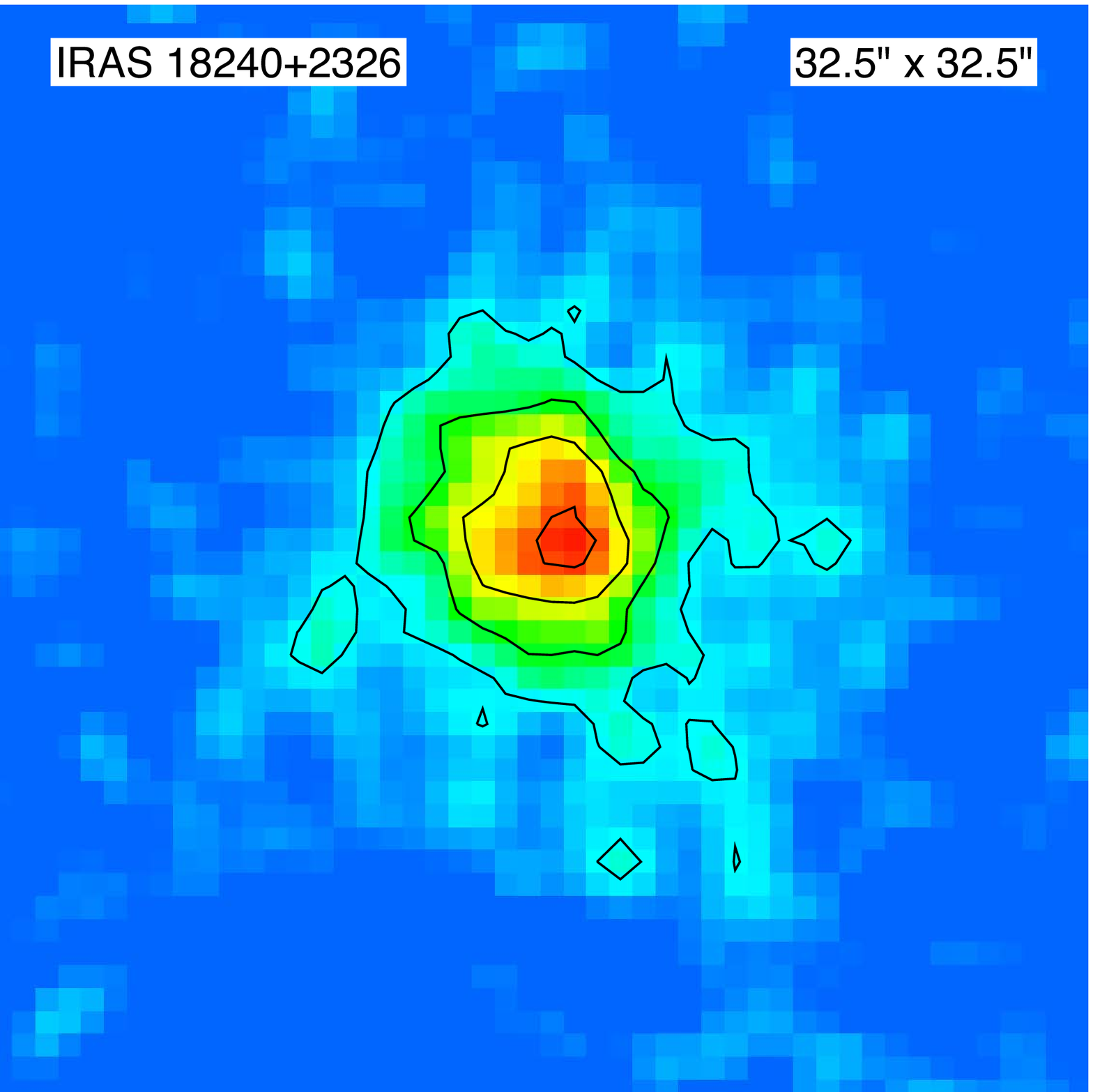}\hspace{0.3cm}
\includegraphics[width=0.3\linewidth]{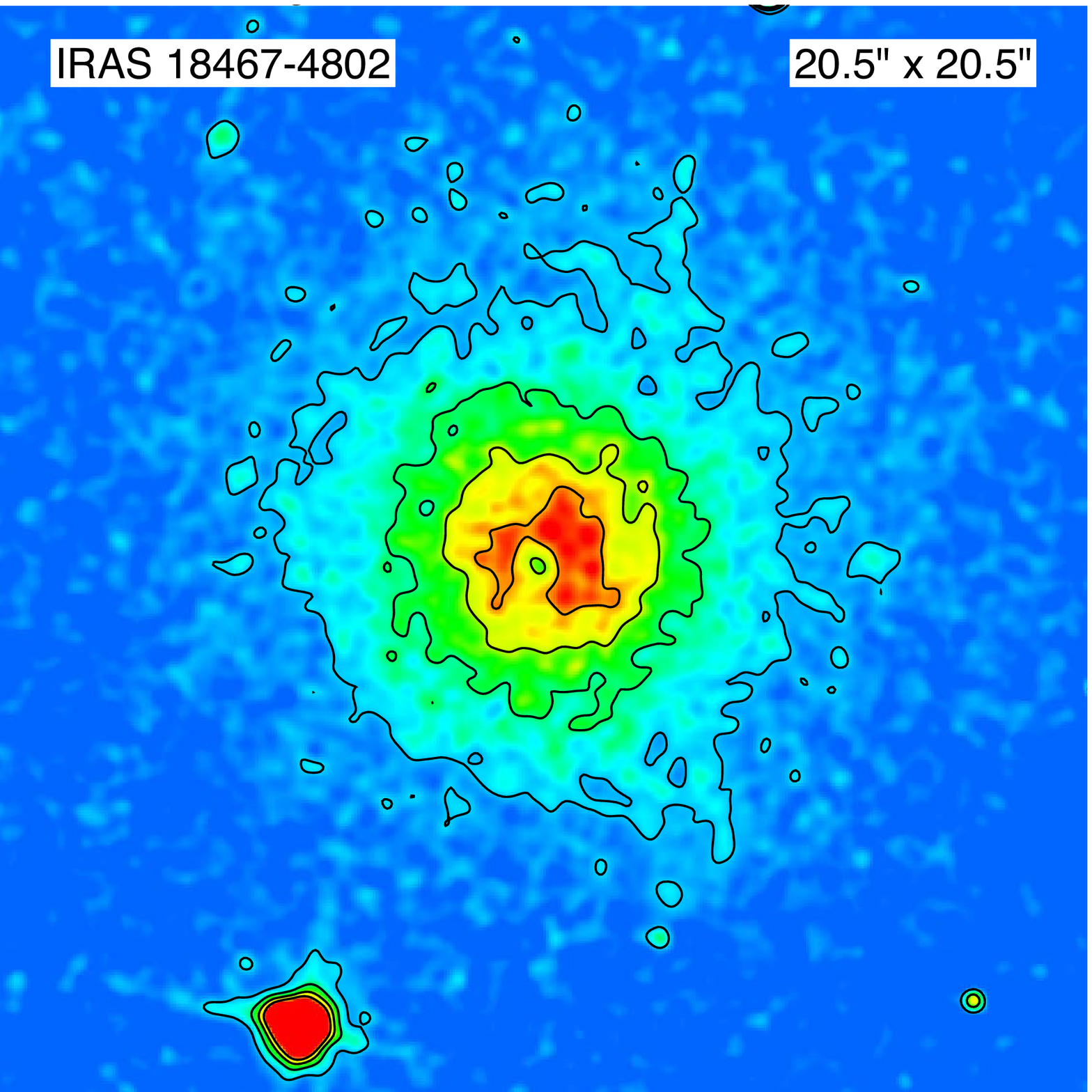}\hspace{0.3cm}
\includegraphics[width=0.3\linewidth]{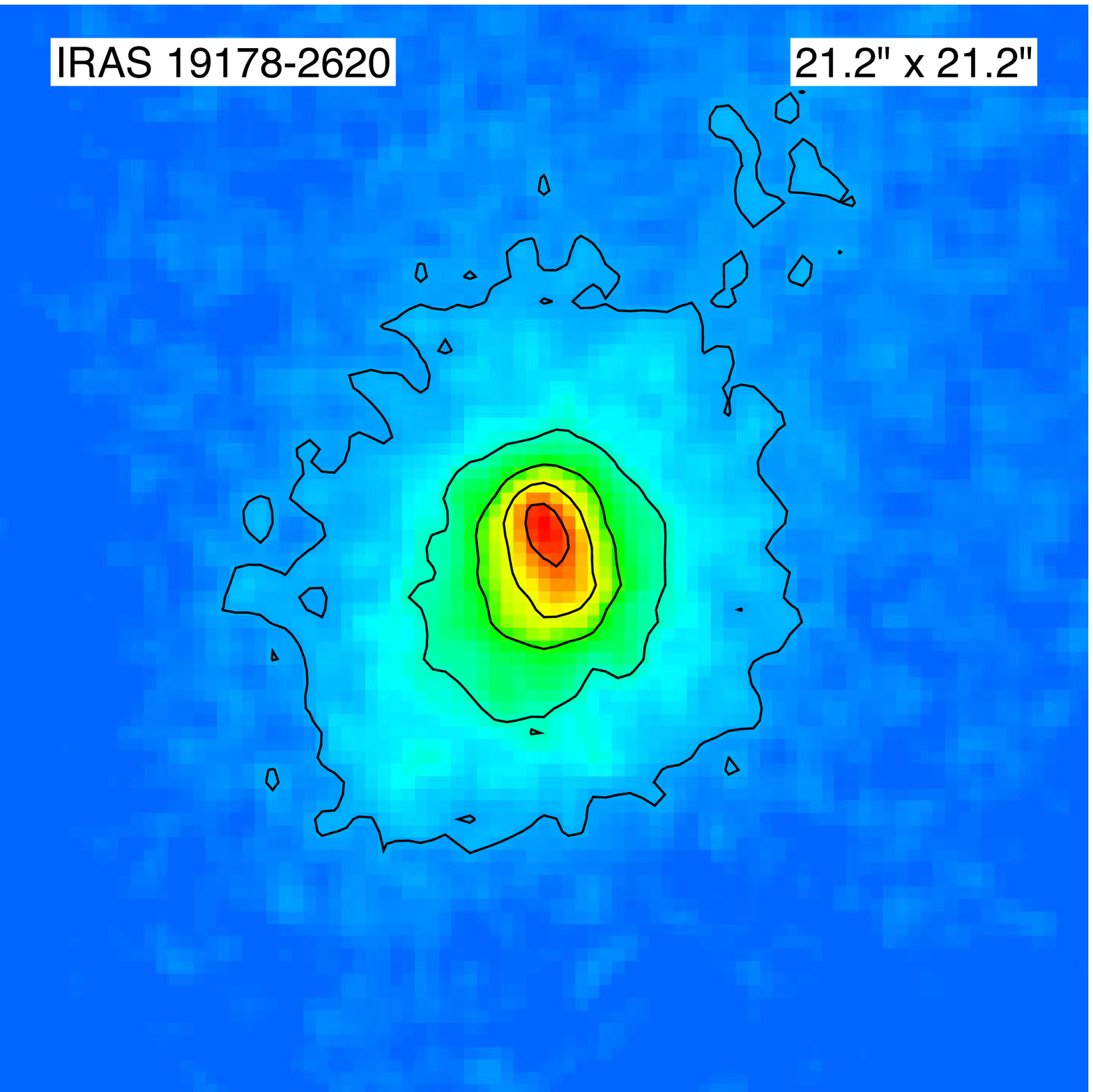}\hspace{0.3cm} }
\vspace{0.6cm}
 \centerline{ 
  \hspace{0.3cm}
  \includegraphics[width=0.3\linewidth]{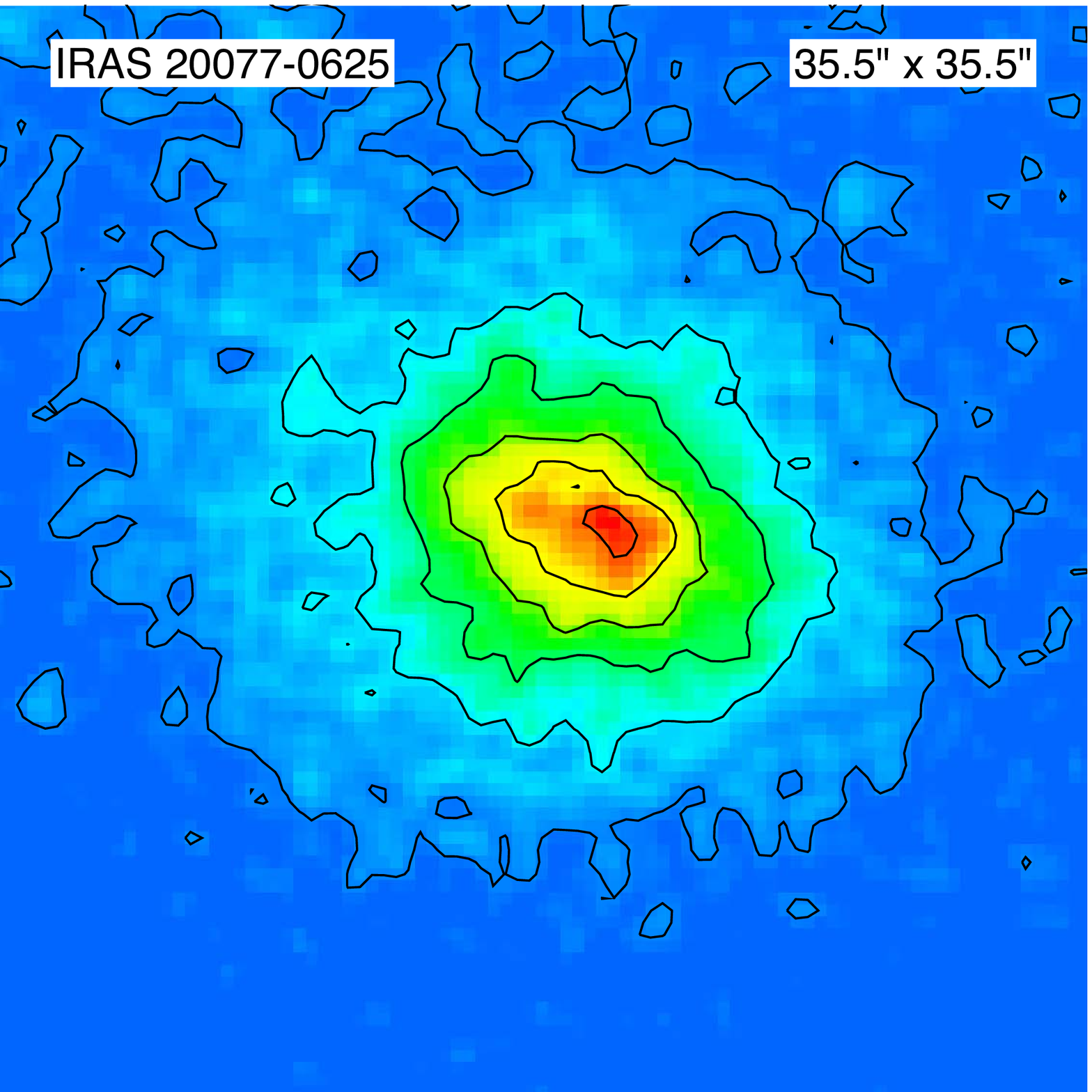}\hspace{0.3cm}
\includegraphics[width=0.3\linewidth]{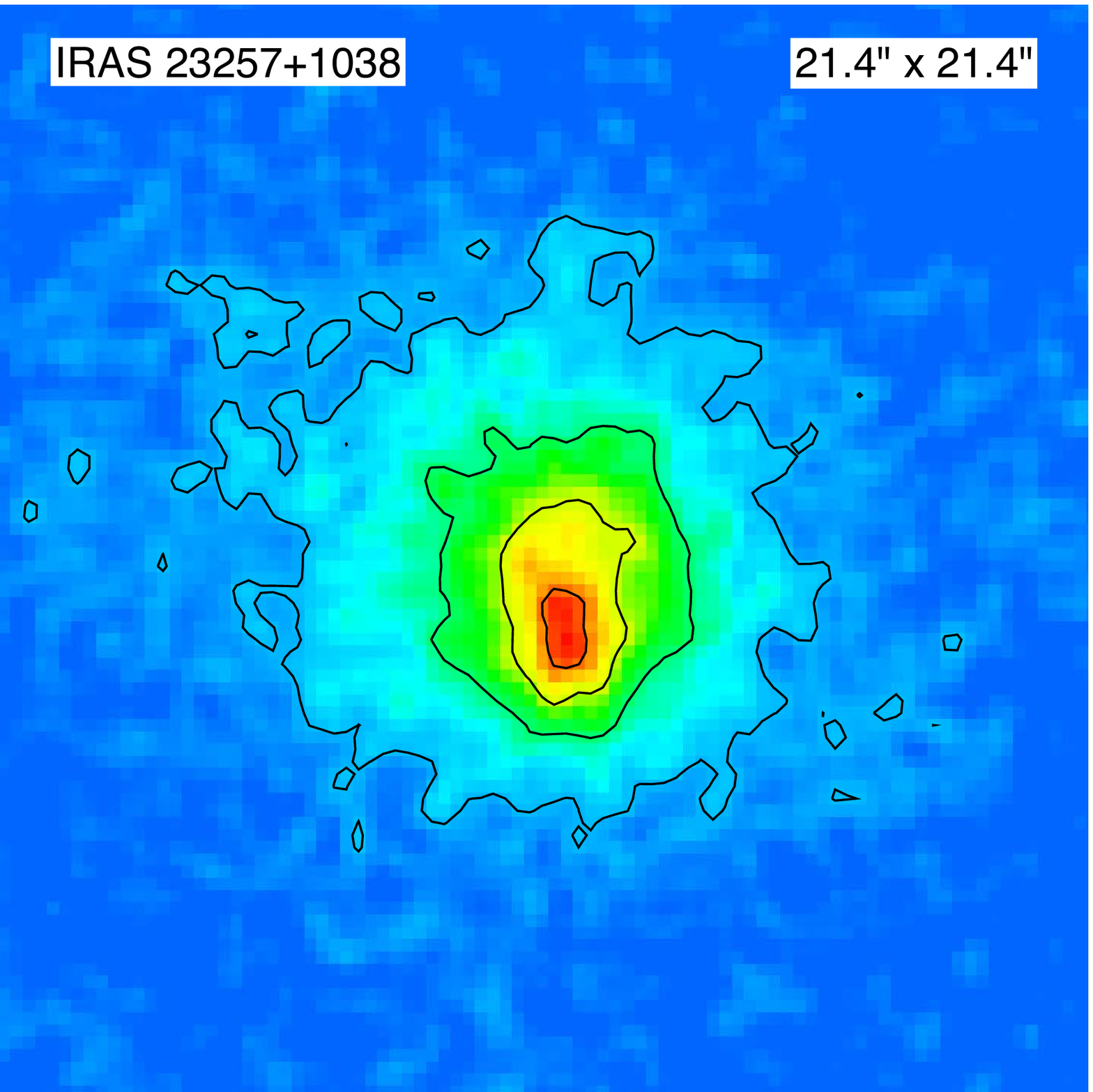}\hspace{0.3cm}
\includegraphics[width=0.3\linewidth]{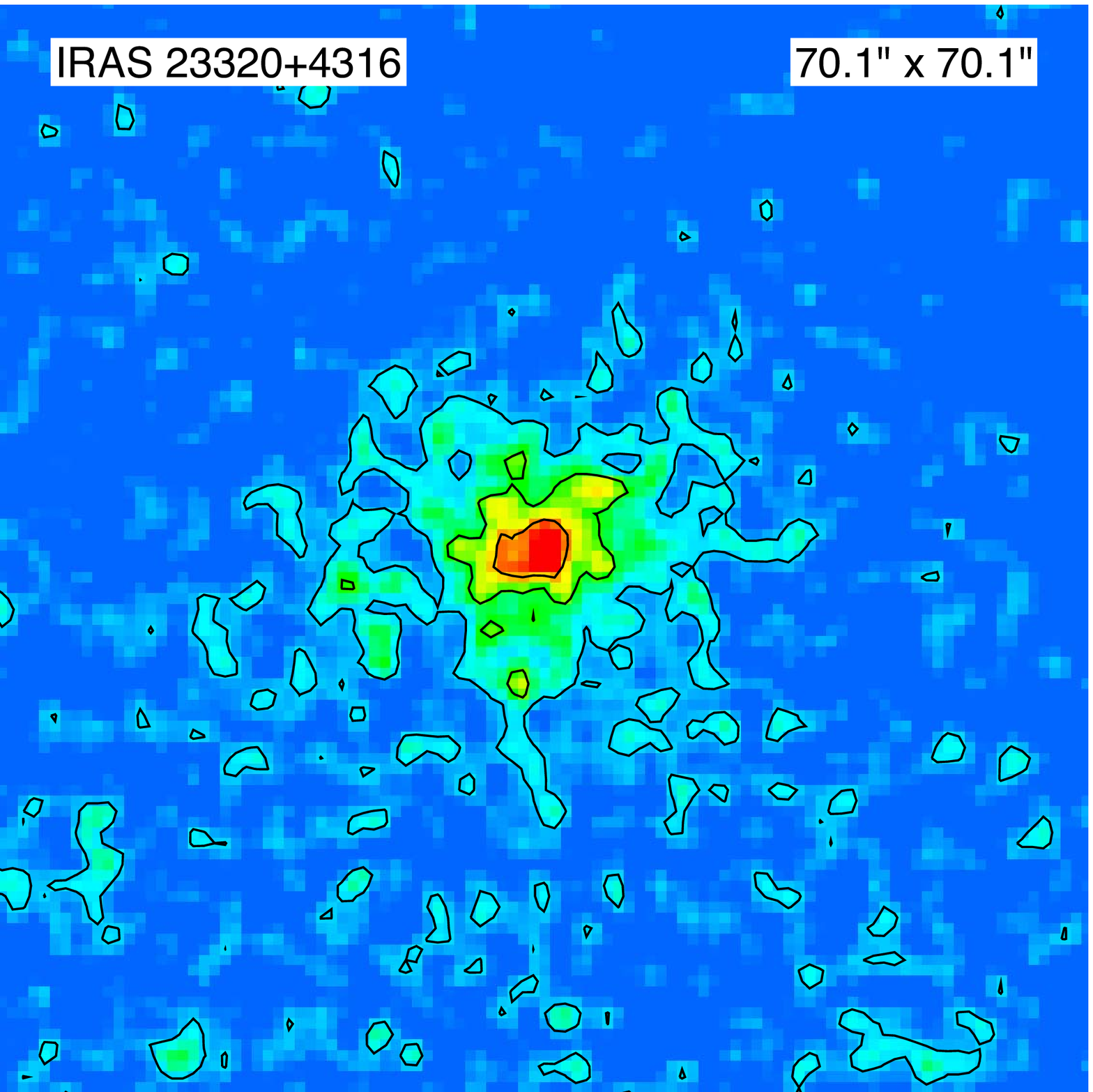}\hspace{0.3cm}  }
\vspace{0.3cm}

  \caption{Continuation of Fig.~5 (on-line only).
Top row, left to right:
IRAS 18240$+$2326, $V$-band;
IRAS 18467$-$4802, $F606W$;
IRAS 19178$-$2620, $B$-band.
Bottom row, left to right: 
IRAS 20077$-$0625, $V$-band; 
IRAS 23257$+$1038, $V$-band; 
IRAS 23320$+$4316, $B$-band.
All images are square, with north to the top and east
to the left.
}
 
\end{figure*}

}   


\section{The images}

\subsection{Presentation of the images}

The main results of the observations are summarized in Table~4.
Column (2) gives the telescope used for the $V$ and $B$ observations, and
column (3) indicates the detection of the envelope and/or the central
star (or the star-illuminated core) at these wavebands.  For the NTT
observations, seven of the envelopes are not detected and are
discussed in the appendix.

Figs.~1--4 present observations of each detected envelope, in three
panels: a gray-scale image of the field in the waveband that best
shows the envelope shape, a close-up contour map of the envelope, and
the azimuthally averaged, radial intensity profile.

In many of the gray-scale images it can be seen that field stars and
galaxies lie along the lines of sight to the envelopes. In order to
make the contour maps, the images are cleaned by removing these
features using the following procedures. For bright stars whose halos
extend over part of the envelope, we fit and subtract circularly
symmetric halo models.  The models are derived from isolated field
stars in the same image, and the fit is made by minimizing residuals
in regions outside the envelope. The cores of these bright stars
are also masked. In the case of IRAS 07454$-$7112, a relatively bright
elliptical galaxy lies close to the AGB star to the South-East, and
this was successfully subtracted using a Sersic galaxy profile (Sersic
\cite{sersic63}). Dimmer stars and small galaxies, are simply masked. All masked
areas are then replaced with local averages. This local average is
determined by considering sectors centered on the AGB star with radial
widths of 1--2\arcsec, and typical ranges in position angle of $\pm
20$--30$\degr$. In this way, the masked areas are filled with
representative local intensities, and Gaussian noise similar to the
sky noise is added in these areas.

For the contour maps in Figs.~1--4, the cleaned images were lightly
smoothed with a Gaussian of width 2 pixels (FWHM) with two exceptions
noted in Sect.~3.5. The gray areas in the figures show the masked
regions as described above. The solid contours are linearly spaced,
and are specified in the figure captions in units of the peak surface
brightness of the envelopes; the equivalent values in mag~arcsec$^{-2}$ are
given in Table~4.  In cases with a strong stellar core, additional
dashed contours are included with a step size equal to the level of
the highest solid contour.  In Figs.~5 and 6 (on-line only), 
the cleaned images  are shown in color, with the contours superposed.


\subsection{General properties}

As described in MH06, deep optical imaging of AGB stars may reveal
one or more of the following: the central star or an unresolved
stellar core; the star-illuminated inner envelope; and the extended
envelope illuminated by the ambient Galactic radiation field. The
typical situation for the thick circumstellar envelopes is that the
star and the star-illuminated core are faint (or not seen) in the $B$ or $V$
bands (because of extinction by the envelope), but the star becomes
dominant at longer wavelengths because of the decreasing opacity of
the dust envelope and the red spectrum of the star.

For the star-illuminated inner envelope, the intensity of scattered
starlight decreases rapidly with angular distance from the center
($\sim \theta^{-3}$, Martin \& Rogers \cite{martin87}). For external
illumination, a faint, extended nebula is seen, with a shallow radial
dependence.  The externally illuminated nebula may also form a plateau
of relatively constant intensity in the central regions where the
optical depth to external radiation is $\ga 1$. Examples of this
(e.g., IRAS 16029$-$3041) may be seen in the radial profiles in
Figs.~1--4.

\subsection{Surface brightness}

Column (4) of Table~4 lists the peak surface brightness in $B$ or $V$
for each circumstellar envelope, based on the photometry described in
Sect.~2.2 and MH06. For cases where the envelope is detected and the
star is of low intensity or not detected, or the envelope plateau is
extended, the measurement is straightforward.  For cases where the
envelope is detected, but the central star is of high intensity, we
estimate the peak intensity by fitting and removing the stellar
contribution at the center by using the profiles of field stars. For
cases where neither the envelope nor the star is detected, we give
upper limits assuming the source size is comparable to the seeing disk
(FWHM). For cases where the star (or stellar core) is detected but not
the envelope, a small source is essentially invisible; for these we
give nominal limits based on the smallest source that would be
detectable: $3\arcsec$ for IRAS 06255$-$4928 and 09521$-$7508 and
$5\arcsec$ for IRAS 05418$-$3224 and 20042$-$4241.

For three envelopes we are able to measure both the $B$ and $V$
surface brightness.  In addition to the $B$ values given in Table~4,
$V = 25.04$~mag~arcsec$^{-2}$ for the C-rich envelope IRAS
09452$+$1330, and $V = 23.43$ and 23.07 mag ~arcsec$^{-2}$, for the
O-rich envelopes IRAS 16029$-$3041 and 17319$-$6234, respectively.

In addition to the measured values, Table~4 also lists an effective
surface brightness in the $B$-band, $B_0$, corrected for interstellar
extinction, and where necessary, transformed from $V$-band
measurements.  The extinction used is from the
NED\footnote{http://ned.ipac.caltech.edu} calculator, assuming all
extinction along each line of sight lies between us and the star. For
envelopes with only $V$-band measurements, we first correct for
extinction, and then adopt the extinction corrected color $(B-V)_0$
appropriate to the chemistry of the envelope. The colors are obtained
from the three envelopes measured in both bands given above. For the
C-rich envelope, $(B-V)_0 = 0.72$~mag, and for the mean of the O-rich
envelopes, $(B-V)_0 = 0.99$~mag. Thus the two types of envelopes are
fairly similar.

\subsection{Envelope shapes}

Inspection of the images in Figs.~1--6 shows that the envelopes exhibit
relatively smooth and symmetric shapes that reveal their large scale
geometry. They range from close to circular to distinctly elliptical.

There are two additional features. First, three envelopes show
elongated central components. These we interpret as the result of the
leakage of light from the central star through polar (or bipolar)
openings in the core of the envelope, as is the case in IRC+10216
(Mauron \& Huggins \cite{mh00}).  Several additional cases are seen on
smaller scales in $HST$ images of the sample (MH06; Sahai \cite{sahai07}).  The
polar cores are noted in Table~4, and details are given in Sect.~3.5.

A second feature of the envelopes of IRAS 16029$-$3041, 17319$-$6234
and 18467$-$4802 is a one-sided asymmetry in the optically thick
central region, which results from asymmetry in the external
illumination. This is most clearly seen as a crescent shape in the
highest level contours. In all three cases the direction of maximum
illumination indicated by the crescent is within 30\degr\ of the
direction of the Galactic plane. The asymmetry is reduced in the
outer, optically thin regions of the envelopes where the observed
intensity is less sensitive to the directionality of the
illumination.

In order to quantify the overall shapes of the envelopes, we fit
ellipses to intensity contours at specified levels. Our procedure
samples an intensity contour at intervals of $10\degr$ in position
angle, omitting regions masked by the presence of field stars, as
described in Sect.~3.1. An ellipse is then fit to the data using a
non-linear least squares model, with the lengths of the semi-major and
semi-minor axes ($a$ and $b$), the center of the ellipse, and the position
angle (P.A.) of the major axis as parameters of the fit. The shape is
then given by the ellipticity $E$, where $E=a/b$. 

An example of the results of this procedure for the case of IRAS
17319$-$6234 is shown in Fig.~7. The figure shows data points sampled at
levels of 20\% and 40\% of the envelope peak intensity, and the fitted
ellipses for which values of $E$ are $1.05\pm0.02$ and $1.09\pm0.03$,
respectively. In spite of gaps in the data because of the masking of
field stars in the image, the shapes are well constrained by the
procedure. In this case the envelope is marginally non-circular,
probably on account of asymmetric illumination seen in the plateau
in the image in Fig.~5.

Using this technique, we characterize the shapes of the
envelopes on the largest size scales allowed by 
the observations in order to minimize the effects of contamination
by the central star (when visible) and the core features noted above.
The results are given in Table~4. For cases with good S/N and resolved 
plateaus, we give the shape measured at 20\% of the peak. For the
other envelopes, which are fainter and less well defined 
at this level, we give the shape at the 40\% of the peak;
Table~4 lists the fitted values of $E$ ($\pm 1 \sigma$),
$a$, and the P.A.\ of the major axis. The expansion time scale,
$t_\mathrm{exp}$ corresponding to $a$ is also given, based on the
distances and expansion velocities given in Table~1. The envelopes are
detected to significantly larger radii in the azimuthally averaged
profiles, as seen in Figs.~1--4. 

The envelope shapes and orientations at higher
intensity levels are generally similar to those in Table~4. 
However there are several exceptions where the inner ellipticity 
is larger and at a different PA, especially in cases which have the core
features noted above. Individual cases are discussed below.

\subsection{Notes on individual objects}
  
\noindent
{\object IRAS 01037$+$1219}  (IRC+10011). The extended envelope is essentially
circular in shape.  At the center a small scale polar core is seen in
the HST/ACS $F816W$ image shown in Fig.~1 of MH06, $\sim0\farcs4$ in
length at P.A.\  $-45\degr$. There is also an asymmetry seen in
CO in the central regions at approximately the same P.A.\
(Castro-Carrizo et al.\ \cite{castro10}).

{\object IRAS 07454$-$7112}  (AFGL 4078). The envelope at low intensities is
approximately circular, but the shape is uncertain because of interference
by field stars. Nearer the center, the shape is more definitely
elliptical at P.A.\ +30\degr. The HST/ACS $F606W$ image of the
star-illuminated center (Sahai \cite{sahai07}) shows a polar core
$\sim0\farcs6$ in extent at P.A.\ $-160\degr$, which aligns with the
central ellipticity.

{\object IRAS 09116$-$2439}  (AFGL 5254).
The envelope is consistent with circular, but the detailed shape is
uncertain because of the large masked area of the field star near the
center.

{\object IRAS 09452$+$1330}  (IRC+10216). In spite of the high S/N of the
image, the low intensity levels are affected by a gradient caused by a
very bright field star outside the image shown (see Mauron \& Huggins
\cite{mh99}). The shape in Table~4 refers to 30\% of the peak intensity.  At
higher levels it is affected by shell structures which are slightly
elliptical, and in the center there is a polar core at P.A.\
+12\degr, illuminated by the star (Skinner et al.\ \cite{skinner98}; Mauron \&
Huggins \cite{mh00}).  Because of the large field, the contours in
Fig.~1 are smoothed to a resolution of $1\farcs6$.

{\object IRAS 10323$-$4611}.  The envelope is elliptical at low intensities
but more circular
towards the center. The HST/ACS $F606W$ image of the star (Sahai \cite{sahai07})
shows a polar core of $\sim0\farcs4$  at  P.A.\ +25\degr.

{\object IRAS 12384$-$4536}.  Moderate but significant ellipticity at all
levels.

{\object IRAS 14591$-$4438}. Low S/N data. Roughly circular at low
intensities, but may be elliptical in the center.

{\object IRAS 16029$-$3041}  (AFGL 1822).
High S/N data. The plateau shows asymmetric illumination, and a
central dip characteristic of an optically thick core.  The small but
significant ellipticity in the extended envelope is likely caused by
the asymmetric illumination.

{\object IRAS 17319$-$6234} (OH~329.8$-$15.8).  High S/N data. The extended
envelope is circular. The plateau shows asymmetry in illumination, and
a central dip characteristic of an optically thick core.

{\object IRAS 18240$+$2326}  (AFGL 2155).
Small but uncertain ellipticity on account of bright, nearby field stars. 

{\object IRAS 18467$-$4802}  (OH\,348.2$-$19.7).
HST/ACS $F606W$ image. The image and contours in Fig.~4 are smoothed
to a resolution of $0\farcs25$. The extended envelope is circular. The plateau
shows asymmetry in illumination, and a central dip characteristic of an
optically thick core.

{\object IRAS 19178$-$2620}  (AFGL 2370).
Highly elliptical envelope. Strong polar core at
P.A. $-$155\degr, inclined to major axis at an angle of $\sim$50\degr. 

{\object IRAS 20077$-$0625}  (AFGL 2514).
Highly elliptical envelope at all intensity levels. Fig.~8 shows a
polar core at the center in the HST/ACS $F606W$ image at P.A.\
$-40\degr$, roughly orthogonal to the major axis.

{\object IRAS 23257$+$1038}  (AFGL 3099).
The extended envelope is circular and centered on the AGB star. There
is a strong polar feature at the center, at P.A.\ $-$175\degr.

{\object IRAS 23320$+$4316}  (AFGL 3116).
Low S/N data. The envelope is roughly circular, with evidence for an
elliptical core at P.A.\ +110\degr.  The HST/ACS $F606W$ image of the
star (Sahai \cite{sahai07}) shows a polar core $\sim0\farcs3$ in extent at
P.A. $-150\degr$, approximately orthogonal to the elliptical core.


\begin{figure}
 \centerline{\includegraphics[width=0.9\linewidth]{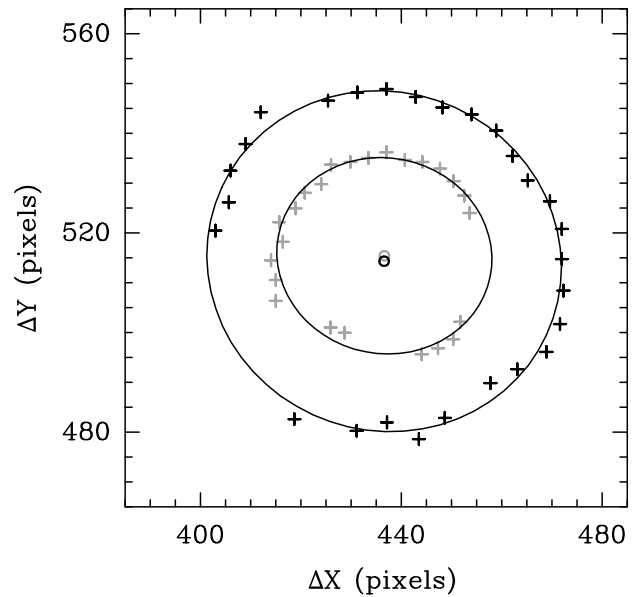} }
  \caption{Contour fitting for IRAS
    17319$-$6234. The inner (grey) and outer (black) crosses sample
    the envelope at $10\degr$ intervals in azimuth and at 20\% and 40\% of
    the peak intensity, respectively. Gaps in the data correspond to
    masked regions of the image. The curves show the fitted ellipses;
    the small inner circles show the fitted centers. 
     }
\end{figure}


\begin{figure}
 \centerline{\includegraphics[width=0.9\linewidth]{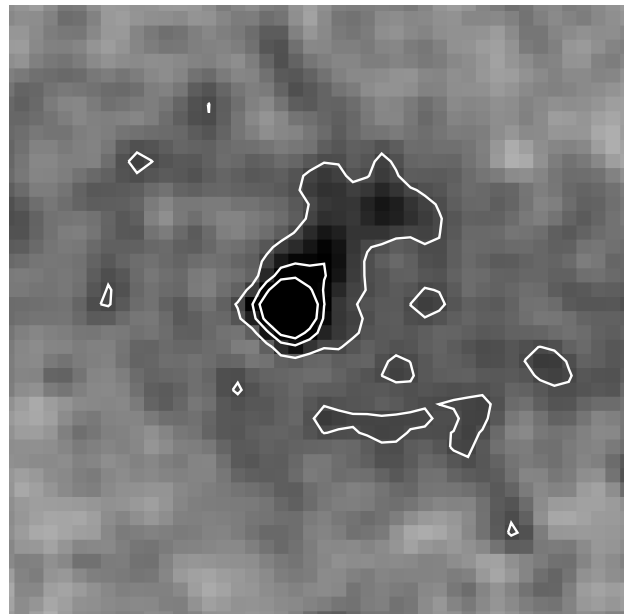} }
  \caption{Polar core at the center of IRAS
     20077$-$0625. HST $F606W$ image. The data have been binned to a
     resolution of $0\farcs1$. Field size $4\farcs3 \times 4\farcs3$. 
     }
\end{figure}


\begin{figure}
 \centerline{\includegraphics[width=0.9\linewidth]{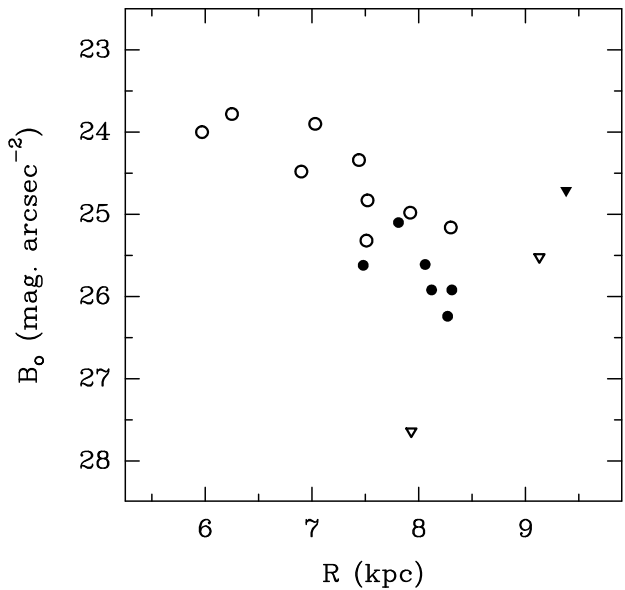} }
  \caption{Surface brightness vs.\ galactocentric radius. Filled and
    open symbols denote C-rich and O-rich envelopes,
    respectively. Circles are detections, triangles upper limits on the
    surface brightness.  
     }
\end{figure}


\section{Discussion}

\subsection{Envelope surface brightness}

\subsubsection{Detectability of the envelopes}

The detectability of an envelope in scattered ambient light depends on
the effective optical depth for scattering, the strength of the
radiation field, and the interstellar extinction along the line of
sight. MH06 describe how the envelope opacity is conveniently
parameterized by an estimate of the angular size of the optically thick
plateau region using:
\begin{displaymath} \Delta\theta = 5\arcsec\,
( \frac{\dot{M}}{10^{-5}\,\mathrm{M_\odot\ yr^{-1}} } ) (
 \frac{V}{10\,\mathrm{km\ s^{-1}}} )^{-1} (
 \frac{d}{1000\,\mathrm{pc}} )^{-1}. \end{displaymath} 
If this exceeds the resolution of the observations, the center of the
envelope should attain maximum surface brightness.
 
Based on the parameters listed in Table~1, all the envelopes observed
have $\Delta \theta \ga0\farcs8$, as given by the above
equation. They should therefore all be partly or fully resolved. If we
consider the envelopes with good measurements (ignoring 4 with
nominal upper limits due to the presence of a bright central star),
there are 15 detections and three upper limits. One of the limits is
sensitive, but it belongs to the envelope with the lowest $\Delta
\theta$, so it it probably just too small to be seen. The two other
non-detections are for the two objects with the highest interstellar
extinction (1.6 and 2.0 mag.\ in $B$) which prevents reaching low
intensity levels. Hence, if allowance is made for extinction, the
characterization in terms of $\Delta\theta$ is a successful guide to
detectability.

\subsubsection{Envelope surface brightness and the interstellar radiation field}

After correction for extinction, the peak surface brightness $B_0$
shows a range of more than 2~mag (see Table~4). The
scatter does not correlate with latitude ($b$) or height above the
Galactic plane ($z$), and is not clearly related to $\Delta\theta$ or
$\dot{M}$. However, it does appear to depend partly on envelope
chemistry and longitude ($l$), and both are found to result from a
remarkable correlation with Galactic radius $R$.

Fig.~9 shows the correlation of $B_0$ with Galactic radius, with $R$
calculated using the stellar distances and coordinates given in
Table~1, and assuming $R=8.0$~kpc for the Sun. It can be seen that the
surface brightness decreases rapidly with $R$ with an approximately
linear relation. Omitting the upper limits for reasons mentioned
above, a least squares fit gives:
\[B_0 = 0.90\,(\pm0.17)\, R +
18.20\,(\pm1.26)\] and a residual rms scatter in $B_0$ of 0.41, in
 units of mag~arcsec$^{-2}$. We interpret this relation to be the
 effect of a radial gradient in the interstellar radiation field. The
 linear dependence in magnitudes translates to an exponential law in
 the mean interstellar intensity. As far as we know, this is the most direct
 measurement so far of the gradient in the solar neighborhood. Our data
suggest that the radiation field decreases by a factor of $\sim$5 between
$R=6$ and $R=8$~kpc. This
is fairly consistent  with the model of Mathis et al. (1983) 
who find a decrease of a factor of 7
between  $R=5$~kpc and the solar radius. A smaller factor of 3 is
favored by the more recent study of Sodroski et al.\ (1997).

Inspection of Fig.~9 shows that the C-rich envelopes (filled symbols)
follow essentially the same relation as the O-rich envelopes (open
symbols). It would appear that the scattering properties of these
different envelopes in the $B$-band are rather similar, and this is
consistent with the similarities in the $B-V$ colors measured in
Sect.~3.3. We therefore interpret the observed difference in the
average brightness of O-rich and C-rich envelopes in terms of their
locations in the Galaxy. In our sample of AGB stars, the C-rich
envelopes lie on average at larger values of $R$, and this is
consistent with the higher relative fraction of carbon stars in the
outer Galaxy (e.g., Jura et al.\ \cite{jura89}). At larger Galactic radii, the
C-rich envelopes are illuminated by a less intense radiation field.


\subsection{Envelope morphology}

\subsubsection{Distribution of observed shapes}

All the envelopes in Figs.~1--4 show relatively smooth radial
profiles except for IRC+10216, which has sub-structure in the form of
multiple arcs.  However, IRC+10216 is much nearer than the other
objects (see Table~1), so these could have similar structure which is
not resolved at the larger distances. None of the envelopes show
evidence for major, single-shell ejections which are seen in a few AGB
envelopes and are thought to be caused by helium shell flashes (e.g.,
Olofsson et al.\ \cite{olofsson00}).

As described in Sect.~3, the large scale shapes of the envelopes are
roughly circular or elliptical, consistent with spherically or axially
symmetric three dimensional structures seen in projection. The
observed shapes are conveniently characterized by the ellipticities
($E$) listed in Table~4; recall, $E=a/b$ $(\ge 1.0)$, where $a$ and $b$ 
are the semi-major and semi-minor axes, respectively.  
The distribution of ellipticities for the
sample (in order of increasing $E$) is shown in Fig.~10, which also
indicates the formal 1-$\sigma$ and 3-$\sigma$ uncertainties. 
As explained in Sect.~3.4, the ellipticities we use
to characterize the distribution refer to the overall shapes of the extended
envelopes. If we use the shapes measured closer to the star (e.g., at 60\% of
the peak intensity), the distribution is similar to that in Fig.~10, but
with somewhat higher ellipticities at the top end due mainly to the contribution of the
core features in a few cases.

From Fig.~10, it can
be seen that deviations of the overall shapes of the envelopes from circular symmetry 
are relatively small.  Approximately half of the
sample (47\%) have $E\la 1.1$; 20\% have $E\ga 1.2$, and the rest are
in between. The observed shapes of axially symmetric envelopes depend of course on
the inclination angles, and they will appear circular when viewed
end-on. However, there is no known bias in the inclinations of the
observed envelopes, so we can assume them to be randomly
oriented, with a median inclination of 60\degr\ to the line of sight,
and rarely seen end-on. On this basis we can draw the following
general conclusions on the 3-dimensional shapes of AGB envelopes: 1. A
significant fraction ($\sim$ half) are close to spherically
symmetric. 2. At least 20\% are significantly flattened. 3. Highly
flattened, disk-like envelopes are relatively rare or
non-existent. The results of Neri et al.\ (\cite{neri98}) based on observations
of AGB envelopes in the millimeter lines of CO are in qualitative
accord with our conclusions.


\begin{figure}
 \centerline{\includegraphics[width=0.9\linewidth]{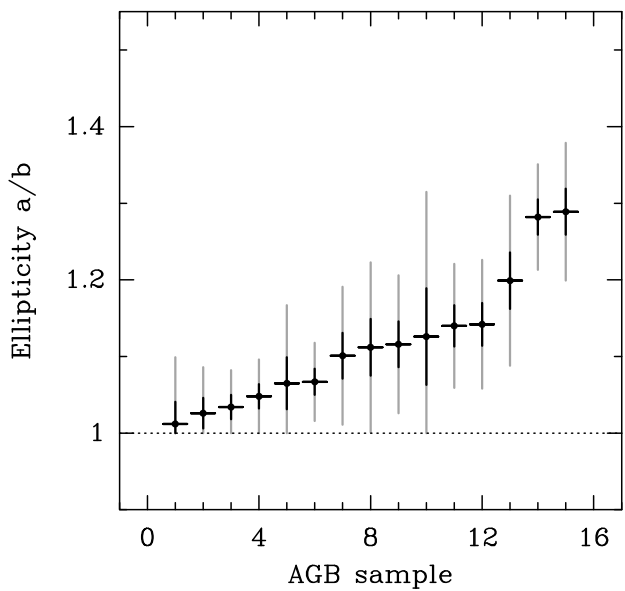} }
  \caption{Distribution of the measured large scale ellipticities of
    the AGB envelopes. The vertical error bars denote the formal 1-$\sigma$
and 3-$\sigma$ uncertainties. }
\end{figure}


\begin{figure*}
 \centerline{\includegraphics[width=0.45\linewidth]{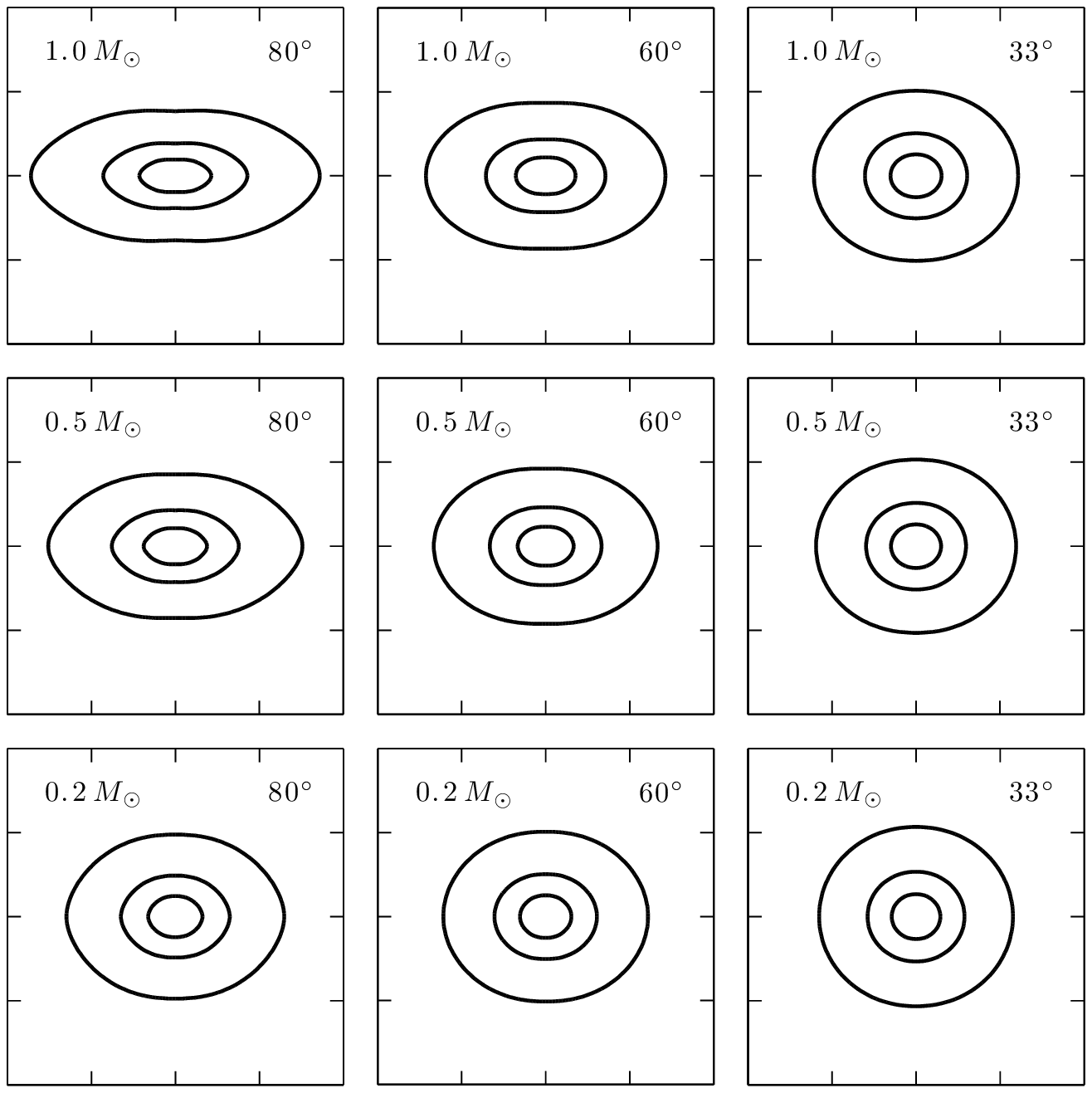}
\hspace{0.3cm}\includegraphics[width=0.45\linewidth]{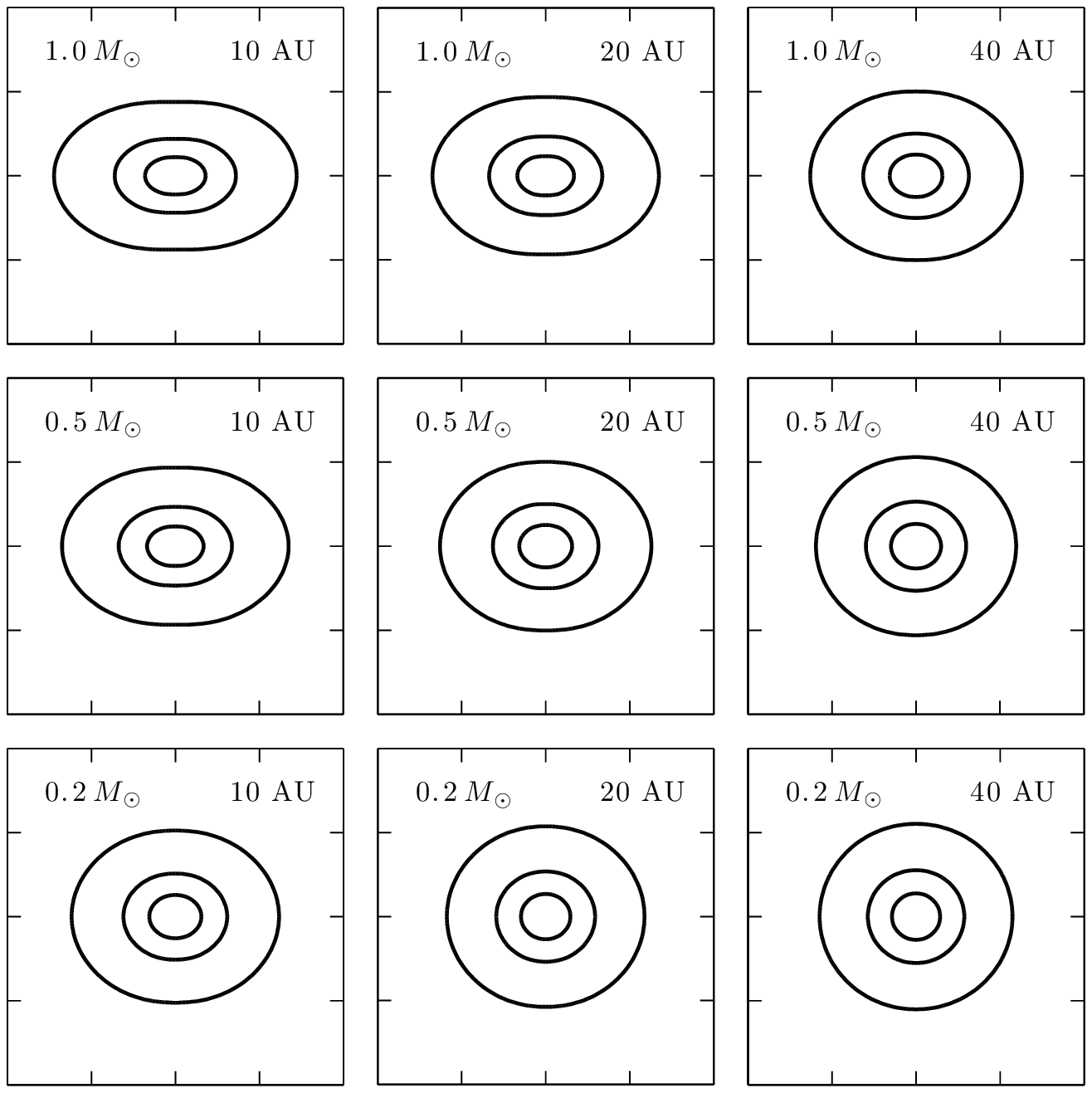} }
  \caption{Contour maps of model images for binary shaping. \emph{Left
  hand panel}: Variation of shape with $M_\mathrm{c}$ and $i$, for
  $d=10$~AU and $V=10$~km~s$^{-1}$. \emph{Right hand panel}:
  Variation of shape with $M_\mathrm{c}$ and $d$ for $i=60\degr$ and
  $V=10$~km~s$^{-1}$. }
 
\end{figure*}


\subsubsection{Shaping mechanisms}

There are two separate categories of shaping that can affect 
the morphology of the circumstellar envelopes: intrinsic shaping by 
or close to the central star, and interaction with the ISM. The latter
can produce significant effects on large angular scales, typified by a 
one-sided bow shock due to the relative motion of the star and the 
surrounding medium.

We first consider whether the ISM interaction is likely to be important for
the observed shapes by comparing estimates of the bow shock stand-off
distance $R_{\rm 0}$ for each object, taking into account the height above
the galactic plane, using equations (2) and (5) of Cox et al.\ (\cite{cox12}),
the stellar data in Table 1, and assuming a typical velocity relative
to the ISM of 30 km~s$^{-1}$ (Feast \& Whitelock \cite{feastwhite00}). We find that
$R_{\rm 0}$ for the envelopes observed ranges from $\sim$ 10$^{18}$ to 10$^{20}$ cm, 
and exceeds the size scale corresponding to the angular radii ($a$) given 
in Table~4 by a median factor of 70, with a range of 30 to 3000. We conclude
that interaction with ISM is unlikely to have a significant influence on the
measured shapes. The observed morphologies are also consistent with this conclusion.

Our understanding of the intrinsic shaping mechanism of AGB envelopes is not
well developed, but the expectation is that envelope shapes are
different for single stars and binaries. Companions to AGB stars are,
however, extremely difficult to detect (Jorissen \& Frankowski \cite{jorissen08})
and it is probable that hidden binaries are included in our sample.

For single AGB stars, which rotate very slowly, the zeroth order
expectation is for spherically symmetric envelopes, although the
extent of deviations caused by magnetic fields or related phenomena is
not known. For AGB stars in binaries, the companion is expected to
influence the shape in several ways. Close-in companions can cause
tidal effects and spin-up, and modify the effective gravity of the
star (e.g., Frankowski \& Tylenda \cite{frankowski02}). However, two additional
effects are likely to be more important overall because they operate
over a much wider range of separations: the flattening of the envelope by
the gravitational field of the companion, and the effects of jets
powered by accretion onto the companion.

\paragraph{Gravitational flattening.} The gravitational
flattening of an envelope towards the orbital plane of a companion is
a prominent feature of gas dynamical simulations of binary AGB systems
(Mastrodemos \& Morris \cite{mastrodemos99}; Gawryszczak et al. \cite{gaw02}). Low mass
companions or wide separations lead to nearly spherical envelopes, and
close-in, high mass companions lead to flattened envelopes.

As a rough guide to the relation between envelope shape and binary
parameters, Huggins et al.\ (\cite{huggins09}) have obtained an approximate
formula for the flattening in terms the companion mass $M_\mathrm{c}$,
the wind speed $V$, and the binary separation $d$, based on the
simulations of Mastrodemos \& Morris (\cite{mastrodemos99}).  They use a highly
simplified, axially symmetric model envelope with a density $\sim
r^{-2}$, and characterized by a shape parameter $K$, where
$K=n_\mathrm{e}(r)/n_\mathrm{p}(r)$ is the ratio of equatorial to
polar density; the relation between $K$ and the binary parameters is
given in their equ.~(4). For an envelope illuminated by external
radiation, they also give a prescription for the observed shape for a
specified inclination of the symmetry axis to the line of sight
($i$). The observed shapes are given in terms of the ratio of column
densities on the major and minor axes,
$N_\mathrm{maj}(\theta)/N_\mathrm{min}(\theta)$, but this can be shown
to be equivalent the ellipticity ($E$) used in this paper.

To illustrate the range of binaries that could give rise to the
observed shapes, Fig.~11 shows contour maps of synthetic images based
on the simplified envelope model. The left hand panel shows the
dependence of the observed shape on $M_\mathrm{c}$ and $i$ for
$d=10$~AU. The right hand panel shows the dependence on $M_\mathrm{c}$
and $d$, for the median inclination ($i=60\degr$). A representative
value of $V=10$~km~s$^{-1}$ is used for all cases.

It can be seen from the figures that the model produces envelope
shapes that are qualitatively similar to those observed, i.e., with
round or approximately elliptical contours. (The pinching effect on
the minor axis of the flattest model is an artifact of the specific
latitude dependence of density in the simple model, and is not
expected to be a feature of real envelopes.) For masses $\la
0.1~M_\odot$, the companion has essentially no effect at any angle,
and the envelopes are very close to circular. For typical inclinations
($i\sim60\degr$), companions with masses from 0.2 to 1.0~M$_\odot$,
significantly affect the envelopes, producing $E\ga 1.2$ for
separations up to 10 and 40~AU, respectively. Thus the range
of observed envelope shapes is consistent with a spectrum of AGB stars
ranging from single stars to binaries with relatively close companions
of moderate mass.

\paragraph{Polar cores and jets.}

The idea that binary companions are responsible for shaping the
non-spherical envelopes is strengthened by the presence of the polar core
asymmetries noted in Sect.~3.  The fact that these elongated regions
of enhanced radiation are seen in the central regions of approximately
half of the sample (Table~4) is striking.

The best studied case is IRC+10216 (IRAS 09452+1330) where the effect
is caused by the leakage of stellar radiation along bipolar axes close
to the star, with one side much brighter than the other because of
extinction (Skinner et al.\ \cite{skinner98}; Mauron \& Huggins \cite{mh00}). Other cases
are probably similar (Sahai \cite{sahai07}), although in the absence of
spectroscopy, localized emission cannot be ruled out.

The polar cores are not expected to be features of the mass loss of
single stars, but they are expected in binary systems. One production
mechanism is the diminution of mass loss along the polar directions
which is an integral part of the binary shaping process, as seen in
the simulations of Mastrodemos \& Morris (\cite{mastrodemos99}). A second, related
mechanism is the sculpting effect of jets from an accretion disk
around the companion (e.g., Soker \& Rappaport \cite{soker00}). The
effects of jets are seen in the Mira system (Josselin et al.\ \cite{josselin00}),
and in a more extreme form in proto-PNe (e.g., Huggins et al.\ \cite{huggins04}). These
mechanisms can act in parallel, and the dominant effect could vary
from object to object.

Since polar cores and gravitational shaping likely arise from similar
circumstances (i.e., the presence of a companion) there is good reason
to suppose that they could be correlated. Our observations provide
evidence for this view.  The three envelopes with the highest
ellipticities all have polar cores (Table~4). The polar cores are
present in about half the sample, somewhat more than the fraction
that are clearly flattened. It could be that the polar cores are formed
by low as well as moderate mass companions, and are therefore more
widely detectable. The relation between these features clearly
warrants further study.

The high frequency of occurrence of polar cores in the observed sample
also places interesting constraints on their physical origin. If they
are intrinsic to envelope shaping, they will be permanent features of
the envelopes. However, if they are caused by jets, the time evolution
needs to be considered. Several cases where jets are suspected are
discussed by Sahai (\cite{sahai07}), and IRAS 01037+1219 (IRC+10011) has been
modeled by Vinkovic et al.\ (\cite{vinkovic04}) in terms of jets which have very
recently turned on ($\le 200$~yr) and are in the process of sculpting
the envelope. Most of the cores observed here are small in radial
extent, and with the high speed of jets, have comparably short time
scales. Since we observe polar cores to be common, it is statistically
unlikely that they have all formed within the last few hundred
years. For the jet interpretation, the jets are likley to be weak or
intermittent, so that the effects are local to the star. In this
picture the more powerful jets seen in proto-PNe would develop later,
when the companion interacts more strongly with the AGB star and
accretes at a higher rate (Huggins \cite{huggins07}).


\begin{figure}
 \centerline{\includegraphics[width=0.9\linewidth]{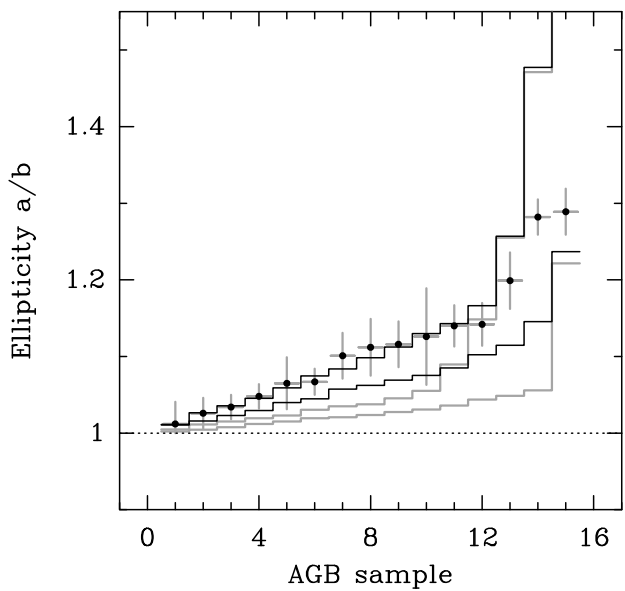} }
 \caption{Comparison of the observed ellipticities of the AGB sample
    with calculated distribution functions. The points are the
    observations, and the vertical error bars denote the 1-$\sigma$
    uncertainties.  The upper and lower gray histograms are population
    synthesis predictions including the measurement errors for case 1
    ($g(q_0)=1$) and case 2 ($ g(q_0) = q_0 ^{-0.9}$), respectively.
    The upper and lower black histograms are population synthesis
    predictions including a residual ellipticity for case 1 and case 2,
    respectively. See text for details.
}
 \end{figure}


\subsubsection{Population synthesis}

The question of binary shaping can be investigated further by
comparing our observations of envelope shapes with the results of a
recent population synthesis of AGB stars by Politano \& Taam
(\cite{politano2011}). For a population consisting of single stars and binaries, they
have calculated the distribution of the gravitational flattening
parameter $K$ of Huggins et al.\ (\cite{huggins09}) described above.  We have used
their results to predict the distribution of the observed shapes of
envelopes, assuming random orientations. Any contribution to the
shaping by jets would be additional. We report the results in Fig.~12,
using the median values of $E$ in 15 quantiles, to represent the
sample of observations.

The population synthesis assumes 50\% of stars form as single stars
and 50\% from in binaries on the main sequence.  For the binaries, we
consider two representative cases for the assumed distribution of
$q_0$, the ratio of the secondary to primary mass on the main
sequence. Case 1 is the standard case of Politano \& Taam (\cite{politano2011}) with
a flat mass distribution, $ g(q_0) = 1 $. Case 2 is a distribution
that favors low mass companions with $g(q_0) = q_0^{-0.9}$.  For case
1, 32\% of the envelopes on the AGB are perturbed by a companion, and
68\% are unperturbed (either single stars, or with distant
companions); for case 2 these numbers are 18\% and 82\%, respectively.

If single stars produce exactly spherical envelopes, both cases 1 and
2 would have more than half the AGB envelopes with nominal
ellipticities of $E=1.0$ (in the absence of measurement errors).  For
realistic comparisons with the observations we include two effects.
The gray histograms in Fig.~12 show the predicted distributions,
where we include by simulation the effect of the average uncertainty
in fitting the ellipses $\sigma = \pm 0.03$. The black histograms in
Fig.~12 show the two cases if we assume a residual ellipticity, 
normally distributed with mean  
1.06 and dispersion $\pm0.06$~(1-$\sigma$) for all envelopes, in quadrature with any binary
perturbation. This residual effect is chosen for illustration, and
could arise from a combination of systematic effects such as residual
asymmetric illumination and plate gradients, and as a natural
dispersion in the true shapes of unperturbed envelopes.

In spite of the extreme simplicity of the underlying envelope model
and the complexity of the population synthesis, it can be seen in
Fig.~12 that the synthesized curves exhibit the qualitative
characteristics of the observed distributions.  The simulations
suggest that the envelopes with ellipticities close to 1.0 are mainly
unperturbed envelopes with a small dispersion in their shapes; and the
less frequent, larger ellipticities are the results of binary
interactions.  There are significant quantitative differences in the
predicted curves for the different assumptions about the binary mass
ratios. It is encouraging that the observational data fall between the
two cases shown, but the observational sample is too small to
discriminate between them at this stage.

From an empirical point of view, if the polar cores indicate binaries
as we suggest, the observation of cores in half the sample would
indicate a somewhat greater percentage of AGB binaries than predicted
by the population synthesis. A larger sample would be important to
investigate this further.

\subsection{AGB stars as precursors of PNe}
The evidence for a hidden population of AGB binaries provided by our
observations is of interest for understanding post-AGB evolution. In
particular, the importance of binaries in shaping planetary nebulae
(PNe) is a subject that is widely debated (De~Marco \cite{demarco09}), and it is
uncertain whether a binary companion is an essential ingredient for PN
formation.  It is observed that 12--21\% of PNe have close binary
companions (Miszalski et al.\ \cite{miszalski09}) and these have evidently passed
through a common envelope phase. Other companions at larger distances
may be present, but have so far proved elusive to observation (e.g.,
Hrivnak et al.\ \cite{hrivnak11}).

The objects in our sample of AGB stars with high mass loss rates are
expected to be the immediate precursors of PNe. We find that
$\sim$20\% of the envelopes already show significant shaping effects
that we attribute to the presence of substantial companions. We also
find evidence for additional, more subtle shaping effects in the form
of polar cores in up 50\% of envelopes, which also probably arise from
companions (including low mass companions).  These numbers are
preliminary because the sample size is small, but they demonstrate
that detailed observations of AGB envelopes have the potential to
characterize aspects of the interactions that contribute to the
shaping of PNe.

\section{Conclusions}

We have carried out a deep imaging survey of 22 AGB stars with high
mass loss rates in order to investigate the geometry of their
circumstellar envelopes. We report the detection of 15 envelopes in
dust-scattered Galactic radiation, and we characterize their
properties in terms of the surface brightness, radial intensity
profile, and observed shape.

We find that the surface brightness of the envelopes shows a rapid
decrease with Galactic radius, which we interpret as a steep gradient
in the interstellar radiation field. As far as we know, this is the
most direct observation of this gradient in the solar neighborhood.

The envelopes show a range of
geometries: approximately half are close to spherically symmetric, and
$\sim$20\% are distinctly elliptical.  We interpret the shapes in
terms of populations of single stars and binaries whose envelopes are
flattened by a companion. The observed distribution of the
ellipticities of the envelopes is qualitatively consistent with the
results of population synthesis models. We also find that
approximately half the sample of envelopes exhibit small-scale, polar
cores which we interpret as the escape of starlight through polar
holes produced by companions.

Our observations of envelope flattening and polar holes point to a
hidden population of binary companions within the circumstellar
envelopes of AGB stars. These companions affect the geometry of the
envelopes on the AGB, and are potentially important guides to the
evolutionary channels that lead to binaries observed in the post-AGB
phase.

\begin{acknowledgements}
 
We acknowledge the use of the MIDAS software from ESO which was used
for the data processing.  The HST data were obtained from the ESA/ESO
ST-ECF Archive Center at Garching, Germany.  This work was supported
in part by NSF grant AST 08-06910 (to PJH).

\end{acknowledgements}




\Online
\begin{appendix} 

\section{Envelope non-detections}

Images of the fields for which the AGB star but not the envelope are
detected are shown in Fig.~A-1. IRAS 06012+0726 (AFGL~865) is not
detected in $V$ or $I$, and is not shown. It is however well detected in
J, H, and K in the 2MASS survey, with $J=13.38$, $H=9.28$, and
$K=6.06$. At a Galactic latitude $b=-7$, this object has the largest
extinction ($A_{\rm V} \sim 1.6$ mag.) based on the NED extinction
calculator, which probably accounts for the non-detection at short
wavelengths.


\begin{figure*}[!ht]
\vspace{0.3cm}
\centerline{
\includegraphics[width=0.27\linewidth]{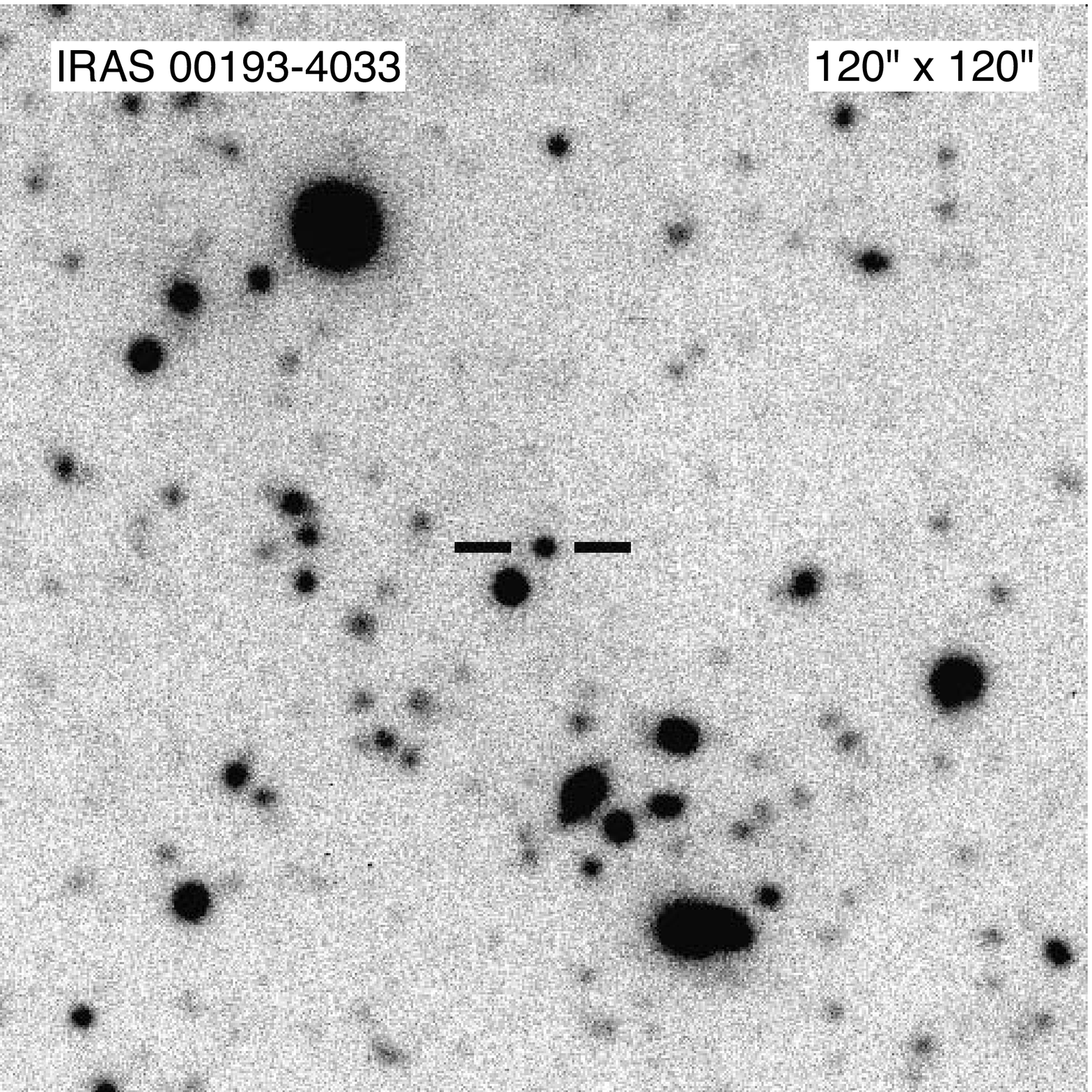}\hspace{0.3cm}
\includegraphics[width=0.27\linewidth]{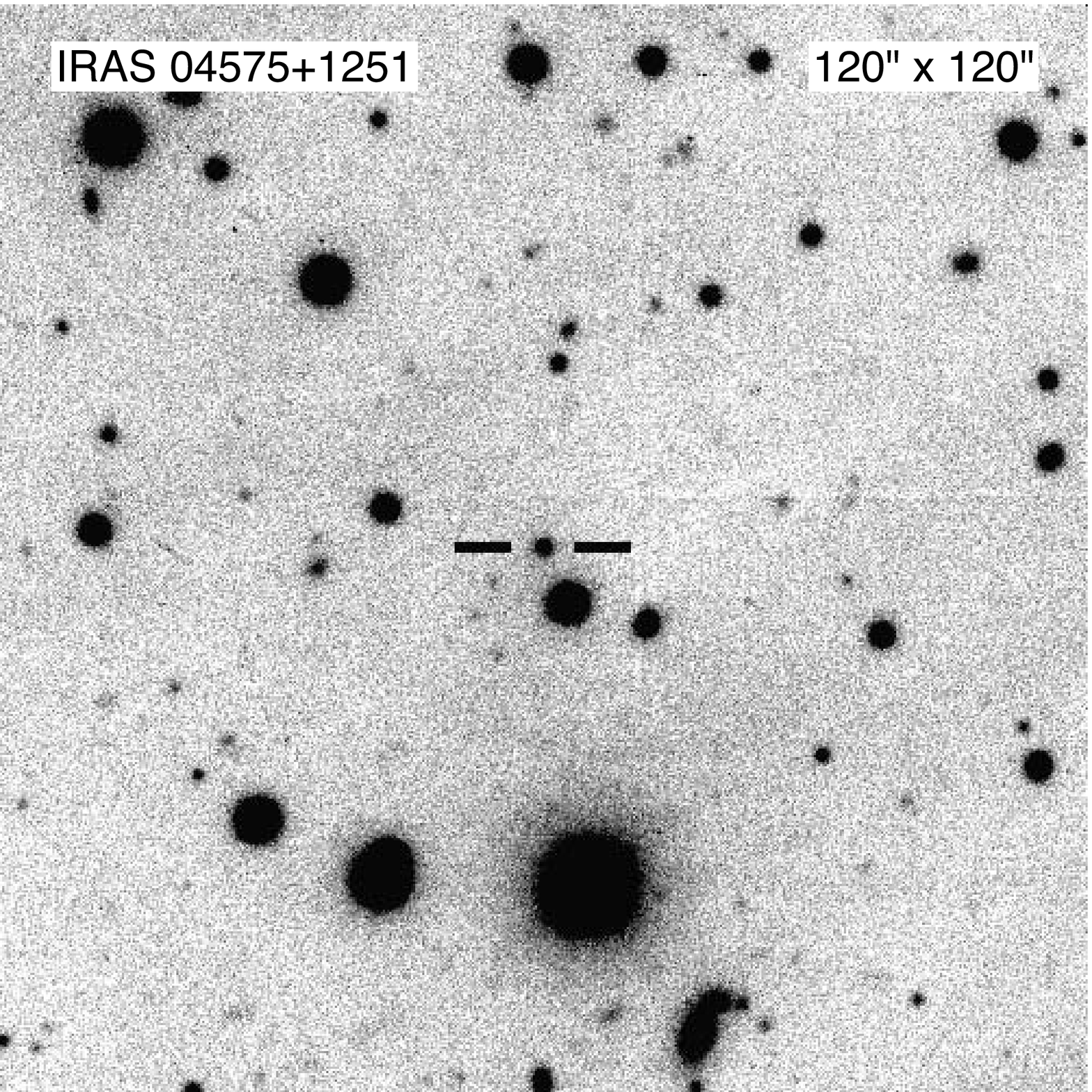}\hspace{0.3cm}
\includegraphics[width=0.27\linewidth]{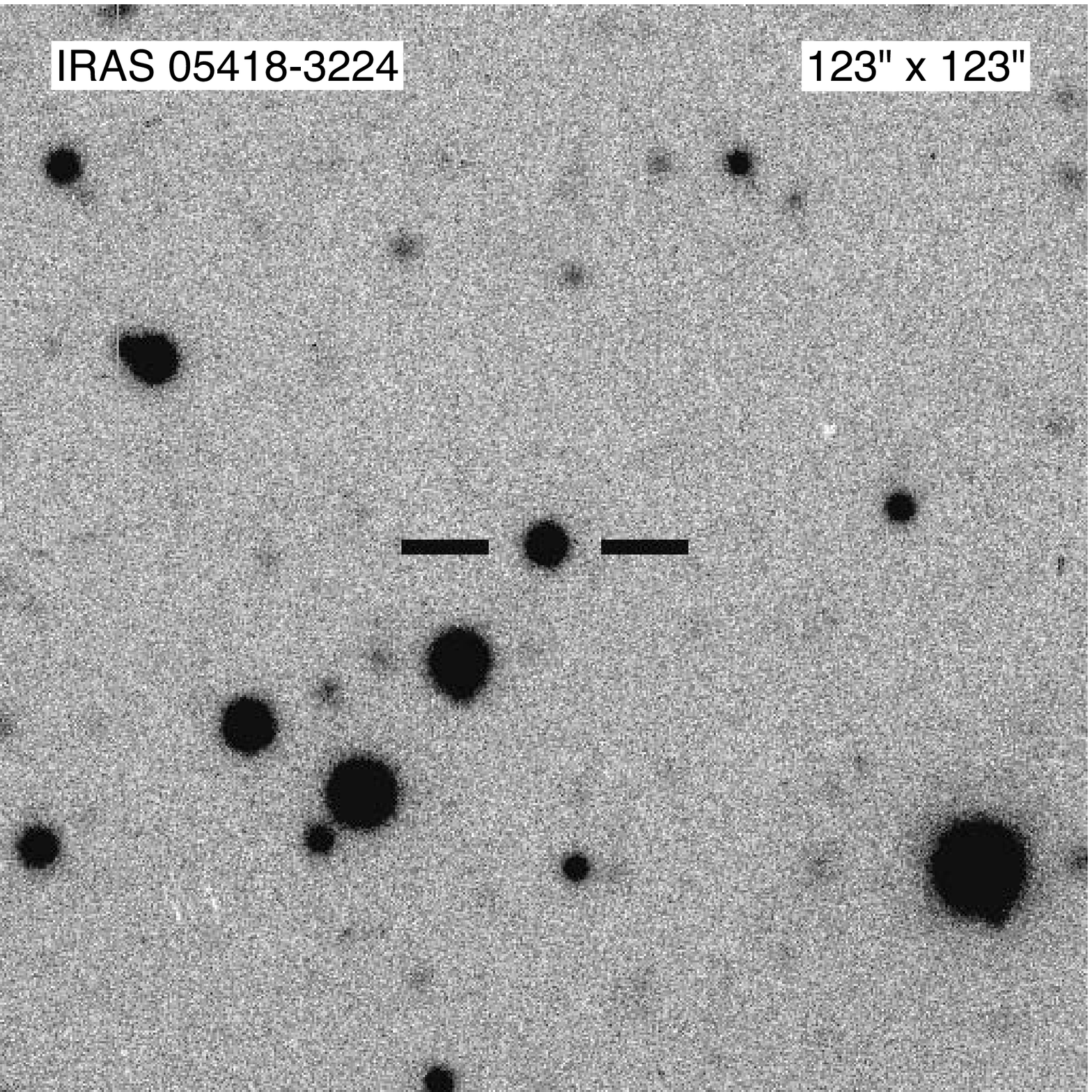} }
\vspace{0.6cm}
\centerline{
\includegraphics[width=0.27\linewidth]{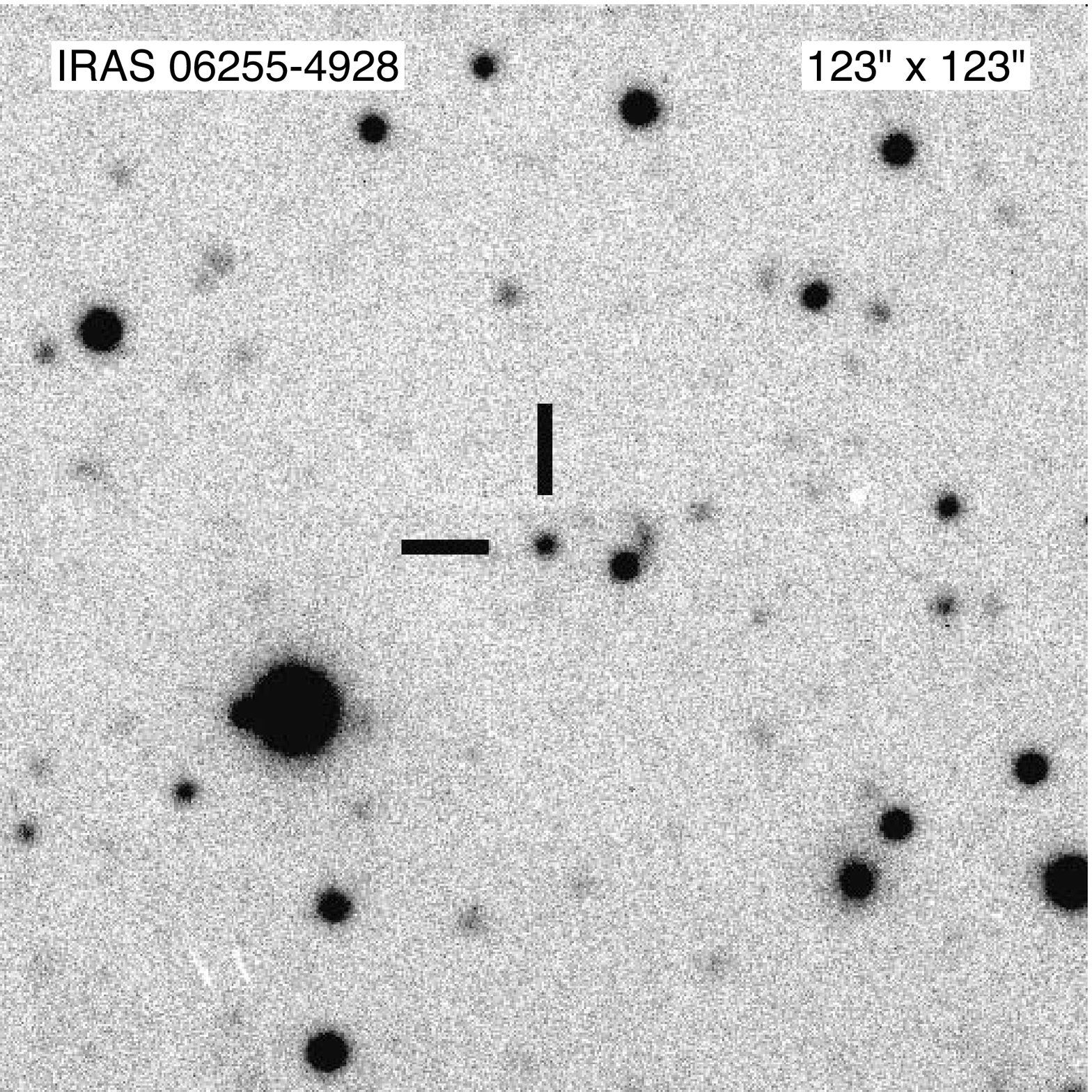}\hspace{0.3cm}
\includegraphics[width=0.27\linewidth]{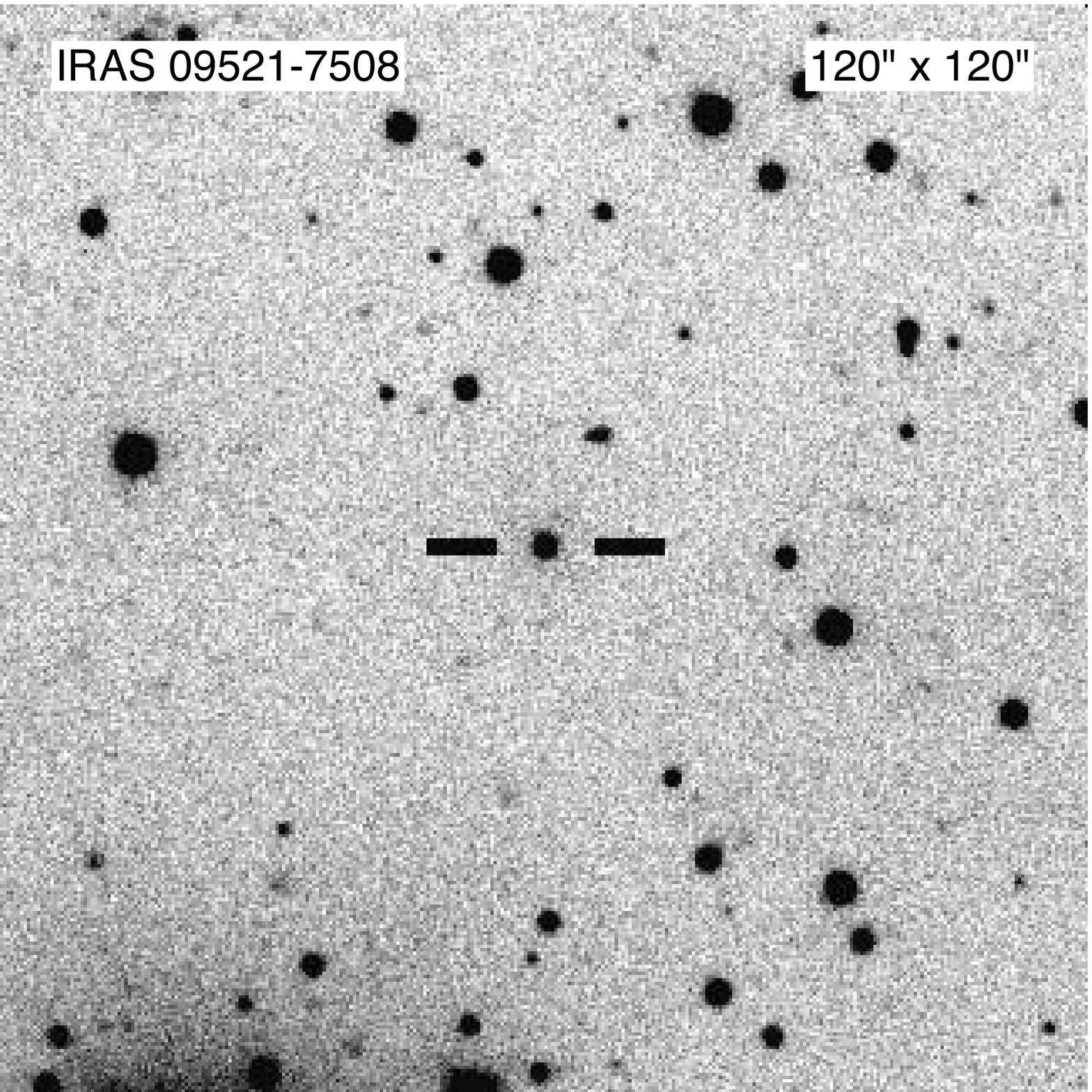}\hspace{0.3cm}
\includegraphics[width=0.27\linewidth]{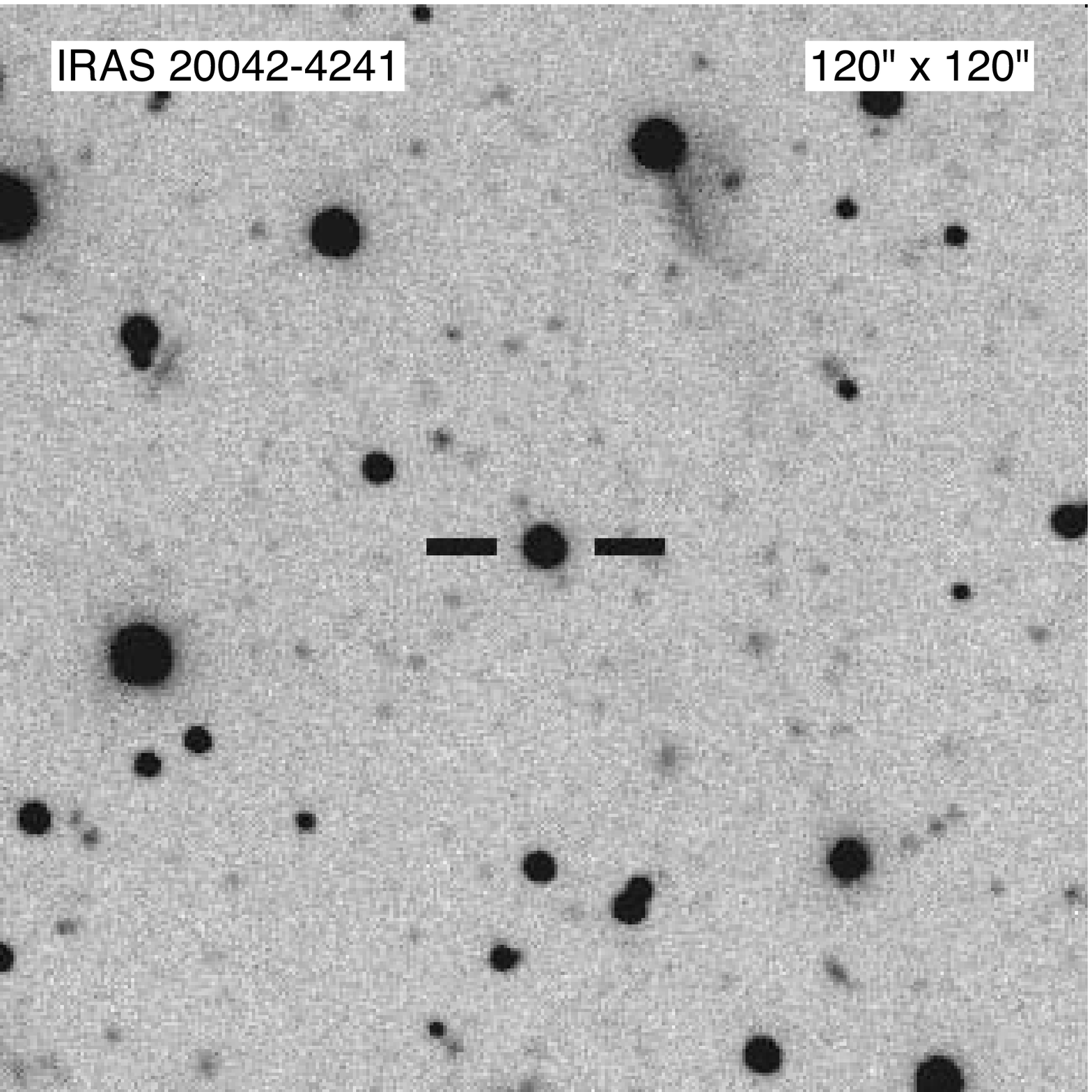} }
\vspace{0.3cm}
\caption { Images of objects with non-detection of the envelope. Ticks
indicate the AGB star.
{\em Top row, left to right}:
IRAS 00193$-$4033 (AFGL 5017), EFOSC2 $V$-band image.
IRAS 04575$+$1251 (AFGL 5134), EFOSC2 $V$-band image.
IRAS 05418$-$3224, EFOSC2 $B$-band image.
{\em Bottom row, left to right}: 
IRAS 06255$-$4928, EFOSC2 $B$-band image.
IRAS 09521$-$7508, EMMI $B$-band image.
IRAS 20042$-$4241, EMMI $B$-band image.}
\end{figure*}

\end{appendix}

\end{document}